\documentclass[paper]{agujournal2019}
\usepackage{url}
\usepackage{lineno}
\usepackage{color}
\usepackage{amsmath,amssymb}
\draftfalse

\newcommand{\ssr}{   {Space Sci. Rev. }}
\newcommand{\pre}{   {Phys Rev. E }}
\newcommand{\jgr}{   {J. Geophys. Res.}}
\newcommand{\grl}{   {Geophys. Res. Lett.}}

\newcommand{\apjl}{   {Astrophys. J. Lett.}}

\newcommand{\aap}{   {Astronomy \& Astrophysics}}

\newcommand{\blue}{\textcolor{black}}

\journalname{JGR: Space Physics}

\begin{document}


\title{Temporal Scales of Electron Precipitation Driven by Whistler-Mode Waves}

\authors{Xiao-Jia Zhang \affil{1,2}, Vassilis Angelopoulos\affil{2}, Anton Artemyev\affil{2}, Didier Mourenas\affil{3,4}, Oleksiy Agapitov\affil{5}, Ethan Tsai\affil{2}, Colin Wilkins\affil{2}}
\affiliation{1}{Department of Physics, University of Texas at Dallas, Richardson, TX, USA}
\affiliation{2}{Department of Earth, Planetary, and Space Sciences, University of California, Los Angeles, USA}
\affiliation{3}{CEA, DAM, DIF, Arpajon, France}
\affiliation{4}{Laboratoire Mati\`ere en Conditions Extr\^emes, Universit\'e Paris-Saclay, CEA, Bruy\`eres-le-Ch\^atel, France}
\affiliation{5}{Space Science Laboratory, University of California, Berkeley, USA}

\correspondingauthor{Xiao-Jia Zhang}{xjzhang@utdallas.edu}

\begin{keypoints}
\item Temporal scales of electron precipitating flux variations are studied with ELFIN CubeSats
\item Main patterns include microbursts of $\sim 1$s, modulated precipitation of $\sim 1$min, and equatorial parameter variations of $\sim 10$min
\item Possible explanations of such time scales are discussed
\end{keypoints}

\begin{abstract}
Electron resonant scattering by whistler-mode waves is one of the most important mechanisms responsible for electron precipitation to the Earth’s atmosphere. The temporal and spatial scales of such precipitation are dictated by properties of their wave source and background plasma characteristics, which control the efficiency of electron resonant scattering. We investigate these scales with measurements from the two low-altitude ELFIN CubeSats that move practically along the same orbit, with along-track separations ranging from seconds to tens of minutes. Conjunctions with the equatorial THEMIS mission are also used to aid our interpretation. We compare the variations in energetic electron precipitation at the same $L$-shells but on successive data collection orbit tracks by the two ELFIN satellites. Variations seen at the smallest inter-satellite separations, those of less than a few seconds, are likely associated with whistler-mode chorus elements or with the scale of chorus wave packets ($0.1 - 1$ s in time and $\sim$ 100 km in space at the equator). Variations between precipitation $L$-shell profiles at intermediate inter-satellite separations, a few seconds to about 1 min, are likely associated with whistler-mode wave power modulations by ultra-low frequency (ULF) waves, i.e., with the wave source region (from a few to tens of seconds to a few minutes in time and $\sim$1000 km in space at the equator). During these two types of variations, consecutive crossings are associated with precipitation $L$-shell profiles very similar to each other. Therefore the spatial and temporal variations at those scales do not change the net electron loss from the outer radiation belt. Variations at the largest range of inter-satellite separations, several minutes to more than 10 min, are likely associated with mesoscale equatorial plasma structures that are affected by convection (at minutes to tens of minutes temporal variations and $\approx$[10$^3$,10$^4$] km spatial scales). The latter type of variations results in appreciable changes in the precipitation $L$-shell profiles and can significantly modify the net electron losses during successive tracks. Thus, such mesoscale variations should be included in simulations of the radiation belt dynamics.
\end{abstract}

\section{Introduction}
Energetic electron scattering by whistler-mode waves is the main mechanism responsible for sub-MeV electron flux depletion deep inside the Earth's inner magnetosphere \cite{Millan&Thorne07,Thorne21:AGU}. There are two main regimes of such scattering: quasi-linear diffusion by low-amplitude, low-coherence waves \cite{bookLyons&Williams,bookSchulz&anzerotti74} and nonlinear resonant transport by intense, highly coherent waves \cite{Hsieh&Omura17:radio_science,Hsieh20,Artemyev22:jgr:DF&ELFIN}. Such a classification in two distinct regimes is oversimplified. Actual wave-particle interactions are more complex, including, e.g., non quasi-linear diffusion by intense waves \cite{Karpman&Shklyar75,Shklyar21}, destruction of nonlinear resonant interactions due to wave-packet amplitude, frequency or phase modulation \cite{Tao13,Mourenas18:jgr,Hiraga&Omura20,An22:Tao,Gan22} and decoherence \cite{Zhang20:grl:frequency,Zhang20:grl:phase}, and breach of the quasi-linear diffusion approximation by highly coherent \cite{Allanson22}, intense \cite{Karimabadi92,Cai20:broading} waves. Many of these effects have been predicted and described theoretically \cite<see reviews by>[and references therein]{Shklyar09:review,Albert13:AGU,Artemyev18:cnsns} based on near-equatorial whistler-mode wave measurements, but their contribution to electron losses relative to classical quasi-linear diffusion and nonlinear resonant transport cannot be easily evaluated from electron flux measurements around the equator, due to the smallness of the loss cone there. However, the temporal and spatial properties of the precipitation resulting from the aforementioned scattering regimes and processes can be assessed from low-altitude satellite measurements, and can be used to differentiate them from each other, as well as to assess their relative contribution to electron losses. There exist three typical spatio-temporal scales of near-equatorial whistler-mode wave-electron resonant interactions: (i) $\sim 10^2$ km, $<1$s scale variations of individual coherent wave-packets, or chorus elements \cite{Santolik04,Agapitov10AnGeo,Agapitov17:grl,Turner17:scales}; (ii) $\sim 10^3$ km, $\sim 1$min scale variations of whistler-mode wave source regions \cite{Agapitov18:correlations}; and (iii) $\sim 10^4$km, $\geq$ a few min scale variations of equatorial gradients of plasma sheet density, and flux of injected electrons \cite<e.g., typical injection scale>[]{Tao11,Runov11jgr}. At a low-altitude, polar-orbiting spacecraft, moving at $\approx$8 km/s, microburst precipitation of a sub-second duration (assumed spatial) would map to a $\lesssim 10^2$ km structure at the equator \cite{OBrien04,Douma19}. Comparison of such measurements at two low-altitude spacecraft with time-resolution of seconds, full $L$-shell coverage and time-separation of a few seconds to many minutes, can reveal equatorial variations of precipitation on scales from a few hundred km to several $\times 10^4$km (depending on $L$-shell), and establish statistical or multi-case study trends of their temporal variations from seconds to minutes. In this study, we use such dual, low-altitude satellite measurements from the ELFIN mission \cite{Angelopoulos20:elfin}. A variety of time separations allow us to investigate the variations of whistler-mode wave-driven electron precipitation on these three spatio-temporal scales.

More specifically, we expect that the smallest spatio-temporal scale of electron precipitation corresponds to the coherence scale of a chorus element \cite{Santolik&Gurnett03,Agapitov17:grl,Turner17:scales}. Chorus waves, nonlinearly generated whistler-mode waves \cite{Nunn05,Demekhov11,Tao20,Omura21:review}, propagate in the form of wave packets forming elements with increasing (rising tone) or decreasing (falling tone) frequency. The duration of such elements does not exceed one second \cite{Teng17,Teng19}, whereas the cross-field coherence of the wave field in such elements does not exceed a few hundreds of km (note that low-altitude spacecraft would cross this scale within less than one second). Thus, the time scale of the resultant microburst precipitation should be within one second, which can only be distinguished by two low-altitude, polar orbiting spacecraft that are separated by a sub-second time difference \cite{Blake&OBrien16,Shumko20}. On the other hand, chorus elements are often generated in series, one after another, and their properties often do not vary significantly from element to element within a few seconds \cite<e.g.,>[]{Santolik03:storm,Li11:grl,Demekhov17}. Therefore, similar bursty precipitation patterns (microburst-type) can be seen by spacecraft separated by a few seconds. In this study, we will examine the pitch-angle (precipitating and trapped fluxes) and energy distributions of such second-scale precipitation events using the highest temporal resolution of ELFIN, a spin-sector cadence of $\sim$0.175 s, at full energy resolution.

We hypothesize that the spatial scale of the whistler-mode wave equatorial excitation (source) region \cite{Agapitov10AnGeo,Aryan16}, from one to a few thousand km, which varies on time scales of minutes (from tens of seconds to several minutes), should be recognizable as 3-30 s precipitation structures which would fluctuate in consecutive precipitation spectra on ELFIN separation time-scales of the order of (but no shorter than) tens of seconds (say, 1 - 10 min). The temporal and spatial variations of the wave source region are mainly attributed to the modulation of the wave's free energy source (anisotropy in the electron velocity distribution) by ULF waves (having wavelengths below $1 R_E$ and periods below a few minutes, as discussed e.g., in \citeA{Li11:modulation1,Li11:modulation2,Watt11,Xia16:ulf&chorus,LiLi22}). Accordingly, we anticipate observations of a stable electron precipitation pattern at ELFIN on a spatial scale of a few to a few tens of seconds, and with similar profiles and intensity on consecutive passes separated by less than 30 seconds. However such profiles of precipitating fluxes should vary (potentially in absolute intensity, location or extent) temporally (from one crossing to the next), in events as the inter-satellite separation increases from 0.5 min to a few minutes. In this study, we will confirm our hypothesis that such 3-30s long, minute-scale precipitation pattern variations are indeed consistent with observations by the two ELFIN satellites.

Whistler-mode chorus waves are typically generated by perpendicularly anisotropic electrons produced by injections \cite{Tao11,Fu14:radiation_belts,Zhang18:whistlers&injections,Frantsuzov22:pop}. The spatial scale of injected electrons is about $\sim 10^4$ km \cite{Nakamura04,Liu13:DF}. These injections also cause large density variations in the background plasma sheet and plasmaspheric plasma \cite{Malaspina14,Malaspina15} on a similar spatial scale. Because the injections are spatially localized, the anisotropic electron flux and associated waves, as well as the ambient density have significant gradients, which will be also manifest in the regions of resonant wave-electron interactions and scattering, affecting the precipitation patterns. The aforementioned spatial structures map to a time scale of precipitation lasting about $\sim 0.5-1.5$ min on a polar orbiting satellite at low altitudes, depending on $L$-shell. Moreover, these injected electrons can persist for a long time, slowly drifting azimuthally towards the dawn \cite{Tao11,Turner17:injection}. Based on the duration of such injections and density variations at the equator (a few minutes), we expect that the region of associated electron precipitation can also be repeated at successive ELFIN passes with inter-satellite separations of a few minutes but no longer. At longer separations, we expect the precipitation patterns of such duration to change significantly. In this study, we confirm this interpretation using examples of such long-lived regions of precipitation and conjunctions with equatorial THEMIS satellites.

The remainder of this paper is structured as follows: Section \ref{sec:observations} describes ELFIN instruments and criteria for event selection; Section \ref{sec:small} shows two examples of consecutive observations by ELFIN satellites with second-scale separations; Section \ref{sec:medium} shows an example of consecutive observations at minute-scale separations, in conjunction with near-equatorial measurements of whistler-mode waves; Section \ref{sec:large} shows observations of ELFIN CubeSats with $\sim$ ten-minute-scale separations, in conjunction with near-equatorial measurements of whistler-mode waves; Section \ref{sec:discussion} discusses the results and summarizes our main findings. Although the main text only includes detailed analysis of a few representative events, the Supporting Information (SI) includes analysis of multiple similar events in each category, supporting our conclusions.

\section{ELFIN Observations}
We examine ELFIN A and B observations with inter-satellite time delay ranging from a few seconds to tens of minutes. The entire list of such events can be found in the Supporting Information, whereas in this section we show typical examples of electron precipitation patterns.
\subsection{Dataset and instruments}\label{sec:observations}
The two identical ELFIN CubeSats, A and B, are equipped with energetic particle detectors (EPD). Here we use the electron detector (EPDE) measurements \cite{Angelopoulos20} covering the $[50, 5000]$ keV energy range with an energy resolution of $\Delta E/E\sim 40\%$ (16 channels), a pitch-angle resolution of $\Delta\alpha\approx 22.5^\circ$, at $\approx$2.8s temporal resolution. \blue{The pitch-angle resolution of $\Delta\alpha\approx 22.5^\circ$ (FWHM) is sufficient to separate the locally trapped (between the local loss-cone angle $\alpha_{LC}\sim65^\circ$ at the ELFIN altitude and $\alpha=90^\circ$) electrons from precipitating (at local pitch-angles $\alpha< \alpha_{LC}\sim65^\circ$) electrons \cite{Angelopoulos20}.} We use three standard data products from this instrument: the spectrum of locally trapped (outside the local bounce loss cone) electron fluxes $j_{trap}(E)$, the spectrum of precipitating (moving toward the ionosphere within the local bounce loss cone) electron fluxes $j_{prec}(E)$, and precipitating-to-trapped flux ratio $j_{prec}/j_{trap}$. The latter ratio gives a natural and physically meaningful measure of the precipitation rate. All events analyzed in this study are characterized by precipitation of 50 keV electrons (as well as of higher energy electrons), indicative of electron scattering into the loss cone by resonant interaction with whistler-mode waves \cite<see details of analysis of whistler-mode driven precipitation events observed by ELFIN in, e.g.,>[]{Mourenas21:jgr:ELFIN,Mourenas22:jgr:ELFIN, Artemyev21:jgr:ducts, Tsai22}.

\subsection{Examples of small-separation events}\label{sec:small}
\begin{figure*}
\centering
\includegraphics[width=0.9\textwidth]{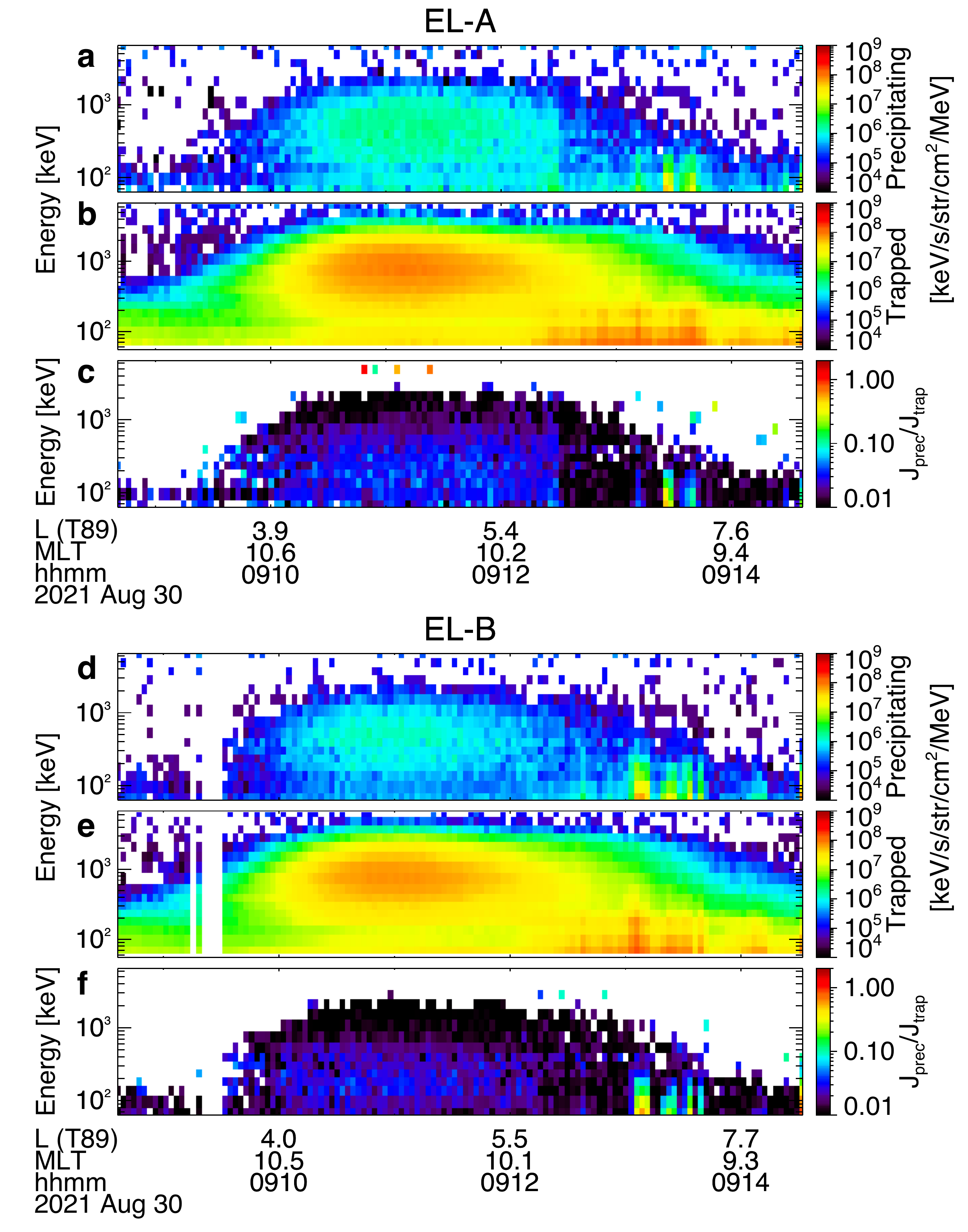}
\caption{The overview of ELFIN A and B observations with a $5.4$s time separation. Panels (a,d) show precipitating electron fluxes, Panels (b,e) show trapped fluxes, Panels (c,f) show the precipitating-to-trapped flux ratio. \label{fig1}}
\end{figure*}

This subsection describes two events with a few seconds of separation between the two ELFIN spacecraft. Figure \ref{fig1} shows an overview of the first event: Panels (a,d) show precipitating electron fluxes, Panels (b,e) show trapped electron fluxes, and Panels (c,f) show the precipitating-to-trapped flux ratio. ELFIN spacecraft move from low $L$-shell to higher $L$-shell at the dayside, and observe the outer radiation belt between 09:10-09:14 UT. This region is characterized by high trapped electron fluxes with energies reaching $3-4$ MeV at $L\in[4,6]$. At the outer edge of the outer radiation belt, $L\sim 7$, both ELFIN A \& B observe bursts of electron precipitation with a time-scale of about one or two ELFIN spins ($\sim 3-6$s) and typical energies below $300$ keV. The precipitating-to-trapped flux ratio reaches $\sim 0.3$ at $\lesssim$100 keV, consistent with intense precipitation, but still below the strong diffusion limit \cite{Kennel69}. The ratio decreases monotonically with energy above 100 keV. These precipitation bursts are recorded over a region of $\Delta L\sim 0.3 - 0.8$, corresponding to the typical scale of plasma sheet injection drifting to the dayside \cite{Tao11}, but each individual burst size is $\Delta L\sim 0.025-0.08$ (and some may be even smaller, because ELFIN's full pitch-angle distribution cadence is limited by its spin; though, sub-spin observations of similar precipitation bursts can be made with examination of individual spin sectors as in \citeA{Zhang22:microbursts}). Thus, the spatial scale of individual precipitation bursts ($160-500$ km, and occasionally less) is comparable to the chorus wave element size \cite{Agapitov17:grl, Agapitov18:correlations}. Based on their duration, these bursts would qualify to be called \blue{low-energy} microburst precipitation. \blue{Such precipitation bursts could be caused by cyclotron resonance interaction with intense, quasi-parallel chorus waves near the equator. But on the dayside during moderately disturbed periods as here (with $Kp=1.3$), field-aligned chorus waves often propagate with relatively mild damping to middle (and even high) latitudes \cite{Meredith12,Agapitov18:jgr}, where cyclotron resonance with intense waves can cause efficient precipitation (with $J_{prec}/J_{trap}>0.1$) of relativistic electrons up to $\sim300-500$ keV \cite{Zhang22:microbursts}. However, during the observed precipitation bursts, efficient precipitation (with $J_{prec}/J_{trap}>0.1$) is limited to low energies ($<200$ keV). It suggests that these precipitation bursts could be} driven by electron scattering via Landau resonance with very-oblique chorus waves, \blue{which is expected to result in a similar low-energy} microburst precipitation \cite{Zhang22:natcom,Artemyev22:jgr:Landau&ELFIN}.

\begin{figure*}
\centering
\includegraphics[width=0.9\textwidth]{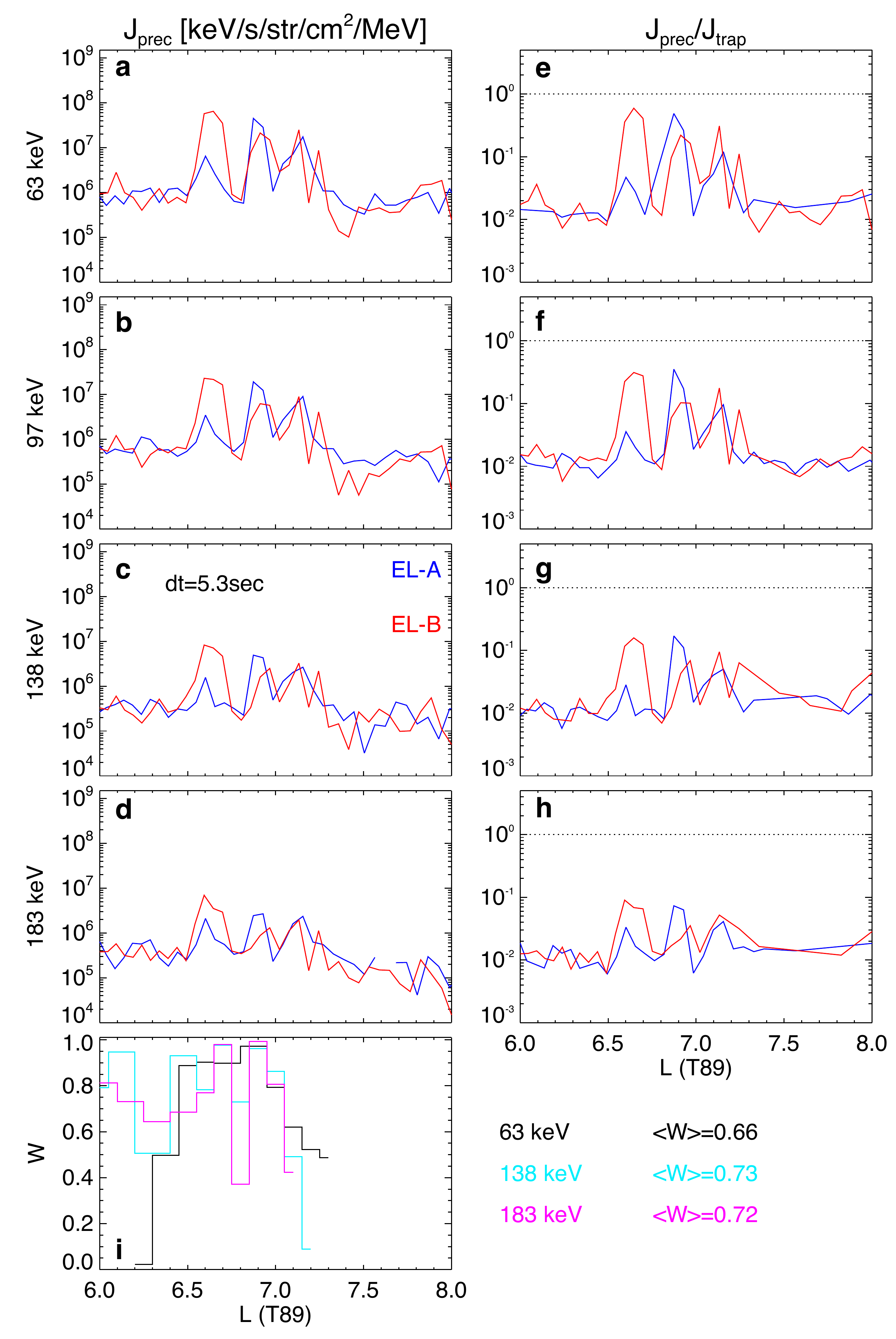}
\caption{(a-d) Precipitating electron fluxes versus $L$-shell for the event from Fig. \ref{fig1}. Four energy channels are shown for ELFIN A (blue) and B (red). (e-h) Same as (a-d) but showing the precipitating-to-trapped flux ratio. (i) Fraction $W$ of precipitating flux variation between ELFIN A \& B measurements due to variation of chorus wave intensity, for selected energies (colors) indicated on the right with the corresponding average $\langle W\rangle$ values. \label{fig2}}
\end{figure*}

To further explore the properties of the precipitation quantitatively, we compare, in Figure \ref{fig2}, the precipitating electron fluxes and precipitating-to-trapped flux ratios measured by ELFIN A and B. We organize the ELFIN A (blue) \& B (red) measured fluxes as a function of $L$-shell. ELFIN A \& B cross the same $L$-shell with just a small and constant time delay of 5.4s. Individual precipitation bursts are clearly seen at both spacecraft, but the precipitating flux magnitude and the precipitating-to-trapped flux ratio are different at the two spacecraft, even with such small time-separation (a few seconds). We examine below the possible causes of these variations.

During the very short time elapsed between the ELFIN A and B measurements at the same location, the cold plasma density is expected to remain unchanged, as well as the latitude of Landau (for very oblique waves) or cyclotron (for quasi-parallel waves) resonance with electrons of a given energy \cite{Kennel&Petschek66, Li13:POES, Mourenas21:jgr:ELFIN, Mourenas22:jgr:ELFIN}. \blue{Note that for nonlinear Landau resonance (the phase trapping effect) the {\it resonance latitude} means the latitude of escape from the resonance, but not the latitude of trapping, because energies of precipitating electrons and latitudes where electrons appear within the loss-cone can be much higher than energies/latitudes of trapping into the resonance \cite<e.g.,>{Agapitov15:grl:acceleration,Zhang22:natcom}. However, for nonlinear Landau resonance, the latitudes of trapping and escape from resonance (as well as initial and final electron energies) are controlled by magnetic field configuration and background plasma density \cite<e.g.,>{Agapitov14:jgr:acceleration,Artemyev22:jgr:Landau&ELFIN} and, thus, are not expected to change within the short time interval between ELFIN A and B measurements. Accordingly, for both cyclotron and Landau resonances,} the chorus wave power at the latitude of resonance is the only quantity affecting the precipitating-to-trapped electron flux ratio $J_{prec}/J_{trap}$, and the latter quantity should be a monotonic function of the former. Therefore, at each $L$, the fraction $W$ of precipitating flux variations $\Delta J_{prec}$ between ELFIN A and B that is directly due to a variation of chorus wave intensity can be evaluated as 
\begin{eqnarray}
W &=& x \;\;\; {\rm for}\,\, 0\leq x\leq 1 , \nonumber\\
W &=& \frac{{ x }}{{ (2x-1) }}\,\, {\rm for}\,\, x<0 ,\,\, {\rm or}\,\, x>1,\label{eq1}\\
{\rm with}\,\, x  &\simeq&{\left( {\frac{{\Delta \left( {J_{prec} /J_{trap} } \right)}}{{J_{prec} /J_{trap} }}} \right)}\cdot \left(\frac{{\Delta J_{prec} }}{{J_{prec} }}\right)^{-1}. \nonumber
\end{eqnarray}
\blue{To obtain this expression of $x$, the total variation of the measured precipitating flux between ELFIN-A and ELFIN-B at the same $L$-shell ($\Delta J_{prec} = J_{prec}(ELFIN-A) - J_{prec}(ELFIN-B)$) has been decomposed into two contributions, $\Delta J_{prec}= \Delta ((J_{prec}/J_{trap})J_{trap}) = \Delta(J_{prec}/J_{trap})\cdot J_{trap} + (J_{prec}/J_{trap})\cdot \Delta J_{trap}$. The first contribution corresponds to the variation of $J_{prec}/J_{trap}$ for a constant $J_{trap}$, i.e., the part of the measured $J_{prec}$ variation which is solely due to variations in the efficiency of wave-induced electron precipitation \cite{Kennel&Petschek66}. The second contribution comes from variations of $J_{trap}$ for a constant $J_{prec}/J_{trap}$, i.e., for a constant efficiency of wave-induced electron precipitation. The ratio of the first contribution to the measured total $\Delta J_{prec}$ variation can be rewritten as}
\blue{
\begin{equation}
\frac{{ \Delta(J_{prec}/J_{trap})\, J_{trap} }}{{ \Delta J_{prec} }} = \frac{{ \Delta(J_{prec}/J_{trap})\, J_{prec} }}{{ \Delta J_{prec}\, (J_{prec}/J_{trap}) }} = x.
\label{eq2}
\end{equation}
Finally, the sum of the absolute values of all contributions to the variation of $J_{prec}$ can be written as $S=|\Delta(J_{prec}/J_{trap})J_{trap}| + |\Delta J_{prec}-\Delta(J_{prec}/J_{trap})J_{trap}|$. Accordingly, the ratio of the contribution from $|\Delta(J_{prec}/J_{trap})J_{trap}|$ over the sum of all contributions is $W=x$ for $0\leq x\leq1$ and $W=x/(2x-1)$ for $x<0$ or $x>1$, as in Equation (\ref{eq1}). $W$ is a quantitative measure of} the portion of $J_{prec}$ variations that is solely attributable to $J_{prec}/J_{trap}$ variations, which are themselves controlled by wave intensity variations. Thus, $W$ shows how significantly $J_{prec}$ varies due to the sole changes in wave intensity. The remaining portion $(1-W)$ of $J_{prec}$ variations between ELFIN A and B is due to a variation of trapped electron fluxes $J_{trap}$ during the event. For a constant $J_{trap}$, the entire variation of the precipitating flux is due to variations in the wave intensity: when $x=1$, $W\to 1$. For a constant $J_{prec}/J_{trap}$ (corresponding to a fixed wave intensity), the entire variation of $J_{prec}$ is due to $J_{trap}$ variations: when $x=0$, $W\to 0$. 

$W$, plotted as a function of $L$ for select energies in Figure \ref{fig2}(i), has high average values for this event, $\langle W\rangle\sim 0.66-0.73$ at $63-183$ keV. During the main bursts of electron precipitation at $L=6.6-6.7$ and $L=6.9$ in Figures \ref{fig2}(a-c), $W$ reaches even higher values, $W\simeq0.80-0.95$. This implies that such bursts are associated with a stable (at least on a time scale of a few seconds) whistler-mode wave source, where the variation of precipitating electron flux is principally due to a variation of the wave intensity within a time scale of a few seconds. This time scale is too short to be due to whistler wave source modulation by dayside compressional ultra-low-frequency waves \cite<such modulation is observed on time scales of tens of seconds to several minutes, see>[]{Zhang20:jgr:ulf}. Therefore, variations between precipitation profiles at second-scale separations should indeed be attributed to wave power variations within individual chorus elements, produced by the whistler-mode chorus wave generation processes themselves \cite<see discussions of such wave power variations in theories of chorus wave generation, e.g., in>[]{Demekhov17,Nunn21}.   

\begin{figure*}
\centering
\includegraphics[width=0.9\textwidth]{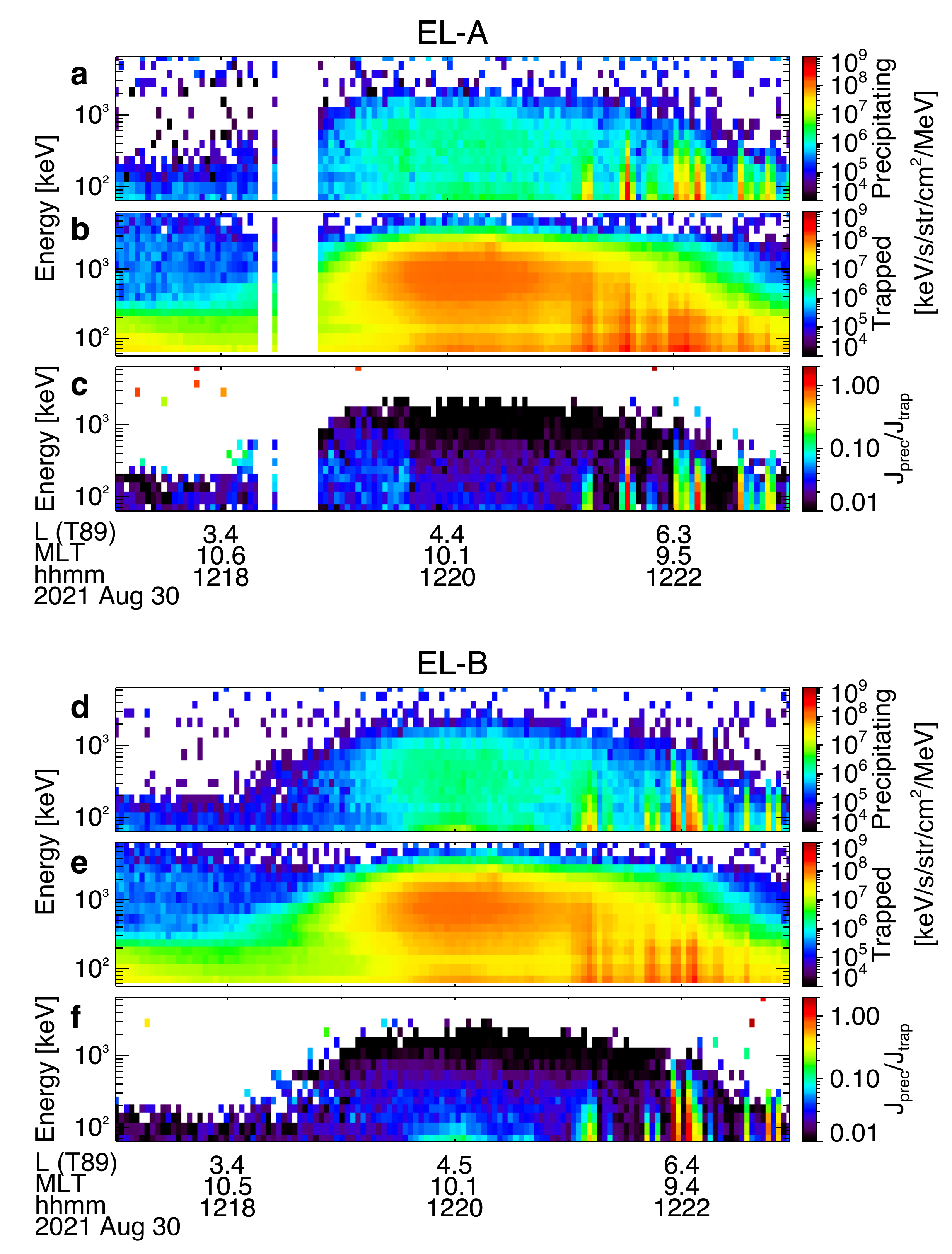}
\caption{The overview of ELFIN A and B observations with $2.8$s separation between spacecraft. Panels (a,d) show precipitating electron fluxes, Panels (b,e) show trapped fluxes, Panels (c,f) show precipitating-to-trapped flux ratio.  \label{fig3}}
\end{figure*}

\begin{figure*}
\centering
\includegraphics[width=0.9\textwidth]{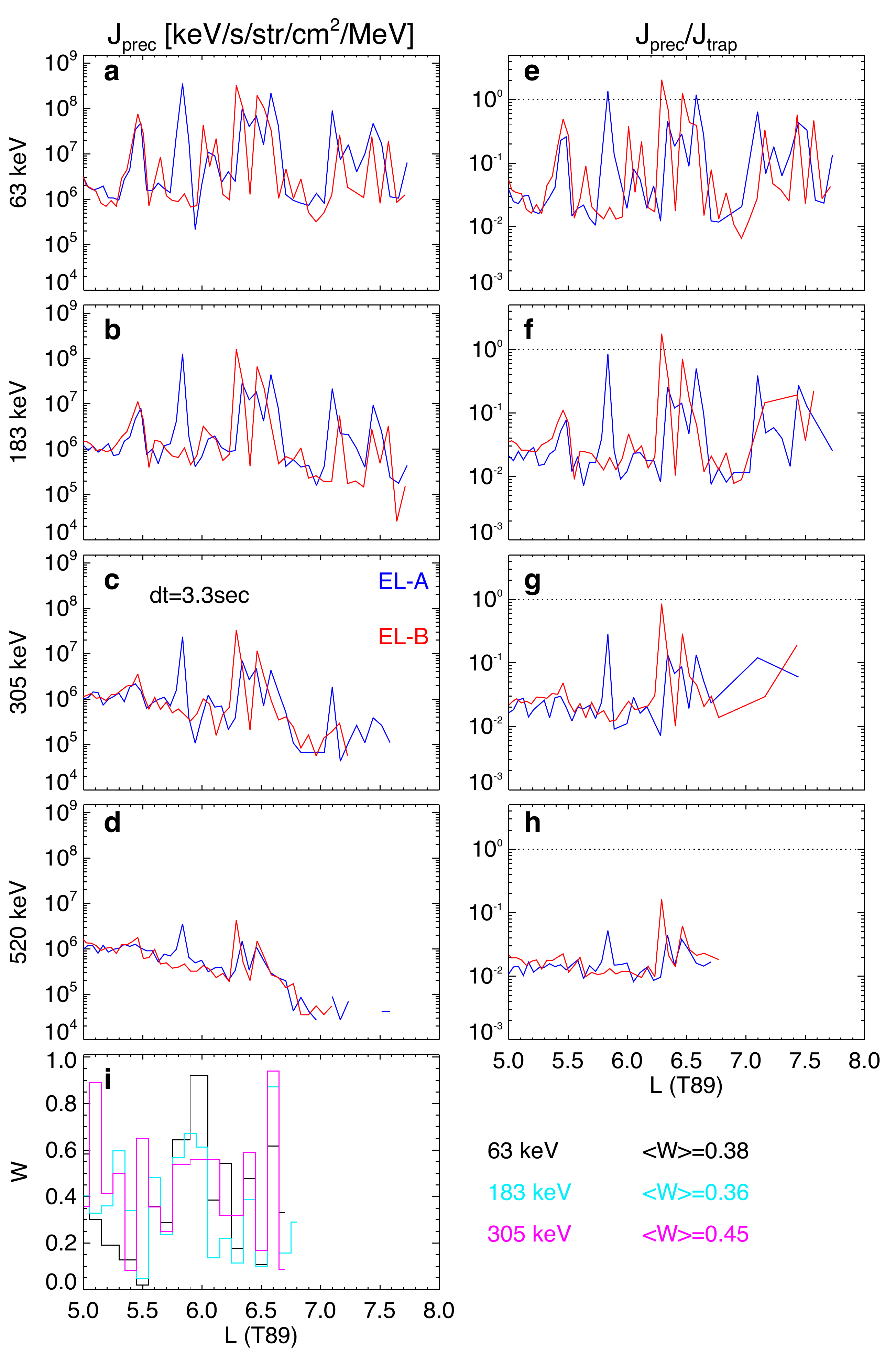}
\caption{ (a-d) Precipitating electron fluxes versus $L$-shell for the event from Fig. \ref{fig3}. Four energy channels are shown for ELFIN A (blue) and B (red). (e-h) Same as (a-d) but showing the precipitating-to-trapped flux ratio. (i) Fraction $W$ of precipitating flux variation between ELFIN A \& B measurements due to variation of chorus wave intensity, for selected energies (colors) indicated on the right with the corresponding average $\langle W\rangle$ values.} \label{fig4}
\end{figure*}

A second event with a small ($\sim 3$s) inter-spacecraft separation is shown in Figures \ref{fig3} and \ref{fig4}, in the same format as Figures \ref{fig1} and \ref{fig2}. Again, the figures depict dayside ELFIN observations exhibiting a series of bursty precipitation outside the plasmapause, at $L>5$, within the outer radiation belt. However, in this event, the energies of precipitating electrons reach 700 keV, i.e., these are relativistic precipitation bursts, resembling ELFIN observations of relativistic microbursts \cite{Zhang22:microbursts}. At the $\sim 3$s inter-spacecraft separation of this event, the spacecraft show similar $L$-shell profiles of precipitating electron fluxes and precipitating flux ratios (Figure \ref{fig4}): one event at $L\sim$5.4 is nearly unchanged after 2.8s, several events at L$>$6 are modified, and one event, at $L\sim$5.8, is only seen on ELFIN A. The fraction $W$ of precipitating electron flux variation between ELFIN A and B that can be ascribed to chorus wave intensity variation is on average lower than during the preceding event, $\langle W\rangle\simeq 0.36-0.45$, but it reaches again high levels $W\sim0.55-0.90$ during $4$ of the main precipitation bursts, at $L\simeq5.8$, $L=6.0$, $L=6.4$, and $L=6.6$ in Figure \ref{fig4}(i).

Based on these two events, we can conclude that bursty electron precipitation can sometimes remain relatively stable over a time-scale of the order of one second \cite{Shumko21}, although that should represent an upper bound for the duration of chorus wave phase coherence. This could be partly due to the $\sim1$s ($\sim$6 spin sector) time-averaging of the measured precipitating flux employed in this paper, which smooths faster variations. This time-averaged precipitating flux should result from wave-particle interactions with at least one full chorus element, including many different \blue{sub-packets.} Therefore, it should be more stable than sub-second bursts produced by individual intense wave packets. It should mainly correspond to scattering by waves with the average amplitude of a chorus rising tone element, within a series of similar elements often lasting a few seconds or more in total \cite{Santolik03:storm, Nunn09, Li11:grl, Demekhov17}. During such time-scale of several seconds, the characteristics of the fine structure of the precipitation pattern measured by ELFIN A \& B, including the multiple bursts distributed in $L\in[5,8]$, indeed remain the same, including the recurrence rate versus $L$ and the energy spectrum of the precipitating electrons. 

\subsection{Example of medium-separation event}\label{sec:medium}
\begin{figure*}
\centering
\includegraphics[width=0.9\textwidth]{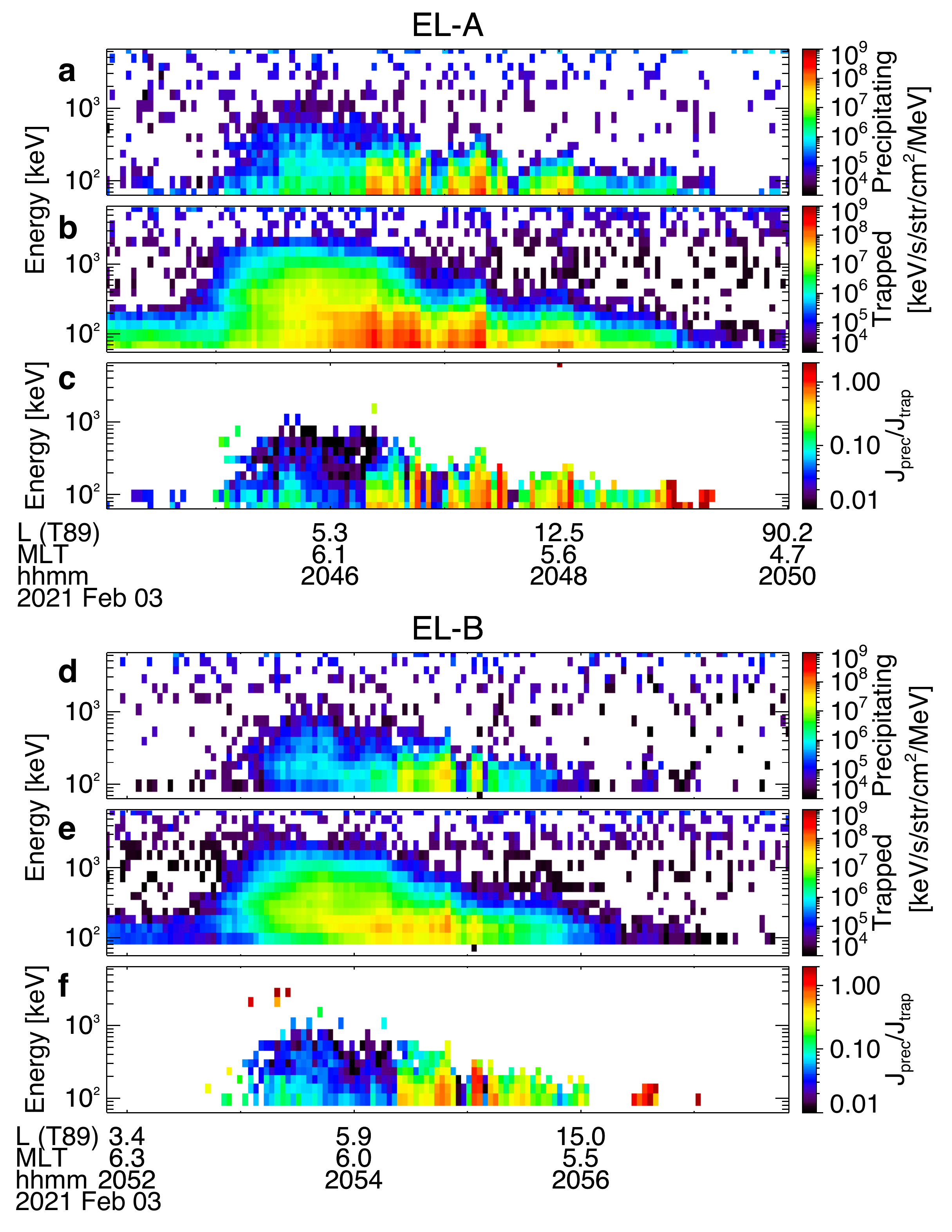}
\caption{Overview of ELFIN A and B observations with $\sim 7.7$min separation between spacecraft. Panels (a,d) show precipitating fluxes, Panels (b,e) show trapped fluxes, Panels (c,f) show precipitating-to-trapped flux ratio. \label{fig5}}
\end{figure*}

The two preceding examples, with only a few seconds of inter-spacecraft separation, established a baseline for further comparisons at larger separations. Figure \ref{fig5} shows an example of such observations with an $\sim8$ min spacecraft separation. Both spacecraft measured multiple precipitation bursts over a significant portion of the outer radiation belt and the inner-edge of the plasma sheet at $L\in[4.5,12]$. The most intense precipitation events are observed at $L>6$. A comparison of ELFIN A and B trapped and precipitating fluxes in this event shows that at $\sim 8$ min delay there is significant evolution of the electron distributions, both above and inside the loss cone: they are both reduced considerably with the elapsed time (at ELFIN B compared to at ELFIN A). However, the energy range of most significant precipitation (based on the precipitation ratio spectrogram in Panels (c,f), does not vary between ELFIN A and B. It has similar profile and remains limited to energies less than $\sim300$ keV.  This suggests that the plasma frequency to gyrofrequency ratio $f_{pe}/f_{ce}$, which controls the latitude of resonance and the energy of precipitating electrons \cite<e.g., see>[]{Agapitov19:fpe, Artemyev22:jgr:Landau&ELFIN, Chen22:microbursts}, probably did not change significantly over $8$ min at most locations.

\begin{figure*}
\centering
\includegraphics[width=0.9\textwidth]{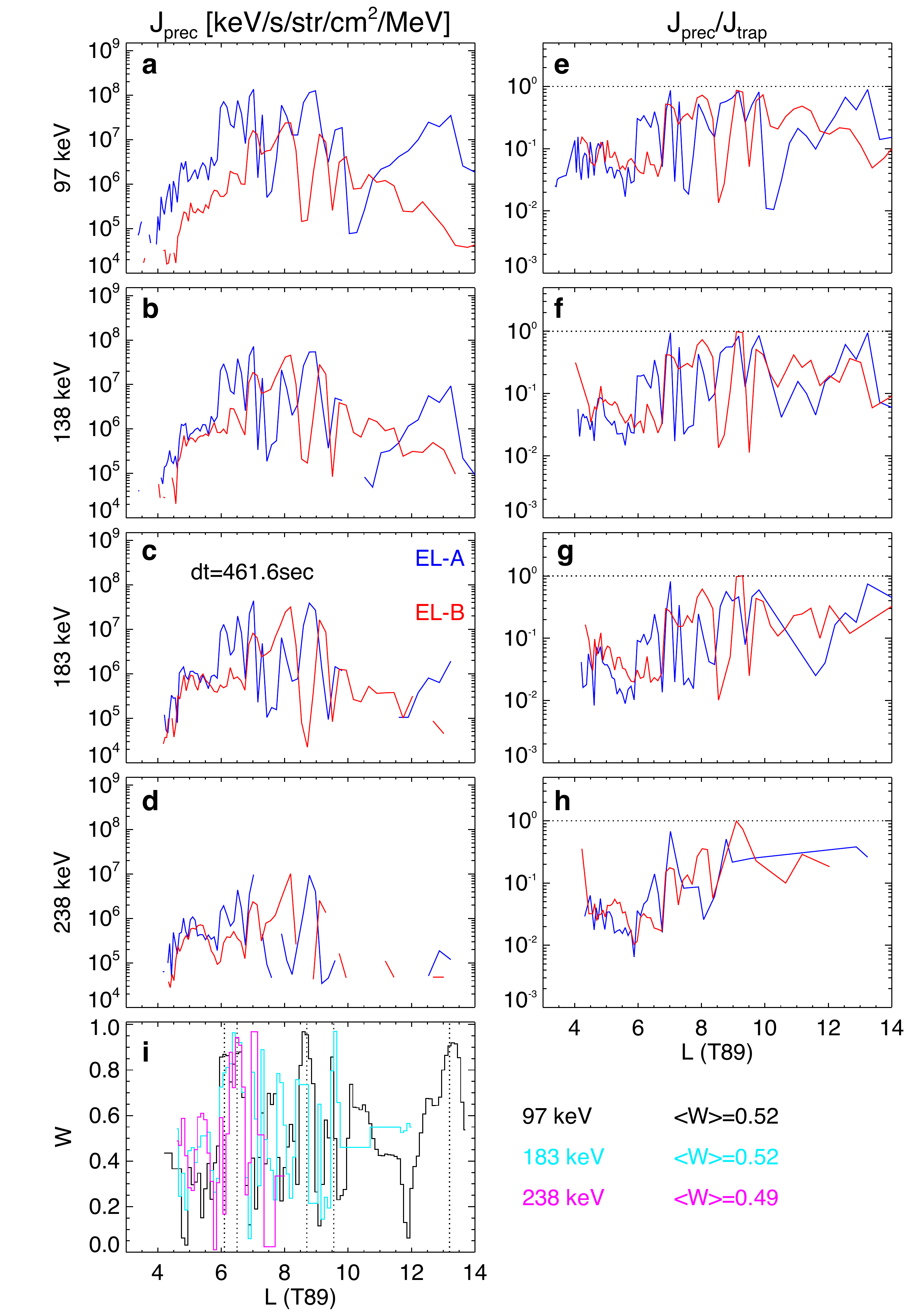}
\caption{(a-d) Precipitating electron fluxes versus $L$-shell for the event from Fig. \ref{fig5}. Four energy channels are shown for ELFIN A (blue) and B (red). (e-h) Same as (a-d) but showing the precipitating-to-trapped flux ratio. (i) Fraction $W$ of precipitating flux variation between ELFIN A \& B measurements due to variation of chorus wave intensity, for selected energies (colors) indicated on the right with the corresponding average $\langle W\rangle$ values. \label{fig6}}
\end{figure*}

Comparing line plots of the precipitating fluxes and precipitating-to-trapped flux ratios at several energies in Figure \ref{fig6}, we see more clearly that the pattern of precipitating fluxes is similar for both spacecraft: spatially quasi-periodic precipitation peaks between $L\in[4.5,14]$. Figure \ref{fig6}(i) further shows that most of these precipitation peaks ($L\sim6.0, 6.5, 8.5, 9.5, 13.$) correspond to a high fraction $W\sim0.7-0.9$ of precipitating flux variation (as opposed to trapped flux variation) attributable to chorus wave intensity variation at the latitude of Landau or cyclotron resonance with precipitating electrons, compared to the moderate average level $\langle W\rangle\sim0.5$ during the entire event. However, these peaks are located at different $L$-shells for ELFIN A and B. Moreover, where ELFIN A observed precipitation peaks ELFIN B occasionally observed local minima of precipitation. 

\begin{figure*}
\centering
\includegraphics[width=0.9\textwidth]{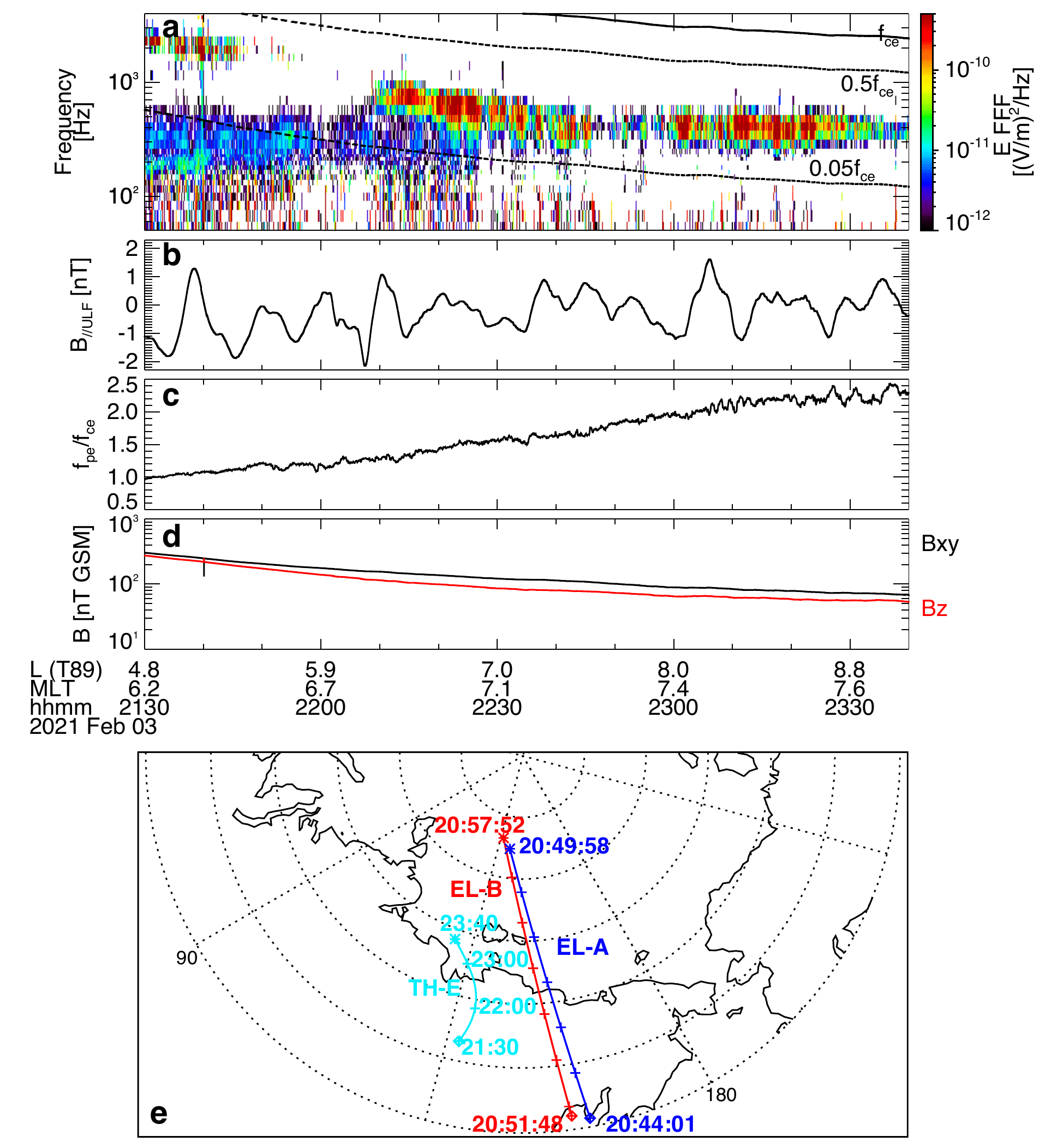}
\caption{THEMIS E equatorial measurements within the $L$-shell, $MLT$ region crossed by ELFIN A\&B: (a) magnetic field spectrum in the whistler-mode frequency range; black curves show fractions of the electron gyrofrequency, (b) ULF fluctuation $B_{\parallel ULF}$ of the parallel component of the background magnetic field, (c) $f_{pe}/f_{ce}$ ratio, (d) $B_z$ and $B_{xy}=\sqrt{B_x^2+B_y^2}$ magnetic fields. The bottom panel, (e), shows projections of ELFIN A\&B orbits and THEMIS E orbit.  \label{fig7}}
\end{figure*}

This is likely an effect of precipitation modulation by ULF waves \cite{Motoba13}. Such waves preferentially originate from the dawn flank magnetopause due to foreshock transients \cite{Hartinger13,Hartinger14}, propagate to lower $L$ \cite{Wright&Elsden20,Klimushkin19,Zhang20:jgr:ulf}, and modulate the equatorial generation of whistler-mode chorus waves \cite{Li11:modulation1,Watt11,Xia16:ulf&chorus,Zhang19:jgr:modulation,LiLi22} responsible for electron precipitation. As a result, precipitation patterns should be quasi-periodic in time and space ($L$), at the typical temporal period of ULF waves (several minutes) and with a spatial period similar to the ULF wavelength ($\sim R_E$). \blue{To verify this conjecture, we plot in Figure \ref{fig7} the conjugate equatorial observations from THEMIS E, D, and A. The whistler-mode chorus waves observed by THEMIS E at $f\sim0.15\,f_{ce}$ (with $f_{ce}$ being the electron gyrofrequency) over a wide $L$-shell range are indeed quasi-periodic in wave power (a temporal effect at the slowly moving THEMIS spacecraft). Similarly, quasi-periodic whistler-mode waves have been measured by THEMIS A, D at the same $L$-shell range, but 30 min ahead of THEMIS E measurements (due to the time difference in their orbits). The wave spectra at THEMIS A, D have low frequency resolutions (6 channels instead of 32 channels for THEMIS E); but presence of quasi-periodic whistler-mode waves at all three THEMIS spacecraft demonstrates that these waves occupy a large MLT, $L$-shell domain and can lead to the electron precipitation bursts at ELFIN. Moreover, the energy range of precipitating electrons (from $300$ keV and down to the lowest energy channel at $50$ keV) agrees well with the expected energy range of electrons resonating with $f\sim0.15\,f_{ce}$ whistler-mode waves \cite<see>[for diffusion rates evaluated based on observed whistler-mode wave characteristics]{Horne13:jgr,Ni14}.}

ULF fluctuations of the parallel (compressional) component of the background magnetic field are simultaneously measured by THEMIS E, reaching amplitudes of $\sim1.5$ nT, with a period of $\sim5-6$ min which is typical of Pc5 waves. A comparison of Figures \ref{fig7}(a,b) shows that most bursts of chorus wave intensity occur at minima of the variations of the parallel component of the magnetic field. This is consistent with a modulation of the distribution of $5-50$ keV electrons by ULF waves that, in turn, modulates the growth of chorus waves below $\sim0.3\,f_{ce}$ \cite{Li11:modulation1, Xia16:ulf&chorus}. This suggests that the precipitation on time-scales of minutes may be due to ULF wave-driven modulation of whistler-mode wave power at the equator.

\subsection{Example of large-separation event}\label{sec:large}
\begin{figure*}
\centering
\includegraphics[width=0.9\textwidth]{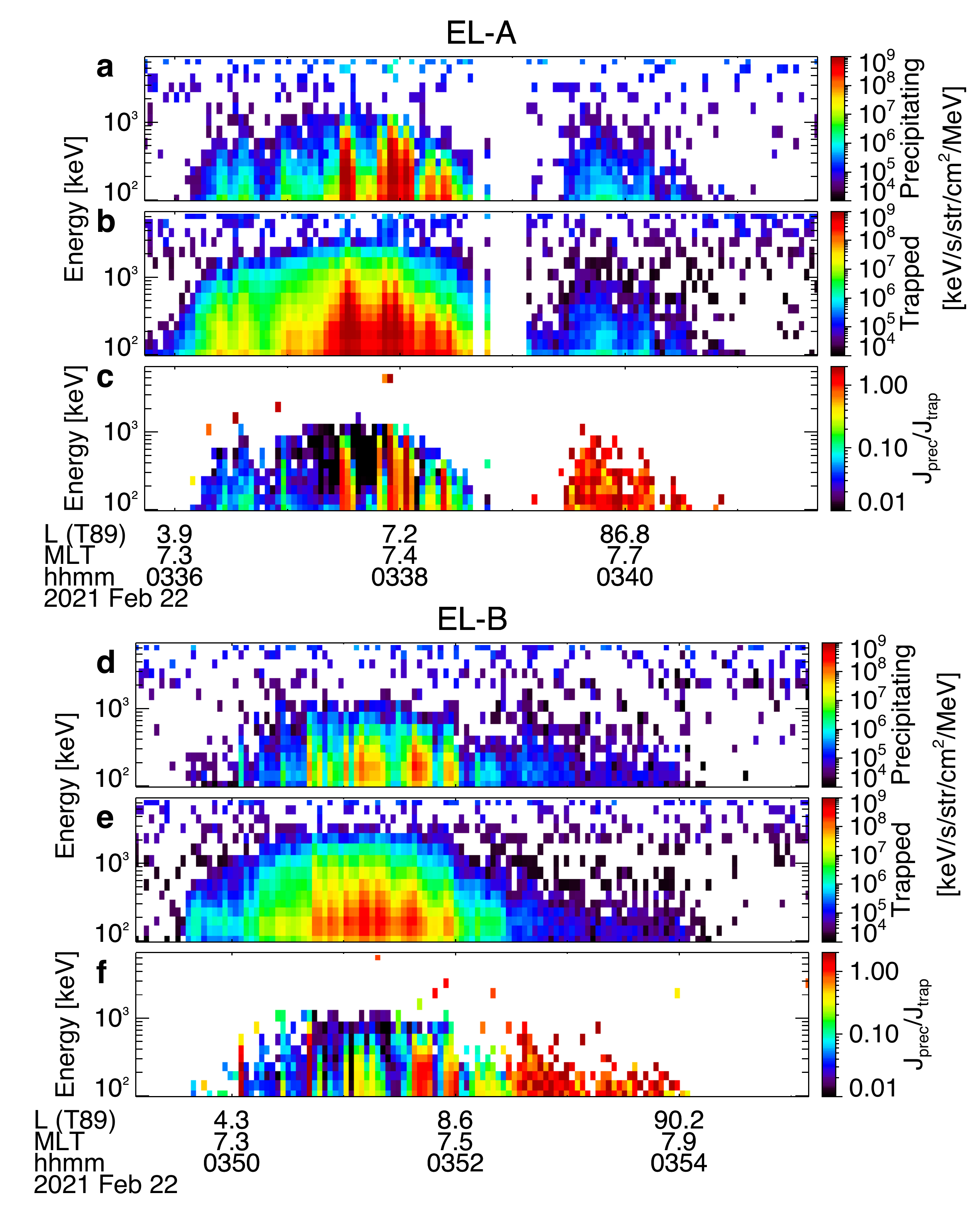}
\caption{The overview of ELFIN A and B observations with $\sim 13.5$min separation between spacecraft. Panels (a,d) show precipitating electron fluxes, Panels (b,e) show trapped fluxes, Panels (c,f) show precipitating-to-trapped flux ratio.\label{fig8}}
\end{figure*}

\begin{figure*}
\centering
\includegraphics[width=0.9\textwidth]{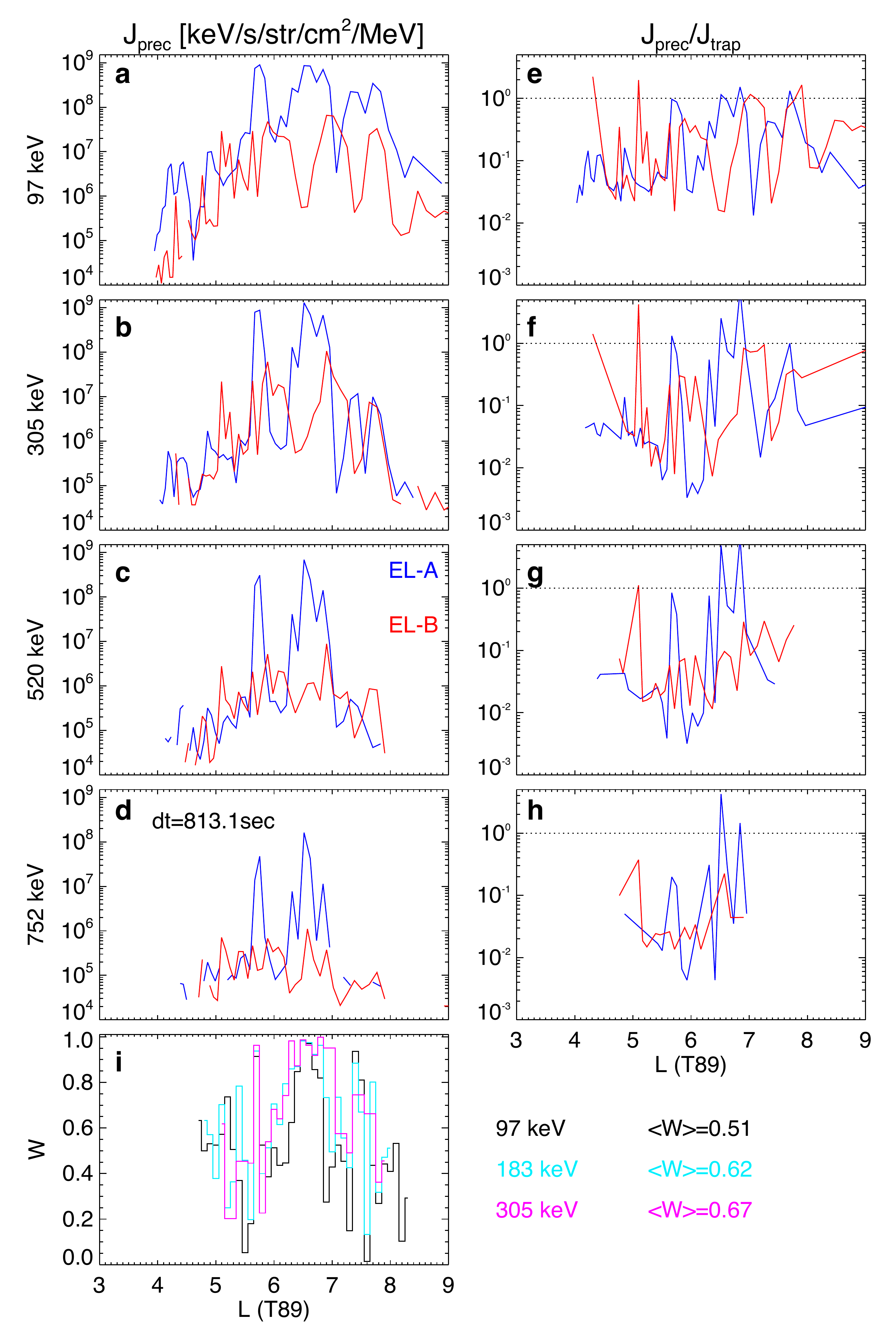}
\caption{(a-d) Precipitating electron fluxes versus $L$-shell for the event from Fig. \ref{fig8}. Four energy channels are shown for ELFIN A (blue) and B (red). (e-h) Same as (a-d) but showing the precipitating-to-trapped flux ratio. (i) Fraction $W$ of precipitating flux variation between ELFIN A \& B measurements due to variation of chorus wave intensity, for selected energies (colors) indicated on the right with the corresponding average $\langle W\rangle$ values. \label{fig9}}
\end{figure*}

Figure \ref{fig8} shows an example of an event with inter-spacecraft separation $\sim 13.5$ min. It shows multiple bursts of precipitation in the outer radiation belt (between the plasmashere at $L\sim 4$ and plasma sheet at $L\geq 8$; the latter is characterized by isotropic precipitation with low level of trapped fluxes). As during the previous event, here too there is a significant decrease of the precipitating flux magnitude from ELFIN A to ELFIN B. But contrary to the previous event, the precipitating-to-trapped flux ratio also exhibits a decrease in time and in the energy of significant precipitation, during the $\sim 13.5$ min elapsed between ELFIN A's to ELFIN B's observations. Figure \ref{fig9} confirms in line-plot format that the precipitating fluxes at ELFIN A are much higher than those at ELFIN B, and this effect increases with energy. Figure \ref{fig9}(i) further shows that the highest peaks of precipitating flux at $L=5.7$ and $L=6.3-6.8$ have a significantly higher value of $W\sim0.95$ compared to the average during the event. This signifies that at these two $L$-shell regions, there is an increase in the fraction of precipitating flux variation which is potentially attributable to chorus wave intensity variation at the latitude of resonance (rather than to trapped flux variation) for higher electron energy $\sim300$ keV than for lower energy $\sim100$ keV. Such variations of both the energy range and efficiency of electron precipitation cannot be explained by a quasi-periodic modulation of whistler-mode waves by ULF waves. Some mesoscale ($\sim 10$ min, $\sim R_E$) variations of the background plasma parameters should be involved. 

\begin{figure*}
\centering
\includegraphics[width=1.0\textwidth]{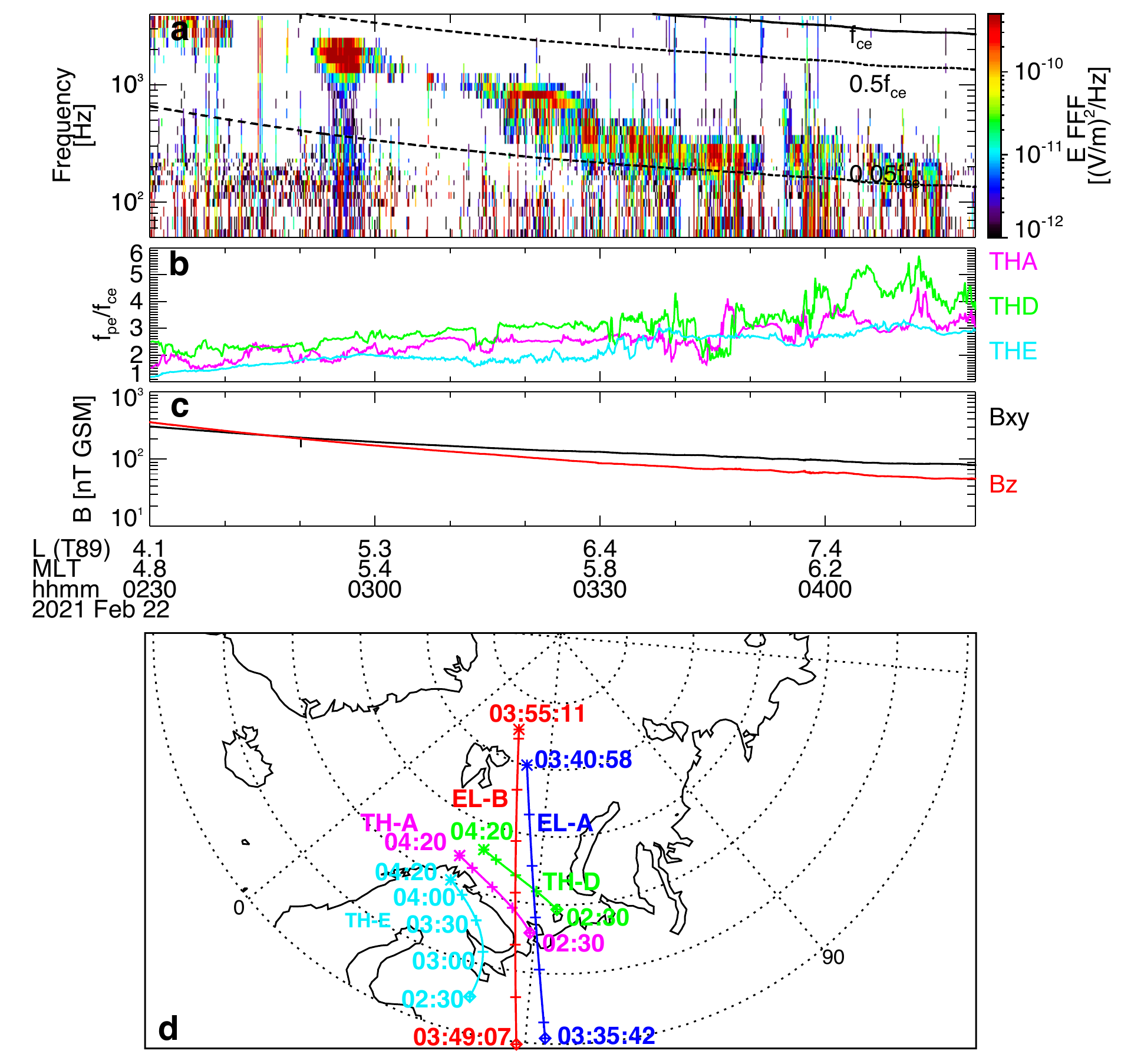}
\caption{THEMIS equatorial measurements within the $L$-shell, $MLT$ region crossed by ELFIN A\&B: (a) magnetic field spectrum in the whistler-mode frequency range measured by THEMIS E; black curves show fractions of the electron gyrofrequency, (b) $f_{pe}/f_{ce}$ ratio for THEMIS A, D, and E, (c) $B_z$ and $B_{xy}=\sqrt{B_x^2+B_y^2}$ magnetic fields \blue{measured by THEMIS E}, (d) projections of ELFIN A\&B orbits and THEMIS A, D, and E orbits. \label{fig10}}
\end{figure*}

\blue{To examine the equatorial plasma and wave variations during the time elapsed between the two ELFIN crossings, we take advantage of measurements from multiple THEMIS probes (THEMIS D, A and E) that were in conjunction with ELFIN at the time, as evidenced by their ionospheric projections in Figure \ref{fig10}(d). THEMIS D, A and E were moving along almost the same orbit (ordered from the leading to the trailing spacecraft) with an MLT separation of $\sim 0.5$ hour and along-track separation of $\sim 30$ min  (also evident in Figure \ref{fig10}(d)). THEMIS E observed continual (though intermittent) power enhancements of whistler-mode waves (Figure \ref{fig10}(a)), accompanied by variations in the background plasma density (not shown). Figure \ref{fig10}(a) shows the high-resolution magnetic field spectrum (32 frequency channels at $\sim [1 {\rm Hz},4 {\rm kHz}]$, at 1s cadence, see \citeA{Cully08:ssr} for details of this $fff$ dataset), which was not available at THEMIS D and A during this interval. The low-resolution magnetic field spectrum (the $fbk$ dataset with 3s cadence, 6 frequency channels at $\sim [1 {\rm Hz},4 {\rm kHz}]$) at THEMIS A shows the same whistler-mode waves as in Fig. \ref{fig10}(a). Therefore, these whistler-mode waves occupy a large MLT sector in space \cite<quite typical for the dawn flank, see>{Meredith12,Agapitov18:jgr} and can explain the electron precipitation observed by ELFIN. Moreover, only whistler-mode waves can resonantly scatter and precipitate electrons over such a wide energy range (from $\sim 1$ MeV and down to the lowest available energy channel of $50$ keV). }

The measured density (not shown) and magnetic field (shown in Figure \ref{fig10}(c) for THEMIS E, and in good agreement with the Tsyganenko magnetic field model) allow us to infer the equatorial magnetic field and to compute the ratio of equatorial plasma frequency to the electron gyrofrequency, $f_{pe}/f_{ce}$ in Figure \ref{fig10}(b). Most variations in that quantity reflect local density variations - they are observed at THEMIS on $\gtrsim$ several minute time scales, which is a mixture of temporal and spatial effects. Like the wave power, $f_{pe}/f_{ce}$ too shows significant variability. The spatial and temporal localization effects can be appreciated from the 3 equatorial satellite measurements but the important point here is that large variations in these parameters can be observed at a single point in time due to plasma density and velocity variations. Since $f_{pe}/f_{ce}$ controls the energy range of electrons resonating with whistler-mode waves, it determines both the wave generation at the equator (hence the observed wave power) \cite<e.g.,>[]{Li11:modulation2} and the resonance energy range of scattering (hence the precipitation efficiency) \cite{Agapitov19:fpe,Artemyev22:jgr:Landau&ELFIN}. Moreover, such mesoscale density variations may cause whistler-mode wave ducting \cite{Hanzelka&Santolik19,Hosseini21:ducts,Ke21:ducts,Chen21:ducting}, allowing them to propagate along the field line without strong damping to higher (middle) latitudes \cite{Bortnik07:landau,Chen13}, where these ducted waves can scatter relativistic electrons \cite<e.g.,>[]{Chen22:microbursts}. The event shown in Figures \ref{fig8}-\ref{fig10} is a good example of strong variations of electron precipitation due to mesoscale variations of the equatorial plasma density and the dynamic and complex phenomena that arise as a result of such variations on timescales greater than several minutes.

Figures S1-S3 in the Supporting Information (SI) show a similar event with ELFIN A and B separated by $\geq 10$ minutes and conjugate observations of meso-scale plasma density (or $f_{pe}/f_{ce}$) variations from THEMIS A, D, E. $L$-shell profiles of precipitating-to-trapped fluxes on ELFIN A and B are quite different, and this difference is larger for higher energy electrons which are scattered and precipitated by the waves farther from the equator. Therefore, this event too supports our conclusions derived from ELFIN and THEMIS observations in Figs. \ref{fig8}-\ref{fig10}. Six additional events in SI present ELFIN A and B measurements with inter-satellite separations $\sim 10$ min, but without conjugate equatorial THEMIS measurements. These events are shown in Figs. S4-S15 and confirm that a $\geq 10$ min separation is sufficiently large to provide quantitatively different net precipitation of energetic electrons. 

\section{Discussion and Conclusions}\label{sec:discussion}

In this study we show variations of electron precipitating fluxes over three different temporal and spatial scales. Comparison of ELFIN A and B measurements at inter-satellite separations from $\sim$ a few seconds to $\sim 15$ min reveals the existence of three types of precipitation patterns and associated variability. Differences on the shortest time-scale of separation, $\lesssim$ 5 s, corresponds to variations of individual precipitation bursts most likely due to variation of whistler-mode waves over the typical scale of chorus elements, or series of such elements. Such short-scale variations are usually observed during electron microburst precipitation events \cite{Blake&OBrien16,Shumko20}. Their fast timescale suggests they are due to electrons precipitating after a single resonant interaction with whistler-mode wave packets, i.e., they are likely due to nonlinear resonant interactions with very intense chorus wave packets quickly scattering electrons into the loss cone \cite{Chen22:microbursts,Zhang22:natcom,Hsieh22}. Our study shows that the spacecraft provide very similar precipitating fluxes and precipitating-to-trapped flux ratios over such a small time delay ($<5$ s) between measurements, even though individual bursts are usually not identical on consecutive passes. The general agreement of the spectra also demonstrates that ELFIN A and B measurements are well inter-calibrated and may be used to quantify precipitation variations on longer time-scales as well. The medium temporal scale of inter-spacecraft separations, $\sim$ a few minutes, reveals ULF wave associated variations of the equatorial whistler-mode source region. ULF waves with period of a few minutes cause the whistler-mode wave power (driven by ULF wave peaks or troughs) to move radially and hence can displace the associated whistler-mode chorus wave driven precipitation bursts in latitude. However, they do not change the intensity of precipitation bursts within a wide $L$-shell range where the ULF waves propagate. In other words, the location of peaks of precipitating-to-trapped flux ratio may change but their magnitude remains approximately the same. The largest inter-satellite separations that we examined, $\gtrsim10$ min, reveal variations of precipitating fluxes which are associated with equatorial mesoscale variations of background plasma and magnetic field characteristics (potentially also related to simultaneous variations of trapped electron fluxes). Contrary to ULF wave-driven variations, these mesoscale variations change the equatorial density and energetic electron flux responsible for for whistler-mode wave generation. Such changes occur due to convection (including corotation) and injections that modify drastically the plasma density and electron flux over timescales of many minutes and spatial scales $\approx 1\,R_E$. This can cause both the intensity of precipitating fluxes and the energy range of precipitating electrons (which depends critically on the equatorial plasma density) to vary significantly within a $\geq 10$ min interval. These three temporal scales are still below the $\sim 1$ hour time scale of substorm dynamics, and thus all three scales should not be related to any significant magnetic field reconfiguration. 

Let us discuss the importance of such temporal variations for electron precipitation. The smallest time-scales ($\approx1$ s) of variations of precipitating fluxes are simply a consequence of the transient nature of chorus wave generation, but likely do not change the total amount of electron losses. Despite the general interest in such variations and their importance for the investigation of the fine wave-particle resonant effects \cite<see discussion in>[]{Shumko21}, their inclusion into radiation belt models remains a topic of debate \cite<see discussion in>[]{Thorne05}. Although the nature of microbursts is likely dominated by nonlinear electron resonant interactions, the long-term effect of these nonlinear interactions can probably be described as a diffusive process with diffusion rates differing somewhat from the quasi-linear ones \cite{Artemyev21:pre,Artemyev22:jgr:diffusion}. Therefore, after being averaged, such multiple microbursts effects, may be incorporated into the classical Fokker-Plank equations via renormalization of electron diffusion rates \cite{Artemyev22:jgr:diffusion}.

The medium-scale variations (likely driven by ULF wave modulation of the equatorial whistler-mode source, see, e.g. \citeA{Li11:modulation1,Xia16:ulf&chorus,Zhang19:jgr:modulation}) are mostly responsible for quasi-periodic spatial and temporal dynamics of the precipitating bursts, but also do not alter the number of such bursts and their intensity within the outer radiation belt. One of the main effects of such variations is the appearance of quasi-periodic pulsations of precipitating electron fluxes and diffuse aurora \cite<see review by>[and references therein]{Nishimura20:ssr}. However, with regard to the net electron losses from the outer radiation belt, these variations can be averaged. The main question raised by such variations is the consequence of whistler-mode modulation by ULF waves for electron losses. Is it a simple spatial variation of the precipitation rate with an average equal to the same net precipitation as in the absence of ULF waves? Or can the ULF wave variation render previously marginally stable electron distributions unstable to wave generation, thus increasing the net precipitation above what it would have been without the ULF waves? In other words, can ULF waves generate whistler-mode waves and the associated electron precipitation from a marginally stable plasma?

The large-scale ($\geq 10$ min) variations of electron precipitation are likely due to mesoscale temporal and spatial variations of the equatorial plasma density and energetic electron fluxes, both responsible for whistler-mode wave generation, and precipitating electron fluxes. Scales of $\sim 1R_E$ are typical for plasma injections \cite{Nakamura02,Liu13:DF} and cold plasma density plumes \cite<e.g.,>{Darrouzet09,Goldstein14}. These spatial structures which evolve (at an approximately fixed $MLT$) over $>10$min due to convection or injections, can be observed in precipitation patterns as differences in precipitating fluxes measured by consecutive ELFIN passes over the same $MLT$ sector. 
Such large-scale precipitation variations should be quite important for the evaluation of net electron losses, because the intensity of precipitation may vary significantly with the combined variations of equatorial plasma density and whistler-mode characteristics. Empirical models of waves and plasma density are constructed independently, and their combination does not include the effect of mesoscale plasma density structures correlated with wave activity. Therefore, such plasma density and wave intensity models will not describe variations of energy range and intensity of electron precipitation observed by ELFIN. This means that Fokker-Plank simulations of energetic electron fluxes may underestimate the role of whistler-mode waves in, e.g., losses of relativistic electrons or acceleration of subrelativistic electrons, as both these processes are strongly dependent on local plasma density and wave characteristics. A new generation of more detailed plasma density models \cite<e.g.,>{Chu17,Zhelavskaya21} may describe some mesoscale density structures, but these models do not include any correlation with wave activity. Recently, however, several statistical studies of electron quasi-linear diffusion rates have started to take more accurately into account the correlations between wave intensity and local plasma density \cite{Agapitov19:fpe, Ross21}. In addition, further investigations of mesoscale wave and background plasma variations and their role for electron precipitation are needed for construction of a next generation wave-plasma models for radiation belt simulations.

\acknowledgments
X.J.Z., A.V.A., and V.A. acknowledge support by NASA awards 80NSSC20K1578, 80NSSC22K0522, NNX14AN68G, and NSF grants NSF-2021749, AGS-1242918, AGS-2019950, AGS-2021749. O.V.A was partially supported by NSF grant number 1914670,  NASA’s Living with a Star (LWS) program (contract 80NSSC20K0218), and NASA contracts 80NNSC19K0848, 80NSSC22K0433, 80NSSC22K0522. We are grateful to NASA's CubeSat Launch Initiative for ELFIN's successful launch. We acknowledge early support of ELFIN project by the AFOSR, under its University Nanosat Program, UNP-8 project, contract FA9453-12-D-0285, and by the California Space Grant program. We acknowledge the critical contributions of numerous staff and students who led to the successful development and operations of the ELFIN mission. The work at UCLA was also supported by NASA contract NAS5-02099 for data analysis from the THEMIS mission. We specifically thank: J. W. Bonnell and F. S. Mozer for the use of EFI data in determining spacecraft potential derived density from THEMIS E, J. P. McFadden for use of ESA data in obtaining electron temperature measurements to incorporate into the procedure for electron density determination from the spacecraft potential, and K. H. Glassmeier, U. Auster and W. Baumjohann for the use of FGM data provided under the lead of the Technical University of Braunschweig and with financial support through the German Ministry for Economy and Technology and the German Center for Aviation and Space (DLR) under contract 50 OC 0302. 

\section*{Open Research} \noindent ELFIN data is available at https://data.elfin.ucla.edu/.

 \noindent THEMIS data is available at \blue{http://themis.ssl.berkeley.edu.}

 \noindent Data analysis was done using SPEDAS V4.1 \cite{Angelopoulos19} available at https://spedas.org/.



\begin{thebibliography}{}

\bibitem [\protect \citeauthoryear {%
{Agapitov}%
\ \protect \BOthers {.}}{%
{Agapitov}%
\ \protect \BOthers {.}}{%
{\protect \APACyear {2014}}%
}]{%
Agapitov14:jgr:acceleration}
\APACinsertmetastar {%
Agapitov14:jgr:acceleration}%
\begin{APACrefauthors}%
{Agapitov}, O\BPBI V.%
, {Artemyev}, A.%
, {Mourenas}, D.%
, {Krasnoselskikh}, V.%
, {Bonnell}, J.%
, {Le Contel}, O.%
\BDBL {}{Angelopoulos}, V.%
\end{APACrefauthors}%
\unskip\
\newblock
\APACrefYearMonthDay{2014}{}{}.
\newblock
{\BBOQ}\APACrefatitle {{The quasi-electrostatic mode of chorus waves and
  electron nonlinear acceleration}} {{The quasi-electrostatic mode of chorus
  waves and electron nonlinear acceleration}}.{\BBCQ}
\newblock
\APACjournalVolNumPages{\jgr}{119}{}{1606--1626}.
\newblock
\begin{APACrefDOI} \doi{10.1002/2013JA019223} \end{APACrefDOI}
\PrintBackRefs{\CurrentBib}

\bibitem [\protect \citeauthoryear {%
{Agapitov}%
, {Artemyev}%
, {Mourenas}%
, {Mozer}%
\BCBL {}\ \BBA {} {Krasnoselskikh}%
}{%
{Agapitov}%
\ \protect \BOthers {.}}{%
{\protect \APACyear {2015}}%
}]{%
Agapitov15:grl:acceleration}
\APACinsertmetastar {%
Agapitov15:grl:acceleration}%
\begin{APACrefauthors}%
{Agapitov}, O\BPBI V.%
, {Artemyev}, A\BPBI V.%
, {Mourenas}, D.%
, {Mozer}, F\BPBI S.%
\BCBL {}\ \BBA {} {Krasnoselskikh}, V.%
\end{APACrefauthors}%
\unskip\
\newblock
\APACrefYearMonthDay{2015}{{\APACmonth{12}}}{}.
\newblock
{\BBOQ}\APACrefatitle {{Nonlinear local parallel acceleration of electrons
  through Landau trapping by oblique whistler mode waves in the outer radiation
  belt}} {{Nonlinear local parallel acceleration of electrons through Landau
  trapping by oblique whistler mode waves in the outer radiation belt}}.{\BBCQ}
\newblock
\APACjournalVolNumPages{\grl}{42}{}{10}.
\newblock
\begin{APACrefDOI} \doi{10.1002/2015GL066887} \end{APACrefDOI}
\PrintBackRefs{\CurrentBib}

\bibitem [\protect \citeauthoryear {%
{Agapitov}%
, {Blum}%
, {Mozer}%
, {Bonnell}%
\BCBL {}\ \BBA {} {Wygant}%
}{%
{Agapitov}%
\ \protect \BOthers {.}}{%
{\protect \APACyear {2017}}%
}]{%
Agapitov17:grl}
\APACinsertmetastar {%
Agapitov17:grl}%
\begin{APACrefauthors}%
{Agapitov}, O\BPBI V.%
, {Blum}, L\BPBI W.%
, {Mozer}, F\BPBI S.%
, {Bonnell}, J\BPBI W.%
\BCBL {}\ \BBA {} {Wygant}, J.%
\end{APACrefauthors}%
\unskip\
\newblock
\APACrefYearMonthDay{2017}{{\APACmonth{03}}}{}.
\newblock
{\BBOQ}\APACrefatitle {{Chorus whistler wave source scales as determined from
  multipoint Van Allen Probe measurements}} {{Chorus whistler wave source
  scales as determined from multipoint Van Allen Probe measurements}}.{\BBCQ}
\newblock
\APACjournalVolNumPages{\grl}{44}{}{2634-2642}.
\newblock
\begin{APACrefDOI} \doi{10.1002/2017GL072701} \end{APACrefDOI}
\PrintBackRefs{\CurrentBib}

\bibitem [\protect \citeauthoryear {%
{Agapitov}%
\ \protect \BOthers {.}}{%
{Agapitov}%
\ \protect \BOthers {.}}{%
{\protect \APACyear {2010}}%
}]{%
Agapitov10AnGeo}
\APACinsertmetastar {%
Agapitov10AnGeo}%
\begin{APACrefauthors}%
{Agapitov}, O\BPBI V.%
, {Krasnoselskikh}, V.%
, {Zaliznyak}, Y.%
, {Angelopoulos}, V.%
, {Le Contel}, O.%
\BCBL {}\ \BBA {} {Rolland}, G.%
\end{APACrefauthors}%
\unskip\
\newblock
\APACrefYearMonthDay{2010}{{\APACmonth{06}}}{}.
\newblock
{\BBOQ}\APACrefatitle {{Chorus source region localization in the Earth's outer
  magnetosphere using THEMIS measurements}} {{Chorus source region localization
  in the Earth's outer magnetosphere using THEMIS measurements}}.{\BBCQ}
\newblock
\APACjournalVolNumPages{Annales Geophysicae}{28}{}{1377-1386}.
\PrintBackRefs{\CurrentBib}

\bibitem [\protect \citeauthoryear {%
{Agapitov}%
, {Mourenas}%
, {Artemyev}%
, {Hospodarsky}%
\BCBL {}\ \BBA {} {Bonnell}%
}{%
{Agapitov}%
\ \protect \BOthers {.}}{%
{\protect \APACyear {2019}}%
}]{%
Agapitov19:fpe}
\APACinsertmetastar {%
Agapitov19:fpe}%
\begin{APACrefauthors}%
{Agapitov}, O\BPBI V.%
, {Mourenas}, D.%
, {Artemyev}, A.%
, {Hospodarsky}, G.%
\BCBL {}\ \BBA {} {Bonnell}, J\BPBI W.%
\end{APACrefauthors}%
\unskip\
\newblock
\APACrefYearMonthDay{2019}{{\APACmonth{06}}}{}.
\newblock
{\BBOQ}\APACrefatitle {{Time Scales for Electron Quasi-linear Diffusion by
  Lower-Band Chorus Waves: The Effects of
  {\ensuremath{\omega}}$_{pe}$/{\ensuremath{\Omega}}$_{ce}$ Dependence on
  Geomagnetic Activity}} {{Time Scales for Electron Quasi-linear Diffusion by
  Lower-Band Chorus Waves: The Effects of
  {\ensuremath{\omega}}$_{pe}$/{\ensuremath{\Omega}}$_{ce}$ Dependence on
  Geomagnetic Activity}}.{\BBCQ}
\newblock
\APACjournalVolNumPages{\grl}{46}{12}{6178-6187}.
\newblock
\begin{APACrefDOI} \doi{10.1029/2019GL083446} \end{APACrefDOI}
\PrintBackRefs{\CurrentBib}

\bibitem [\protect \citeauthoryear {%
{Agapitov}%
, {Mourenas}%
, {Artemyev}%
, {Mozer}%
, {Bonnell}%
\BCBL {}\ \protect \BOthers {.}}{%
{Agapitov}%
, {Mourenas}%
, {Artemyev}%
, {Mozer}%
, {Bonnell}%
\BCBL {}\ \protect \BOthers {.}}{%
{\protect \APACyear {2018}}%
}]{%
Agapitov18:correlations}
\APACinsertmetastar {%
Agapitov18:correlations}%
\begin{APACrefauthors}%
{Agapitov}, O\BPBI V.%
, {Mourenas}, D.%
, {Artemyev}, A.%
, {Mozer}, F\BPBI S.%
, {Bonnell}, J\BPBI W.%
, {Angelopoulos}, V.%
\BDBL {}{Krasnoselskikh}, V.%
\end{APACrefauthors}%
\unskip\
\newblock
\APACrefYearMonthDay{2018}{{\APACmonth{10}}}{}.
\newblock
{\BBOQ}\APACrefatitle {{Spatial Extent and Temporal Correlation of Chorus and
  Hiss: Statistical Results From Multipoint THEMIS Observations}} {{Spatial
  Extent and Temporal Correlation of Chorus and Hiss: Statistical Results From
  Multipoint THEMIS Observations}}.{\BBCQ}
\newblock
\APACjournalVolNumPages{Journal of Geophysical Research (Space
  Physics)}{123}{10}{8317-8330}.
\newblock
\begin{APACrefDOI} \doi{10.1029/2018JA025725} \end{APACrefDOI}
\PrintBackRefs{\CurrentBib}

\bibitem [\protect \citeauthoryear {%
{Agapitov}%
, {Mourenas}%
, {Artemyev}%
, {Mozer}%
, {Hospodarsky}%
\BCBL {}\ \protect \BOthers {.}}{%
{Agapitov}%
, {Mourenas}%
, {Artemyev}%
, {Mozer}%
, {Hospodarsky}%
\BCBL {}\ \protect \BOthers {.}}{%
{\protect \APACyear {2018}}%
}]{%
Agapitov18:jgr}
\APACinsertmetastar {%
Agapitov18:jgr}%
\begin{APACrefauthors}%
{Agapitov}, O\BPBI V.%
, {Mourenas}, D.%
, {Artemyev}, A\BPBI V.%
, {Mozer}, F\BPBI S.%
, {Hospodarsky}, G.%
, {Bonnell}, J.%
\BCBL {}\ \BBA {} {Krasnoselskikh}, V.%
\end{APACrefauthors}%
\unskip\
\newblock
\APACrefYearMonthDay{2018}{{\APACmonth{01}}}{}.
\newblock
{\BBOQ}\APACrefatitle {{Synthetic Empirical Chorus Wave Model From Combined Van
  Allen Probes and Cluster Statistics}} {{Synthetic Empirical Chorus Wave Model
  From Combined Van Allen Probes and Cluster Statistics}}.{\BBCQ}
\newblock
\APACjournalVolNumPages{Journal of Geophysical Research (Space
  Physics)}{123}{1}{297-314}.
\newblock
\begin{APACrefDOI} \doi{10.1002/2017JA024843} \end{APACrefDOI}
\PrintBackRefs{\CurrentBib}

\bibitem [\protect \citeauthoryear {%
{Albert}%
, {Tao}%
\BCBL {}\ \BBA {} {Bortnik}%
}{%
{Albert}%
\ \protect \BOthers {.}}{%
{\protect \APACyear {2013}}%
}]{%
Albert13:AGU}
\APACinsertmetastar {%
Albert13:AGU}%
\begin{APACrefauthors}%
{Albert}, J\BPBI M.%
, {Tao}, X.%
\BCBL {}\ \BBA {} {Bortnik}, J.%
\end{APACrefauthors}%
\unskip\
\newblock
\APACrefYearMonthDay{2013}{}{}.
\newblock
{\BBOQ}\APACrefatitle {{Aspects of Nonlinear Wave-Particle Interactions}}
  {{Aspects of Nonlinear Wave-Particle Interactions}}.{\BBCQ}
\newblock
\BIn{} D.~{Summers}, I\BPBI U.~{Mann}, D\BPBI N.~{Baker}\BCBL {}\ \BBA {}
  M.~{Schulz}\ (\BEDS), \APACrefbtitle {Dynamics of the Earth's Radiation Belts
  and Inner Magnetosphere.} {Dynamics of the earth's radiation belts and inner
  magnetosphere.}
\newblock
\begin{APACrefDOI} \doi{10.1029/2012GM001324} \end{APACrefDOI}
\PrintBackRefs{\CurrentBib}

\bibitem [\protect \citeauthoryear {%
{Allanson}%
, {Thomas}%
, {Watt}%
\BCBL {}\ \BBA {} {Neukirch}%
}{%
{Allanson}%
\ \protect \BOthers {.}}{%
{\protect \APACyear {2022}}%
}]{%
Allanson22}
\APACinsertmetastar {%
Allanson22}%
\begin{APACrefauthors}%
{Allanson}, O.%
, {Thomas}, E.%
, {Watt}, C\BPBI E\BPBI J.%
\BCBL {}\ \BBA {} {Neukirch}, T.%
\end{APACrefauthors}%
\unskip\
\newblock
\APACrefYearMonthDay{2022}{}{}.
\newblock
{\BBOQ}\APACrefatitle {{Weak Turbulence and Quasilinear Diffusion for
  Relativistic Wave-Particle Interactions Via a Markov Approach}} {{Weak
  Turbulence and Quasilinear Diffusion for Relativistic Wave-Particle
  Interactions Via a Markov Approach}}.{\BBCQ}
\newblock
\APACjournalVolNumPages{Frontiers in Physics}{8:805699}{}{}.
\newblock
\begin{APACrefDOI} \doi{10.3389/fspas.2021.805699} \end{APACrefDOI}
\PrintBackRefs{\CurrentBib}

\bibitem [\protect \citeauthoryear {%
{An}%
, {Wu}%
\BCBL {}\ \BBA {} {Tao}%
}{%
{An}%
\ \protect \BOthers {.}}{%
{\protect \APACyear {2022}}%
}]{%
An22:Tao}
\APACinsertmetastar {%
An22:Tao}%
\begin{APACrefauthors}%
{An}, Z.%
, {Wu}, Y.%
\BCBL {}\ \BBA {} {Tao}, X.%
\end{APACrefauthors}%
\unskip\
\newblock
\APACrefYearMonthDay{2022}{{\APACmonth{02}}}{}.
\newblock
{\BBOQ}\APACrefatitle {{Electron Dynamics in a Chorus Wave Field Generated From
  Particle-In-Cell Simulations}} {{Electron Dynamics in a Chorus Wave Field
  Generated From Particle-In-Cell Simulations}}.{\BBCQ}
\newblock
\APACjournalVolNumPages{\grl}{49}{3}{e97778}.
\newblock
\begin{APACrefDOI} \doi{10.1029/2022GL097778} \end{APACrefDOI}
\PrintBackRefs{\CurrentBib}

\bibitem [\protect \citeauthoryear {%
{Angelopoulos}%
, {Artemyev}%
, {Phan}%
\BCBL {}\ \BBA {} {Miyashita}%
}{%
{Angelopoulos}%
, {Artemyev}%
\BCBL {}\ \protect \BOthers {.}}{%
{\protect \APACyear {2020}}%
}]{%
Angelopoulos20}
\APACinsertmetastar {%
Angelopoulos20}%
\begin{APACrefauthors}%
{Angelopoulos}, V.%
, {Artemyev}, A.%
, {Phan}, T\BPBI D.%
\BCBL {}\ \BBA {} {Miyashita}, Y.%
\end{APACrefauthors}%
\unskip\
\newblock
\APACrefYearMonthDay{2020}{{\APACmonth{01}}}{}.
\newblock
{\BBOQ}\APACrefatitle {{Near-Earth magnetotail reconnection powers space
  storms}} {{Near-Earth magnetotail reconnection powers space storms}}.{\BBCQ}
\newblock
\APACjournalVolNumPages{Nature Physics}{16}{3}{317-321}.
\newblock
\begin{APACrefDOI} \doi{10.1038/s41567-019-0749-4} \end{APACrefDOI}
\PrintBackRefs{\CurrentBib}

\bibitem [\protect \citeauthoryear {%
{Angelopoulos}%
\ \protect \BOthers {.}}{%
{Angelopoulos}%
\ \protect \BOthers {.}}{%
{\protect \APACyear {2019}}%
}]{%
Angelopoulos19}
\APACinsertmetastar {%
Angelopoulos19}%
\begin{APACrefauthors}%
{Angelopoulos}, V.%
, {Cruce}, P.%
, {Drozdov}, A.%
, {Grimes}, E\BPBI W.%
, {Hatzigeorgiu}, N.%
, {King}, D\BPBI A.%
\BDBL {}{Schroeder}, P.%
\end{APACrefauthors}%
\unskip\
\newblock
\APACrefYearMonthDay{2019}{{\APACmonth{01}}}{}.
\newblock
{\BBOQ}\APACrefatitle {{The Space Physics Environment Data Analysis System
  (SPEDAS)}} {{The Space Physics Environment Data Analysis System
  (SPEDAS)}}.{\BBCQ}
\newblock
\APACjournalVolNumPages{\ssr}{215}{}{9}.
\newblock
\begin{APACrefDOI} \doi{10.1007/s11214-018-0576-4} \end{APACrefDOI}
\PrintBackRefs{\CurrentBib}

\bibitem [\protect \citeauthoryear {%
{Angelopoulos}%
, {Tsai}%
\BCBL {}\ \protect \BOthers {.}}{%
{Angelopoulos}%
, {Tsai}%
\BCBL {}\ \protect \BOthers {.}}{%
{\protect \APACyear {2020}}%
}]{%
Angelopoulos20:elfin}
\APACinsertmetastar {%
Angelopoulos20:elfin}%
\begin{APACrefauthors}%
{Angelopoulos}, V.%
, {Tsai}, E.%
, {Bingley}, L.%
, {Shaffer}, C.%
, {Turner}, D\BPBI L.%
, {Runov}, A.%
\BDBL {}{Zhang}, G\BPBI Y.%
\end{APACrefauthors}%
\unskip\
\newblock
\APACrefYearMonthDay{2020}{{\APACmonth{07}}}{}.
\newblock
{\BBOQ}\APACrefatitle {{The ELFIN Mission}} {{The ELFIN Mission}}.{\BBCQ}
\newblock
\APACjournalVolNumPages{\ssr}{216}{5}{103}.
\newblock
\begin{APACrefDOI} \doi{10.1007/s11214-020-00721-7} \end{APACrefDOI}
\PrintBackRefs{\CurrentBib}

\bibitem [\protect \citeauthoryear {%
{Artemyev}%
, {Demekhov}%
\BCBL {}\ \protect \BOthers {.}}{%
{Artemyev}%
, {Demekhov}%
\BCBL {}\ \protect \BOthers {.}}{%
{\protect \APACyear {2021}}%
}]{%
Artemyev21:jgr:ducts}
\APACinsertmetastar {%
Artemyev21:jgr:ducts}%
\begin{APACrefauthors}%
{Artemyev}, A\BPBI V.%
, {Demekhov}, A\BPBI G.%
, {Zhang}, X\BPBI J.%
, {Angelopoulos}, V.%
, {Mourenas}, D.%
, {Fedorenko}, Y\BPBI V.%
\BDBL {}{Shinohara}, I.%
\end{APACrefauthors}%
\unskip\
\newblock
\APACrefYearMonthDay{2021}{{\APACmonth{11}}}{}.
\newblock
{\BBOQ}\APACrefatitle {{Role of Ducting in Relativistic Electron Loss by
  Whistler-Mode Wave Scattering}} {{Role of Ducting in Relativistic Electron
  Loss by Whistler-Mode Wave Scattering}}.{\BBCQ}
\newblock
\APACjournalVolNumPages{Journal of Geophysical Research (Space
  Physics)}{126}{11}{e29851}.
\newblock
\begin{APACrefDOI} \doi{10.1029/2021JA029851} \end{APACrefDOI}
\PrintBackRefs{\CurrentBib}

\bibitem [\protect \citeauthoryear {%
{Artemyev}%
, {Mourenas}%
, {Zhang}%
\BCBL {}\ \BBA {} {Vainchtein}%
}{%
{Artemyev}%
, {Mourenas}%
\BCBL {}\ \protect \BOthers {.}}{%
{\protect \APACyear {2022}}%
}]{%
Artemyev22:jgr:diffusion}
\APACinsertmetastar {%
Artemyev22:jgr:diffusion}%
\begin{APACrefauthors}%
{Artemyev}, A\BPBI V.%
, {Mourenas}, D.%
, {Zhang}, X\BPBI J.%
\BCBL {}\ \BBA {} {Vainchtein}, D.%
\end{APACrefauthors}%
\unskip\
\newblock
\APACrefYearMonthDay{2022}{{\APACmonth{09}}}{}.
\newblock
{\BBOQ}\APACrefatitle {{On the Incorporation of Nonlinear Resonant
  Wave-Particle Interactions Into Radiation Belt Models}} {{On the
  Incorporation of Nonlinear Resonant Wave-Particle Interactions Into Radiation
  Belt Models}}.{\BBCQ}
\newblock
\APACjournalVolNumPages{Journal of Geophysical Research (Space
  Physics)}{127}{9}{e30853}.
\newblock
\begin{APACrefDOI} \doi{10.1029/2022JA030853} \end{APACrefDOI}
\PrintBackRefs{\CurrentBib}

\bibitem [\protect \citeauthoryear {%
{Artemyev}%
, {Neishtadt}%
\BCBL {}\ \BBA {} {Angelopoulos}%
}{%
{Artemyev}%
, {Neishtadt}%
\BCBL {}\ \BBA {} {Angelopoulos}%
}{%
{\protect \APACyear {2022}}%
}]{%
Artemyev22:jgr:DF&ELFIN}
\APACinsertmetastar {%
Artemyev22:jgr:DF&ELFIN}%
\begin{APACrefauthors}%
{Artemyev}, A\BPBI V.%
, {Neishtadt}, A\BPBI I.%
\BCBL {}\ \BBA {} {Angelopoulos}, V.%
\end{APACrefauthors}%
\unskip\
\newblock
\APACrefYearMonthDay{2022}{{\APACmonth{04}}}{}.
\newblock
{\BBOQ}\APACrefatitle {{On the Role of Whistler-Mode Waves in Electron
  Interaction With Dipolarizing Flux Bundles}} {{On the Role of Whistler-Mode
  Waves in Electron Interaction With Dipolarizing Flux Bundles}}.{\BBCQ}
\newblock
\APACjournalVolNumPages{Journal of Geophysical Research (Space
  Physics)}{127}{4}{e30265}.
\newblock
\begin{APACrefDOI} \doi{10.1029/2022JA030265} \end{APACrefDOI}
\PrintBackRefs{\CurrentBib}

\bibitem [\protect \citeauthoryear {%
{Artemyev}%
\ \protect \BOthers {.}}{%
{Artemyev}%
\ \protect \BOthers {.}}{%
{\protect \APACyear {2018}}%
}]{%
Artemyev18:cnsns}
\APACinsertmetastar {%
Artemyev18:cnsns}%
\begin{APACrefauthors}%
{Artemyev}, A\BPBI V.%
, {Neishtadt}, A\BPBI I.%
, {Vainchtein}, D\BPBI L.%
, {Vasiliev}, A\BPBI A.%
, {Vasko}, I\BPBI Y.%
\BCBL {}\ \BBA {} {Zelenyi}, L\BPBI M.%
\end{APACrefauthors}%
\unskip\
\newblock
\APACrefYearMonthDay{2018}{{\APACmonth{12}}}{}.
\newblock
{\BBOQ}\APACrefatitle {{Trapping (capture) into resonance and scattering on
  resonance: Summary of results for space plasma systems}} {{Trapping (capture)
  into resonance and scattering on resonance: Summary of results for space
  plasma systems}}.{\BBCQ}
\newblock
\APACjournalVolNumPages{Communications in Nonlinear Science and Numerical
  Simulations}{65}{}{111-160}.
\newblock
\begin{APACrefDOI} \doi{10.1016/j.cnsns.2018.05.004} \end{APACrefDOI}
\PrintBackRefs{\CurrentBib}

\bibitem [\protect \citeauthoryear {%
{Artemyev}%
, {Neishtadt}%
, {Vasiliev}%
\BCBL {}\ \BBA {} {Mourenas}%
}{%
{Artemyev}%
, {Neishtadt}%
\BCBL {}\ \protect \BOthers {.}}{%
{\protect \APACyear {2021}}%
}]{%
Artemyev21:pre}
\APACinsertmetastar {%
Artemyev21:pre}%
\begin{APACrefauthors}%
{Artemyev}, A\BPBI V.%
, {Neishtadt}, A\BPBI I.%
, {Vasiliev}, A\BPBI A.%
\BCBL {}\ \BBA {} {Mourenas}, D.%
\end{APACrefauthors}%
\unskip\
\newblock
\APACrefYearMonthDay{2021}{{\APACmonth{11}}}{}.
\newblock
{\BBOQ}\APACrefatitle {{Transitional regime of electron resonant interaction
  with whistler-mode waves in inhomogeneous space plasma}} {{Transitional
  regime of electron resonant interaction with whistler-mode waves in
  inhomogeneous space plasma}}.{\BBCQ}
\newblock
\APACjournalVolNumPages{\pre}{104}{5}{055203}.
\newblock
\begin{APACrefDOI} \doi{10.1103/PhysRevE.104.055203} \end{APACrefDOI}
\PrintBackRefs{\CurrentBib}

\bibitem [\protect \citeauthoryear {%
{Artemyev}%
, {Zhang}%
\BCBL {}\ \protect \BOthers {.}}{%
{Artemyev}%
, {Zhang}%
\BCBL {}\ \protect \BOthers {.}}{%
{\protect \APACyear {2022}}%
}]{%
Artemyev22:jgr:Landau&ELFIN}
\APACinsertmetastar {%
Artemyev22:jgr:Landau&ELFIN}%
\begin{APACrefauthors}%
{Artemyev}, A\BPBI V.%
, {Zhang}, X\BPBI J.%
, {Zou}, Y.%
, {Mourenas}, D.%
, {Angelopoulos}, V.%
, {Vainchtein}, D.%
\BDBL {}{Wilkins}, C.%
\end{APACrefauthors}%
\unskip\
\newblock
\APACrefYearMonthDay{2022}{{\APACmonth{06}}}{}.
\newblock
{\BBOQ}\APACrefatitle {{On the Nature of Intense Sub-Relativistic Electron
  Precipitation}} {{On the Nature of Intense Sub-Relativistic Electron
  Precipitation}}.{\BBCQ}
\newblock
\APACjournalVolNumPages{Journal of Geophysical Research (Space
  Physics)}{127}{6}{e30571}.
\newblock
\begin{APACrefDOI} \doi{10.1029/2022JA030571} \end{APACrefDOI}
\PrintBackRefs{\CurrentBib}

\bibitem [\protect \citeauthoryear {%
{Aryan}%
, {Sibeck}%
, {Balikhin}%
, {Agapitov}%
\BCBL {}\ \BBA {} {Kletzing}%
}{%
{Aryan}%
\ \protect \BOthers {.}}{%
{\protect \APACyear {2016}}%
}]{%
Aryan16}
\APACinsertmetastar {%
Aryan16}%
\begin{APACrefauthors}%
{Aryan}, H.%
, {Sibeck}, D.%
, {Balikhin}, M.%
, {Agapitov}, O.%
\BCBL {}\ \BBA {} {Kletzing}, C.%
\end{APACrefauthors}%
\unskip\
\newblock
\APACrefYearMonthDay{2016}{{\APACmonth{08}}}{}.
\newblock
{\BBOQ}\APACrefatitle {{Observation of chorus waves by the Van Allen Probes:
  Dependence on solar wind parameters and scale size}} {{Observation of chorus
  waves by the Van Allen Probes: Dependence on solar wind parameters and scale
  size}}.{\BBCQ}
\newblock
\APACjournalVolNumPages{Journal of Geophysical Research (Space
  Physics)}{121}{8}{7608-7621}.
\newblock
\begin{APACrefDOI} \doi{10.1002/2016JA022775} \end{APACrefDOI}
\PrintBackRefs{\CurrentBib}

\bibitem [\protect \citeauthoryear {%
{Blake}%
\ \BBA {} {O'Brien}%
}{%
{Blake}%
\ \BBA {} {O'Brien}%
}{%
{\protect \APACyear {2016}}%
}]{%
Blake&OBrien16}
\APACinsertmetastar {%
Blake&OBrien16}%
\begin{APACrefauthors}%
{Blake}, J\BPBI B.%
\BCBT {}\ \BBA {} {O'Brien}, T\BPBI P.%
\end{APACrefauthors}%
\unskip\
\newblock
\APACrefYearMonthDay{2016}{{\APACmonth{04}}}{}.
\newblock
{\BBOQ}\APACrefatitle {{Observations of small-scale latitudinal structure in
  energetic electron precipitation}} {{Observations of small-scale latitudinal
  structure in energetic electron precipitation}}.{\BBCQ}
\newblock
\APACjournalVolNumPages{Journal of Geophysical Research (Space
  Physics)}{121}{4}{3031-3035}.
\newblock
\begin{APACrefDOI} \doi{10.1002/2015JA021815} \end{APACrefDOI}
\PrintBackRefs{\CurrentBib}

\bibitem [\protect \citeauthoryear {%
{Bortnik}%
, {Thorne}%
, {Meredith}%
\BCBL {}\ \BBA {} {Santolik}%
}{%
{Bortnik}%
\ \protect \BOthers {.}}{%
{\protect \APACyear {2007}}%
}]{%
Bortnik07:landau}
\APACinsertmetastar {%
Bortnik07:landau}%
\begin{APACrefauthors}%
{Bortnik}, J.%
, {Thorne}, R\BPBI M.%
, {Meredith}, N\BPBI P.%
\BCBL {}\ \BBA {} {Santolik}, O.%
\end{APACrefauthors}%
\unskip\
\newblock
\APACrefYearMonthDay{2007}{{\APACmonth{08}}}{}.
\newblock
{\BBOQ}\APACrefatitle {{Ray tracing of penetrating chorus and its implications
  for the radiation belts}} {{Ray tracing of penetrating chorus and its
  implications for the radiation belts}}.{\BBCQ}
\newblock
\APACjournalVolNumPages{\grl}{34}{}{L15109}.
\newblock
\begin{APACrefDOI} \doi{10.1029/2007GL030040} \end{APACrefDOI}
\PrintBackRefs{\CurrentBib}

\bibitem [\protect \citeauthoryear {%
{Cai}%
, {Wu}%
\BCBL {}\ \BBA {} {Tao}%
}{%
{Cai}%
\ \protect \BOthers {.}}{%
{\protect \APACyear {2020}}%
}]{%
Cai20:broading}
\APACinsertmetastar {%
Cai20:broading}%
\begin{APACrefauthors}%
{Cai}, B.%
, {Wu}, Y.%
\BCBL {}\ \BBA {} {Tao}, X.%
\end{APACrefauthors}%
\unskip\
\newblock
\APACrefYearMonthDay{2020}{{\APACmonth{06}}}{}.
\newblock
{\BBOQ}\APACrefatitle {{Effects of Nonlinear Resonance Broadening on
  Interactions Between Electrons and Whistler Mode Waves}} {{Effects of
  Nonlinear Resonance Broadening on Interactions Between Electrons and Whistler
  Mode Waves}}.{\BBCQ}
\newblock
\APACjournalVolNumPages{\grl}{47}{11}{e87991}.
\newblock
\begin{APACrefDOI} \doi{10.1029/2020GL087991} \end{APACrefDOI}
\PrintBackRefs{\CurrentBib}

\bibitem [\protect \citeauthoryear {%
L.~{Chen}%
, {Thorne}%
, {Li}%
\BCBL {}\ \BBA {} {Bortnik}%
}{%
L.~{Chen}%
\ \protect \BOthers {.}}{%
{\protect \APACyear {2013}}%
}]{%
Chen13}
\APACinsertmetastar {%
Chen13}%
\begin{APACrefauthors}%
{Chen}, L.%
, {Thorne}, R\BPBI M.%
, {Li}, W.%
\BCBL {}\ \BBA {} {Bortnik}, J.%
\end{APACrefauthors}%
\unskip\
\newblock
\APACrefYearMonthDay{2013}{{\APACmonth{03}}}{}.
\newblock
{\BBOQ}\APACrefatitle {{Modeling the wave normal distribution of chorus waves}}
  {{Modeling the wave normal distribution of chorus waves}}.{\BBCQ}
\newblock
\APACjournalVolNumPages{\jgr}{118}{}{1074-1088}.
\newblock
\begin{APACrefDOI} \doi{10.1029/2012JA018343} \end{APACrefDOI}
\PrintBackRefs{\CurrentBib}

\bibitem [\protect \citeauthoryear {%
L.~{Chen}%
\ \protect \BOthers {.}}{%
L.~{Chen}%
\ \protect \BOthers {.}}{%
{\protect \APACyear {2022}}%
}]{%
Chen22:microbursts}
\APACinsertmetastar {%
Chen22:microbursts}%
\begin{APACrefauthors}%
{Chen}, L.%
, {Zhang}, X\BHBI J.%
, {Artemyev}, A.%
, {Angelopoulos}, V.%
, {Tsai}, E.%
, {Wilkins}, C.%
\BCBL {}\ \BBA {} {Horne}, R\BPBI B.%
\end{APACrefauthors}%
\unskip\
\newblock
\APACrefYearMonthDay{2022}{{\APACmonth{03}}}{}.
\newblock
{\BBOQ}\APACrefatitle {{Ducted Chorus Waves Cause Sub-Relativistic and
  Relativistic Electron Microbursts}} {{Ducted Chorus Waves Cause
  Sub-Relativistic and Relativistic Electron Microbursts}}.{\BBCQ}
\newblock
\APACjournalVolNumPages{\grl}{49}{5}{e97559}.
\newblock
\begin{APACrefDOI} \doi{10.1029/2021GL097559} \end{APACrefDOI}
\PrintBackRefs{\CurrentBib}

\bibitem [\protect \citeauthoryear {%
R.~{Chen}%
\ \protect \BOthers {.}}{%
R.~{Chen}%
\ \protect \BOthers {.}}{%
{\protect \APACyear {2021}}%
}]{%
Chen21:ducting}
\APACinsertmetastar {%
Chen21:ducting}%
\begin{APACrefauthors}%
{Chen}, R.%
, {Gao}, X.%
, {Lu}, Q.%
, {Chen}, L.%
, {Tsurutani}, B\BPBI T.%
, {Li}, W.%
\BDBL {}{Wang}, S.%
\end{APACrefauthors}%
\unskip\
\newblock
\APACrefYearMonthDay{2021}{{\APACmonth{04}}}{}.
\newblock
{\BBOQ}\APACrefatitle {{In Situ Observations of Whistler Mode Chorus Waves
  Guided by Density Ducts}} {{In Situ Observations of Whistler Mode Chorus
  Waves Guided by Density Ducts}}.{\BBCQ}
\newblock
\APACjournalVolNumPages{Journal of Geophysical Research (Space
  Physics)}{126}{4}{e28814}.
\newblock
\begin{APACrefDOI} \doi{10.1029/2020JA028814} \end{APACrefDOI}
\PrintBackRefs{\CurrentBib}

\bibitem [\protect \citeauthoryear {%
{Chu}%
\ \protect \BOthers {.}}{%
{Chu}%
\ \protect \BOthers {.}}{%
{\protect \APACyear {2017}}%
}]{%
Chu17}
\APACinsertmetastar {%
Chu17}%
\begin{APACrefauthors}%
{Chu}, X.%
, {Bortnik}, J.%
, {Li}, W.%
, {Ma}, Q.%
, {Denton}, R.%
, {Yue}, C.%
\BDBL {}{Menietti}, J.%
\end{APACrefauthors}%
\unskip\
\newblock
\APACrefYearMonthDay{2017}{{\APACmonth{09}}}{}.
\newblock
{\BBOQ}\APACrefatitle {{A neural network model of three-dimensional dynamic
  electron density in the inner magnetosphere}} {{A neural network model of
  three-dimensional dynamic electron density in the inner
  magnetosphere}}.{\BBCQ}
\newblock
\APACjournalVolNumPages{Journal of Geophysical Research (Space
  Physics)}{122}{9}{9183-9197}.
\newblock
\begin{APACrefDOI} \doi{10.1002/2017JA024464} \end{APACrefDOI}
\PrintBackRefs{\CurrentBib}

\bibitem [\protect \citeauthoryear {%
{Cully}%
, {Ergun}%
, {Stevens}%
, {Nammari}%
\BCBL {}\ \BBA {} {Westfall}%
}{%
{Cully}%
\ \protect \BOthers {.}}{%
{\protect \APACyear {2008}}%
}]{%
Cully08:ssr}
\APACinsertmetastar {%
Cully08:ssr}%
\begin{APACrefauthors}%
{Cully}, C\BPBI M.%
, {Ergun}, R\BPBI E.%
, {Stevens}, K.%
, {Nammari}, A.%
\BCBL {}\ \BBA {} {Westfall}, J.%
\end{APACrefauthors}%
\unskip\
\newblock
\APACrefYearMonthDay{2008}{{\APACmonth{12}}}{}.
\newblock
{\BBOQ}\APACrefatitle {{The THEMIS Digital Fields Board}} {{The THEMIS Digital
  Fields Board}}.{\BBCQ}
\newblock
\APACjournalVolNumPages{\ssr}{141}{}{343-355}.
\newblock
\begin{APACrefDOI} \doi{10.1007/s11214-008-9417-1} \end{APACrefDOI}
\PrintBackRefs{\CurrentBib}

\bibitem [\protect \citeauthoryear {%
{Darrouzet}%
\ \protect \BOthers {.}}{%
{Darrouzet}%
\ \protect \BOthers {.}}{%
{\protect \APACyear {2009}}%
}]{%
Darrouzet09}
\APACinsertmetastar {%
Darrouzet09}%
\begin{APACrefauthors}%
{Darrouzet}, F.%
, {Gallagher}, D\BPBI L.%
, {Andr{\'e}}, N.%
, {Carpenter}, D\BPBI L.%
, {Dandouras}, I.%
, {D{\'e}cr{\'e}au}, P\BPBI M\BPBI E.%
\BDBL {}{Tu}, J.%
\end{APACrefauthors}%
\unskip\
\newblock
\APACrefYearMonthDay{2009}{{\APACmonth{05}}}{}.
\newblock
{\BBOQ}\APACrefatitle {{Plasmaspheric Density Structures and Dynamics:
  Properties Observed by the CLUSTER and IMAGE Missions}} {{Plasmaspheric
  Density Structures and Dynamics: Properties Observed by the CLUSTER and IMAGE
  Missions}}.{\BBCQ}
\newblock
\APACjournalVolNumPages{\ssr}{145}{}{55-106}.
\newblock
\begin{APACrefDOI} \doi{10.1007/s11214-008-9438-9} \end{APACrefDOI}
\PrintBackRefs{\CurrentBib}

\bibitem [\protect \citeauthoryear {%
{Demekhov}%
}{%
{Demekhov}%
}{%
{\protect \APACyear {2011}}%
}]{%
Demekhov11}
\APACinsertmetastar {%
Demekhov11}%
\begin{APACrefauthors}%
{Demekhov}, A\BPBI G.%
\end{APACrefauthors}%
\unskip\
\newblock
\APACrefYearMonthDay{2011}{{\APACmonth{04}}}{}.
\newblock
{\BBOQ}\APACrefatitle {{Generation of VLF emissions with the increasing and
  decreasing frequency in the magnetosperic cyclotron maser in the backward
  wave oscillator regime}} {{Generation of VLF emissions with the increasing
  and decreasing frequency in the magnetosperic cyclotron maser in the backward
  wave oscillator regime}}.{\BBCQ}
\newblock
\APACjournalVolNumPages{Radiophysics and Quantum Electronics}{53}{}{609-622}.
\newblock
\begin{APACrefDOI} \doi{10.1007/s11141-011-9256-x} \end{APACrefDOI}
\PrintBackRefs{\CurrentBib}

\bibitem [\protect \citeauthoryear {%
{Demekhov}%
, {Taubenschuss}%
\BCBL {}\ \BBA {} {Santol{\'{\i}}k}%
}{%
{Demekhov}%
\ \protect \BOthers {.}}{%
{\protect \APACyear {2017}}%
}]{%
Demekhov17}
\APACinsertmetastar {%
Demekhov17}%
\begin{APACrefauthors}%
{Demekhov}, A\BPBI G.%
, {Taubenschuss}, U.%
\BCBL {}\ \BBA {} {Santol{\'{\i}}k}, O.%
\end{APACrefauthors}%
\unskip\
\newblock
\APACrefYearMonthDay{2017}{{\APACmonth{01}}}{}.
\newblock
{\BBOQ}\APACrefatitle {{Simulation of VLF chorus emissions in the magnetosphere
  and comparison with THEMIS spacecraft data}} {{Simulation of VLF chorus
  emissions in the magnetosphere and comparison with THEMIS spacecraft
  data}}.{\BBCQ}
\newblock
\APACjournalVolNumPages{\jgr}{122}{}{166-184}.
\newblock
\begin{APACrefDOI} \doi{10.1002/2016JA023057} \end{APACrefDOI}
\PrintBackRefs{\CurrentBib}

\bibitem [\protect \citeauthoryear {%
{Douma}%
\ \protect \BOthers {.}}{%
{Douma}%
\ \protect \BOthers {.}}{%
{\protect \APACyear {2019}}%
}]{%
Douma19}
\APACinsertmetastar {%
Douma19}%
\begin{APACrefauthors}%
{Douma}, E.%
, {Rodger}, C\BPBI J.%
, {Blum}, L\BPBI W.%
, {O'Brien}, T\BPBI P.%
, {Clilverd}, M\BPBI A.%
\BCBL {}\ \BBA {} {Blake}, J\BPBI B.%
\end{APACrefauthors}%
\unskip\
\newblock
\APACrefYearMonthDay{2019}{{\APACmonth{07}}}{}.
\newblock
{\BBOQ}\APACrefatitle {{Characteristics of Relativistic Microburst Intensity
  From SAMPEX Observations}} {{Characteristics of Relativistic Microburst
  Intensity From SAMPEX Observations}}.{\BBCQ}
\newblock
\APACjournalVolNumPages{Journal of Geophysical Research (Space
  Physics)}{124}{7}{5627-5640}.
\newblock
\begin{APACrefDOI} \doi{10.1029/2019JA026757} \end{APACrefDOI}
\PrintBackRefs{\CurrentBib}

\bibitem [\protect \citeauthoryear {%
{Frantsuzov}%
, {Artemyev}%
, {Shustov}%
\BCBL {}\ \BBA {} {Zhang}%
}{%
{Frantsuzov}%
\ \protect \BOthers {.}}{%
{\protect \APACyear {2022}}%
}]{%
Frantsuzov22:pop}
\APACinsertmetastar {%
Frantsuzov22:pop}%
\begin{APACrefauthors}%
{Frantsuzov}, V\BPBI A.%
, {Artemyev}, A\BPBI V.%
, {Shustov}, P\BPBI I.%
\BCBL {}\ \BBA {} {Zhang}, X\BPBI J.%
\end{APACrefauthors}%
\unskip\
\newblock
\APACrefYearMonthDay{2022}{{\APACmonth{05}}}{}.
\newblock
{\BBOQ}\APACrefatitle {{Marginal stability of whistler-mode waves in plasma
  with multiple electron populations}} {{Marginal stability of whistler-mode
  waves in plasma with multiple electron populations}}.{\BBCQ}
\newblock
\APACjournalVolNumPages{Physics of Plasmas}{29}{5}{052901}.
\newblock
\begin{APACrefDOI} \doi{10.1063/5.0085953FULL: /proj/ads/abstracts/}
  \end{APACrefDOI}
\PrintBackRefs{\CurrentBib}

\bibitem [\protect \citeauthoryear {%
{Fu}%
\ \protect \BOthers {.}}{%
{Fu}%
\ \protect \BOthers {.}}{%
{\protect \APACyear {2014}}%
}]{%
Fu14:radiation_belts}
\APACinsertmetastar {%
Fu14:radiation_belts}%
\begin{APACrefauthors}%
{Fu}, X.%
, {Cowee}, M\BPBI M.%
, {Friedel}, R\BPBI H.%
, {Funsten}, H\BPBI O.%
, {Gary}, S\BPBI P.%
, {Hospodarsky}, G\BPBI B.%
\BDBL {}{Winske}, D.%
\end{APACrefauthors}%
\unskip\
\newblock
\APACrefYearMonthDay{2014}{{\APACmonth{10}}}{}.
\newblock
{\BBOQ}\APACrefatitle {{Whistler anisotropy instabilities as the source of
  banded chorus: Van Allen Probes observations and particle-in-cell
  simulations}} {{Whistler anisotropy instabilities as the source of banded
  chorus: Van Allen Probes observations and particle-in-cell
  simulations}}.{\BBCQ}
\newblock
\APACjournalVolNumPages{Journal of Geophysical Research (Space
  Physics)}{119}{}{8288-8298}.
\newblock
\begin{APACrefDOI} \doi{10.1002/2014JA020364} \end{APACrefDOI}
\PrintBackRefs{\CurrentBib}

\bibitem [\protect \citeauthoryear {%
{Gan}%
, {Li}%
, {Ma}%
, {Artemyev}%
\BCBL {}\ \BBA {} {Albert}%
}{%
{Gan}%
\ \protect \BOthers {.}}{%
{\protect \APACyear {2022}}%
}]{%
Gan22}
\APACinsertmetastar {%
Gan22}%
\begin{APACrefauthors}%
{Gan}, L.%
, {Li}, W.%
, {Ma}, Q.%
, {Artemyev}, A\BPBI V.%
\BCBL {}\ \BBA {} {Albert}, J\BPBI M.%
\end{APACrefauthors}%
\unskip\
\newblock
\APACrefYearMonthDay{2022}{{\APACmonth{05}}}{}.
\newblock
{\BBOQ}\APACrefatitle {{Dependence of Nonlinear Effects on Whistler-Mode Wave
  Bandwidth and Amplitude: A Perspective From Diffusion Coefficients}}
  {{Dependence of Nonlinear Effects on Whistler-Mode Wave Bandwidth and
  Amplitude: A Perspective From Diffusion Coefficients}}.{\BBCQ}
\newblock
\APACjournalVolNumPages{Journal of Geophysical Research (Space
  Physics)}{127}{5}{e30063}.
\newblock
\begin{APACrefDOI} \doi{10.1029/2021JA030063} \end{APACrefDOI}
\PrintBackRefs{\CurrentBib}

\bibitem [\protect \citeauthoryear {%
{Goldstein}%
\ \protect \BOthers {.}}{%
{Goldstein}%
\ \protect \BOthers {.}}{%
{\protect \APACyear {2014}}%
}]{%
Goldstein14}
\APACinsertmetastar {%
Goldstein14}%
\begin{APACrefauthors}%
{Goldstein}, J.%
, {de Pascuale}, S.%
, {Kletzing}, C.%
, {Kurth}, W.%
, {Genestreti}, K\BPBI J.%
, {Skoug}, R\BPBI M.%
\BDBL {}{Spence}, H.%
\end{APACrefauthors}%
\unskip\
\newblock
\APACrefYearMonthDay{2014}{{\APACmonth{09}}}{}.
\newblock
{\BBOQ}\APACrefatitle {{Simulation of Van Allen Probes plasmapause encounters}}
  {{Simulation of Van Allen Probes plasmapause encounters}}.{\BBCQ}
\newblock
\APACjournalVolNumPages{\jgr}{119}{}{7464-7484}.
\newblock
\begin{APACrefDOI} \doi{10.1002/2014JA020252} \end{APACrefDOI}
\PrintBackRefs{\CurrentBib}

\bibitem [\protect \citeauthoryear {%
{Hanzelka}%
\ \BBA {} {Santol{\'\i}k}%
}{%
{Hanzelka}%
\ \BBA {} {Santol{\'\i}k}%
}{%
{\protect \APACyear {2019}}%
}]{%
Hanzelka&Santolik19}
\APACinsertmetastar {%
Hanzelka&Santolik19}%
\begin{APACrefauthors}%
{Hanzelka}, M.%
\BCBT {}\ \BBA {} {Santol{\'\i}k}, O.%
\end{APACrefauthors}%
\unskip\
\newblock
\APACrefYearMonthDay{2019}{{\APACmonth{06}}}{}.
\newblock
{\BBOQ}\APACrefatitle {{Effects of Ducting on Whistler Mode Chorus or Exohiss
  in the Outer Radiation Belt}} {{Effects of Ducting on Whistler Mode Chorus or
  Exohiss in the Outer Radiation Belt}}.{\BBCQ}
\newblock
\APACjournalVolNumPages{\grl}{46}{11}{5735-5745}.
\newblock
\begin{APACrefDOI} \doi{10.1029/2019GL083115} \end{APACrefDOI}
\PrintBackRefs{\CurrentBib}

\bibitem [\protect \citeauthoryear {%
{Hartinger}%
, {Turner}%
, {Plaschke}%
, {Angelopoulos}%
\BCBL {}\ \BBA {} {Singer}%
}{%
{Hartinger}%
\ \protect \BOthers {.}}{%
{\protect \APACyear {2013}}%
}]{%
Hartinger13}
\APACinsertmetastar {%
Hartinger13}%
\begin{APACrefauthors}%
{Hartinger}, M\BPBI D.%
, {Turner}, D\BPBI L.%
, {Plaschke}, F.%
, {Angelopoulos}, V.%
\BCBL {}\ \BBA {} {Singer}, H.%
\end{APACrefauthors}%
\unskip\
\newblock
\APACrefYearMonthDay{2013}{{\APACmonth{01}}}{}.
\newblock
{\BBOQ}\APACrefatitle {{The role of transient ion foreshock phenomena in
  driving Pc5 ULF wave activity}} {{The role of transient ion foreshock
  phenomena in driving Pc5 ULF wave activity}}.{\BBCQ}
\newblock
\APACjournalVolNumPages{\jgr}{118}{}{299-312}.
\newblock
\begin{APACrefDOI} \doi{10.1029/2012JA018349} \end{APACrefDOI}
\PrintBackRefs{\CurrentBib}

\bibitem [\protect \citeauthoryear {%
{Hartinger}%
, {Welling}%
, {Viall}%
, {Moldwin}%
\BCBL {}\ \BBA {} {Ridley}%
}{%
{Hartinger}%
\ \protect \BOthers {.}}{%
{\protect \APACyear {2014}}%
}]{%
Hartinger14}
\APACinsertmetastar {%
Hartinger14}%
\begin{APACrefauthors}%
{Hartinger}, M\BPBI D.%
, {Welling}, D.%
, {Viall}, N\BPBI M.%
, {Moldwin}, M\BPBI B.%
\BCBL {}\ \BBA {} {Ridley}, A.%
\end{APACrefauthors}%
\unskip\
\newblock
\APACrefYearMonthDay{2014}{{\APACmonth{10}}}{}.
\newblock
{\BBOQ}\APACrefatitle {{The effect of magnetopause motion on fast mode
  resonance}} {{The effect of magnetopause motion on fast mode
  resonance}}.{\BBCQ}
\newblock
\APACjournalVolNumPages{\jgr}{119}{}{8212-8227}.
\newblock
\begin{APACrefDOI} \doi{10.1002/2014JA020401} \end{APACrefDOI}
\PrintBackRefs{\CurrentBib}

\bibitem [\protect \citeauthoryear {%
{Hiraga}%
\ \BBA {} {Omura}%
}{%
{Hiraga}%
\ \BBA {} {Omura}%
}{%
{\protect \APACyear {2020}}%
}]{%
Hiraga&Omura20}
\APACinsertmetastar {%
Hiraga&Omura20}%
\begin{APACrefauthors}%
{Hiraga}, R.%
\BCBT {}\ \BBA {} {Omura}, Y.%
\end{APACrefauthors}%
\unskip\
\newblock
\APACrefYearMonthDay{2020}{{\APACmonth{02}}}{}.
\newblock
{\BBOQ}\APACrefatitle {{Acceleration mechanism of radiation belt electrons
  through interaction with multi-subpacket chorus waves}} {{Acceleration
  mechanism of radiation belt electrons through interaction with
  multi-subpacket chorus waves}}.{\BBCQ}
\newblock
\APACjournalVolNumPages{Earth, Planets, and Space}{72}{1}{21}.
\newblock
\begin{APACrefDOI} \doi{10.1186/s40623-020-1134-3} \end{APACrefDOI}
\PrintBackRefs{\CurrentBib}

\bibitem [\protect \citeauthoryear {%
{Horne}%
\ \protect \BOthers {.}}{%
{Horne}%
\ \protect \BOthers {.}}{%
{\protect \APACyear {2013}}%
}]{%
Horne13:jgr}
\APACinsertmetastar {%
Horne13:jgr}%
\begin{APACrefauthors}%
{Horne}, R\BPBI B.%
, {Kersten}, T.%
, {Glauert}, S\BPBI A.%
, {Meredith}, N\BPBI P.%
, {Boscher}, D.%
, {Sicard-Piet}, A.%
\BDBL {}{Li}, W.%
\end{APACrefauthors}%
\unskip\
\newblock
\APACrefYearMonthDay{2013}{{\APACmonth{10}}}{}.
\newblock
{\BBOQ}\APACrefatitle {{A new diffusion matrix for whistler mode chorus waves}}
  {{A new diffusion matrix for whistler mode chorus waves}}.{\BBCQ}
\newblock
\APACjournalVolNumPages{\jgr}{118}{}{6302-6318}.
\newblock
\begin{APACrefDOI} \doi{10.1002/jgra.50594} \end{APACrefDOI}
\PrintBackRefs{\CurrentBib}

\bibitem [\protect \citeauthoryear {%
{Hosseini}%
, {Agapitov}%
, {Harid}%
\BCBL {}\ \BBA {} {Go{\l}kowski}%
}{%
{Hosseini}%
\ \protect \BOthers {.}}{%
{\protect \APACyear {2021}}%
}]{%
Hosseini21:ducts}
\APACinsertmetastar {%
Hosseini21:ducts}%
\begin{APACrefauthors}%
{Hosseini}, P.%
, {Agapitov}, O.%
, {Harid}, V.%
\BCBL {}\ \BBA {} {Go{\l}kowski}, M.%
\end{APACrefauthors}%
\unskip\
\newblock
\APACrefYearMonthDay{2021}{{\APACmonth{03}}}{}.
\newblock
{\BBOQ}\APACrefatitle {{Evidence of Small Scale Plasma Irregularity Effects on
  Whistler Mode Chorus Propagation}} {{Evidence of Small Scale Plasma
  Irregularity Effects on Whistler Mode Chorus Propagation}}.{\BBCQ}
\newblock
\APACjournalVolNumPages{\grl}{48}{5}{e92850}.
\newblock
\begin{APACrefDOI} \doi{10.1029/2021GL092850} \end{APACrefDOI}
\PrintBackRefs{\CurrentBib}

\bibitem [\protect \citeauthoryear {%
{Hsieh}%
, {Kubota}%
\BCBL {}\ \BBA {} {Omura}%
}{%
{Hsieh}%
\ \protect \BOthers {.}}{%
{\protect \APACyear {2020}}%
}]{%
Hsieh20}
\APACinsertmetastar {%
Hsieh20}%
\begin{APACrefauthors}%
{Hsieh}, Y\BHBI K.%
, {Kubota}, Y.%
\BCBL {}\ \BBA {} {Omura}, Y.%
\end{APACrefauthors}%
\unskip\
\newblock
\APACrefYearMonthDay{2020}{}{}.
\newblock
{\BBOQ}\APACrefatitle {Nonlinear Evolution of Radiation Belt Electron Fluxes
  Interacting With Oblique Whistler Mode Chorus Emissions} {Nonlinear evolution
  of radiation belt electron fluxes interacting with oblique whistler mode
  chorus emissions}.{\BBCQ}
\newblock
\APACjournalVolNumPages{Journal of Geophysical Research: Space
  Physics}{}{}{e2019JA027465}.
\newblock
\begin{APACrefURL}
  \url{https://agupubs.onlinelibrary.wiley.com/doi/abs/10.1029/2019JA027465}
  \end{APACrefURL}
\newblock
\APACrefnote{e2019JA027465 2019JA027465}
\newblock
\begin{APACrefDOI} \doi{10.1029/2019JA027465} \end{APACrefDOI}
\PrintBackRefs{\CurrentBib}

\bibitem [\protect \citeauthoryear {%
{Hsieh}%
\ \BBA {} {Omura}%
}{%
{Hsieh}%
\ \BBA {} {Omura}%
}{%
{\protect \APACyear {2017}}%
}]{%
Hsieh&Omura17:radio_science}
\APACinsertmetastar {%
Hsieh&Omura17:radio_science}%
\begin{APACrefauthors}%
{Hsieh}, Y\BHBI K.%
\BCBT {}\ \BBA {} {Omura}, Y.%
\end{APACrefauthors}%
\unskip\
\newblock
\APACrefYearMonthDay{2017}{}{}.
\newblock
{\BBOQ}\APACrefatitle {Study of Wave-Particle Interactions for Whistler Mode
  Waves at Oblique Angles by Utilizing the Gyroaveraging Method} {Study of
  wave-particle interactions for whistler mode waves at oblique angles by
  utilizing the gyroaveraging method}.{\BBCQ}
\newblock
\APACjournalVolNumPages{Radio Science}{52}{10}{1268--1281}.
\newblock
\begin{APACrefURL} \url{http://dx.doi.org/10.1002/2017RS006245}
  \end{APACrefURL}
\newblock
\APACrefnote{2017RS006245}
\newblock
\begin{APACrefDOI} \doi{10.1002/2017RS006245} \end{APACrefDOI}
\PrintBackRefs{\CurrentBib}

\bibitem [\protect \citeauthoryear {%
{Hsieh}%
, {Omura}%
\BCBL {}\ \BBA {} {Kubota}%
}{%
{Hsieh}%
\ \protect \BOthers {.}}{%
{\protect \APACyear {2022}}%
}]{%
Hsieh22}
\APACinsertmetastar {%
Hsieh22}%
\begin{APACrefauthors}%
{Hsieh}, Y\BHBI K.%
, {Omura}, Y.%
\BCBL {}\ \BBA {} {Kubota}, Y.%
\end{APACrefauthors}%
\unskip\
\newblock
\APACrefYearMonthDay{2022}{{\APACmonth{01}}}{}.
\newblock
{\BBOQ}\APACrefatitle {{Energetic Electron Precipitation Induced by Oblique
  Whistler Mode Chorus Emissions}} {{Energetic Electron Precipitation Induced
  by Oblique Whistler Mode Chorus Emissions}}.{\BBCQ}
\newblock
\APACjournalVolNumPages{Journal of Geophysical Research (Space
  Physics)}{127}{1}{e29583}.
\newblock
\begin{APACrefDOI} \doi{10.1029/2021JA029583} \end{APACrefDOI}
\PrintBackRefs{\CurrentBib}

\bibitem [\protect \citeauthoryear {%
{Karimabadi}%
, {Krauss-Varban}%
\BCBL {}\ \BBA {} {Terasawa}%
}{%
{Karimabadi}%
\ \protect \BOthers {.}}{%
{\protect \APACyear {1992}}%
}]{%
Karimabadi92}
\APACinsertmetastar {%
Karimabadi92}%
\begin{APACrefauthors}%
{Karimabadi}, H.%
, {Krauss-Varban}, D.%
\BCBL {}\ \BBA {} {Terasawa}, T.%
\end{APACrefauthors}%
\unskip\
\newblock
\APACrefYearMonthDay{1992}{{\APACmonth{09}}}{}.
\newblock
{\BBOQ}\APACrefatitle {{Physics of pitch angle scattering and velocity
  diffusion 1. Theory}} {{Physics of pitch angle scattering and velocity
  diffusion 1. Theory}}.{\BBCQ}
\newblock
\APACjournalVolNumPages{\jgr}{97}{A9}{13853-13864}.
\newblock
\begin{APACrefDOI} \doi{10.1029/92JA00997} \end{APACrefDOI}
\PrintBackRefs{\CurrentBib}

\bibitem [\protect \citeauthoryear {%
{Karpman}%
\ \BBA {} {Shklyar}%
}{%
{Karpman}%
\ \BBA {} {Shklyar}%
}{%
{\protect \APACyear {1975}}%
}]{%
Karpman&Shklyar75}
\APACinsertmetastar {%
Karpman&Shklyar75}%
\begin{APACrefauthors}%
{Karpman}, V\BPBI I.%
\BCBT {}\ \BBA {} {Shklyar}, D\BPBI R.%
\end{APACrefauthors}%
\unskip\
\newblock
\APACrefYearMonthDay{1975}{}{}.
\newblock
{\BBOQ}\APACrefatitle {{Nonlinear Landau damping in an inhomogeneous plasma}}
  {{Nonlinear Landau damping in an inhomogeneous plasma}}.{\BBCQ}
\newblock
\APACjournalVolNumPages{Sov. JETP}{40}{}{53-56}.
\PrintBackRefs{\CurrentBib}

\bibitem [\protect \citeauthoryear {%
{Ke}%
\ \protect \BOthers {.}}{%
{Ke}%
\ \protect \BOthers {.}}{%
{\protect \APACyear {2021}}%
}]{%
Ke21:ducts}
\APACinsertmetastar {%
Ke21:ducts}%
\begin{APACrefauthors}%
{Ke}, Y.%
, {Chen}, L.%
, {Gao}, X.%
, {Lu}, Q.%
, {Wang}, X.%
, {Chen}, R.%
\BDBL {}{Wang}, S.%
\end{APACrefauthors}%
\unskip\
\newblock
\APACrefYearMonthDay{2021}{{\APACmonth{04}}}{}.
\newblock
{\BBOQ}\APACrefatitle {{Whistler Mode Waves Trapped by Density Irregularities
  in the Earth's Magnetosphere}} {{Whistler Mode Waves Trapped by Density
  Irregularities in the Earth's Magnetosphere}}.{\BBCQ}
\newblock
\APACjournalVolNumPages{\grl}{48}{7}{e92305}.
\newblock
\begin{APACrefDOI} \doi{10.1029/2020GL092305} \end{APACrefDOI}
\PrintBackRefs{\CurrentBib}

\bibitem [\protect \citeauthoryear {%
{Kennel}%
}{%
{Kennel}%
}{%
{\protect \APACyear {1969}}%
}]{%
Kennel69}
\APACinsertmetastar {%
Kennel69}%
\begin{APACrefauthors}%
{Kennel}, C\BPBI F.%
\end{APACrefauthors}%
\unskip\
\newblock
\APACrefYearMonthDay{1969}{}{}.
\newblock
{\BBOQ}\APACrefatitle {{Consequences of a magnetospheric plasma.}}
  {{Consequences of a magnetospheric plasma.}}{\BBCQ}
\newblock
\APACjournalVolNumPages{Reviews of Geophysics and Space Physics}{7}{}{379-419}.
\newblock
\begin{APACrefDOI} \doi{10.1029/RG007i001p00379} \end{APACrefDOI}
\PrintBackRefs{\CurrentBib}

\bibitem [\protect \citeauthoryear {%
{Kennel}%
\ \BBA {} {Petschek}%
}{%
{Kennel}%
\ \BBA {} {Petschek}%
}{%
{\protect \APACyear {1966}}%
}]{%
Kennel&Petschek66}
\APACinsertmetastar {%
Kennel&Petschek66}%
\begin{APACrefauthors}%
{Kennel}, C\BPBI F.%
\BCBT {}\ \BBA {} {Petschek}, H\BPBI E.%
\end{APACrefauthors}%
\unskip\
\newblock
\APACrefYearMonthDay{1966}{{\APACmonth{01}}}{}.
\newblock
{\BBOQ}\APACrefatitle {{Limit on Stably Trapped Particle Fluxes}} {{Limit on
  Stably Trapped Particle Fluxes}}.{\BBCQ}
\newblock
\APACjournalVolNumPages{\jgr}{71}{}{1-28}.
\PrintBackRefs{\CurrentBib}

\bibitem [\protect \citeauthoryear {%
{Klimushkin}%
, {Mager}%
, {Zong}%
\BCBL {}\ \BBA {} {Glassmeier}%
}{%
{Klimushkin}%
\ \protect \BOthers {.}}{%
{\protect \APACyear {2019}}%
}]{%
Klimushkin19}
\APACinsertmetastar {%
Klimushkin19}%
\begin{APACrefauthors}%
{Klimushkin}, D\BPBI Y.%
, {Mager}, P\BPBI N.%
, {Zong}, Q.%
\BCBL {}\ \BBA {} {Glassmeier}, K\BHBI H.%
\end{APACrefauthors}%
\unskip\
\newblock
\APACrefYearMonthDay{2019}{Apr}{}.
\newblock
{\BBOQ}\APACrefatitle {{Alfv{\'e}n Wave Generation by a Compact Source Moving
  on the Magnetopause: Asymptotic Solution}} {{Alfv{\'e}n Wave Generation by a
  Compact Source Moving on the Magnetopause: Asymptotic Solution}}.{\BBCQ}
\newblock
\APACjournalVolNumPages{Journal of Geophysical Research (Space
  Physics)}{124}{4}{2720-2735}.
\newblock
\begin{APACrefDOI} \doi{10.1029/2018JA025801} \end{APACrefDOI}
\PrintBackRefs{\CurrentBib}

\bibitem [\protect \citeauthoryear {%
L.~{Li}%
\ \protect \BOthers {.}}{%
L.~{Li}%
\ \protect \BOthers {.}}{%
{\protect \APACyear {2022}}%
}]{%
LiLi22}
\APACinsertmetastar {%
LiLi22}%
\begin{APACrefauthors}%
{Li}, L.%
, {Omura}, Y.%
, {Zhou}, X\BHBI Z.%
, {Zong}, Q\BHBI G.%
, {Rankin}, R.%
, {Yue}, C.%
\BCBL {}\ \BBA {} {Fu}, S\BHBI Y.%
\end{APACrefauthors}%
\unskip\
\newblock
\APACrefYearMonthDay{2022}{{\APACmonth{05}}}{}.
\newblock
{\BBOQ}\APACrefatitle {{Nonlinear Wave Growth Analysis of Chorus Emissions
  Modulated by ULF Waves}} {{Nonlinear Wave Growth Analysis of Chorus Emissions
  Modulated by ULF Waves}}.{\BBCQ}
\newblock
\APACjournalVolNumPages{\grl}{49}{10}{e97978}.
\newblock
\begin{APACrefDOI} \doi{10.1029/2022GL097978} \end{APACrefDOI}
\PrintBackRefs{\CurrentBib}

\bibitem [\protect \citeauthoryear {%
W.~{Li}%
, {Bortnik}%
\BCBL {}\ \protect \BOthers {.}}{%
W.~{Li}%
, {Bortnik}%
\BCBL {}\ \protect \BOthers {.}}{%
{\protect \APACyear {2011}}%
}]{%
Li11:modulation2}
\APACinsertmetastar {%
Li11:modulation2}%
\begin{APACrefauthors}%
{Li}, W.%
, {Bortnik}, J.%
, {Thorne}, R\BPBI M.%
, {Nishimura}, Y.%
, {Angelopoulos}, V.%
\BCBL {}\ \BBA {} {Chen}, L.%
\end{APACrefauthors}%
\unskip\
\newblock
\APACrefYearMonthDay{2011}{{\APACmonth{06}}}{}.
\newblock
{\BBOQ}\APACrefatitle {{Modulation of whistler mode chorus waves: 2. Role of
  density variations}} {{Modulation of whistler mode chorus waves: 2. Role of
  density variations}}.{\BBCQ}
\newblock
\APACjournalVolNumPages{\jgr}{116}{}{A06206}.
\newblock
\begin{APACrefDOI} \doi{10.1029/2010JA016313} \end{APACrefDOI}
\PrintBackRefs{\CurrentBib}

\bibitem [\protect \citeauthoryear {%
W.~{Li}%
\ \protect \BOthers {.}}{%
W.~{Li}%
\ \protect \BOthers {.}}{%
{\protect \APACyear {2013}}%
}]{%
Li13:POES}
\APACinsertmetastar {%
Li13:POES}%
\begin{APACrefauthors}%
{Li}, W.%
, {Ni}, B.%
, {Thorne}, R\BPBI M.%
, {Bortnik}, J.%
, {Green}, J\BPBI C.%
, {Kletzing}, C\BPBI A.%
\BDBL {}{Hospodarsky}, G\BPBI B.%
\end{APACrefauthors}%
\unskip\
\newblock
\APACrefYearMonthDay{2013}{{\APACmonth{09}}}{}.
\newblock
{\BBOQ}\APACrefatitle {{Constructing the global distribution of chorus wave
  intensity using measurements of electrons by the POES satellites and waves by
  the Van Allen Probes}} {{Constructing the global distribution of chorus wave
  intensity using measurements of electrons by the POES satellites and waves by
  the Van Allen Probes}}.{\BBCQ}
\newblock
\APACjournalVolNumPages{\grl}{40}{}{4526-4532}.
\newblock
\begin{APACrefDOI} \doi{10.1002/grl.50920} \end{APACrefDOI}
\PrintBackRefs{\CurrentBib}

\bibitem [\protect \citeauthoryear {%
W.~{Li}%
, {Thorne}%
, {Bortnik}%
, {Nishimura}%
\BCBL {}\ \BBA {} {Angelopoulos}%
}{%
W.~{Li}%
, {Thorne}%
, {Bortnik}%
, {Nishimura}%
\BCBL {}\ \BBA {} {Angelopoulos}%
}{%
{\protect \APACyear {2011}}%
}]{%
Li11:modulation1}
\APACinsertmetastar {%
Li11:modulation1}%
\begin{APACrefauthors}%
{Li}, W.%
, {Thorne}, R\BPBI M.%
, {Bortnik}, J.%
, {Nishimura}, Y.%
\BCBL {}\ \BBA {} {Angelopoulos}, V.%
\end{APACrefauthors}%
\unskip\
\newblock
\APACrefYearMonthDay{2011}{{\APACmonth{06}}}{}.
\newblock
{\BBOQ}\APACrefatitle {{Modulation of whistler mode chorus waves: 1. Role of
  compressional Pc4-5 pulsations}} {{Modulation of whistler mode chorus waves:
  1. Role of compressional Pc4-5 pulsations}}.{\BBCQ}
\newblock
\APACjournalVolNumPages{\jgr}{116}{}{A06205}.
\newblock
\begin{APACrefDOI} \doi{10.1029/2010JA016312} \end{APACrefDOI}
\PrintBackRefs{\CurrentBib}

\bibitem [\protect \citeauthoryear {%
W.~{Li}%
, {Thorne}%
, {Bortnik}%
, {Shprits}%
\BCBL {}\ \protect \BOthers {.}}{%
W.~{Li}%
, {Thorne}%
, {Bortnik}%
, {Shprits}%
\BCBL {}\ \protect \BOthers {.}}{%
{\protect \APACyear {2011}}%
}]{%
Li11:grl}
\APACinsertmetastar {%
Li11:grl}%
\begin{APACrefauthors}%
{Li}, W.%
, {Thorne}, R\BPBI M.%
, {Bortnik}, J.%
, {Shprits}, Y\BPBI Y.%
, {Nishimura}, Y.%
, {Angelopoulos}, V.%
\BDBL {}{Bonnell}, J\BPBI W.%
\end{APACrefauthors}%
\unskip\
\newblock
\APACrefYearMonthDay{2011}{{\APACmonth{07}}}{}.
\newblock
{\BBOQ}\APACrefatitle {{Typical properties of rising and falling tone chorus
  waves}} {{Typical properties of rising and falling tone chorus
  waves}}.{\BBCQ}
\newblock
\APACjournalVolNumPages{\grl}{38}{}{14103}.
\newblock
\begin{APACrefDOI} \doi{10.1029/2011GL047925} \end{APACrefDOI}
\PrintBackRefs{\CurrentBib}

\bibitem [\protect \citeauthoryear {%
{Liu}%
, {Angelopoulos}%
, {Runov}%
\BCBL {}\ \BBA {} {Zhou}%
}{%
{Liu}%
\ \protect \BOthers {.}}{%
{\protect \APACyear {2013}}%
}]{%
Liu13:DF}
\APACinsertmetastar {%
Liu13:DF}%
\begin{APACrefauthors}%
{Liu}, J.%
, {Angelopoulos}, V.%
, {Runov}, A.%
\BCBL {}\ \BBA {} {Zhou}, X\BHBI Z.%
\end{APACrefauthors}%
\unskip\
\newblock
\APACrefYearMonthDay{2013}{{\APACmonth{05}}}{}.
\newblock
{\BBOQ}\APACrefatitle {{On the current sheets surrounding dipolarizing flux
  bundles in the magnetotail: The case for wedgelets}} {{On the current sheets
  surrounding dipolarizing flux bundles in the magnetotail: The case for
  wedgelets}}.{\BBCQ}
\newblock
\APACjournalVolNumPages{\jgr}{118}{}{2000-2020}.
\newblock
\begin{APACrefDOI} \doi{10.1002/jgra.50092} \end{APACrefDOI}
\PrintBackRefs{\CurrentBib}

\bibitem [\protect \citeauthoryear {%
{Lyons}%
\ \BBA {} {Williams}%
}{%
{Lyons}%
\ \BBA {} {Williams}%
}{%
{\protect \APACyear {1984}}%
}]{%
bookLyons&Williams}
\APACinsertmetastar {%
bookLyons&Williams}%
\begin{APACrefauthors}%
{Lyons}, L\BPBI R.%
\BCBT {}\ \BBA {} {Williams}, D\BPBI J.%
\end{APACrefauthors}%
\unskip\
\newblock
\APACrefYear{1984}.
\newblock
\APACrefbtitle {{Quantitative aspects of magnetospheric physics.}}
  {{Quantitative aspects of magnetospheric physics.}}\ ({Lyons, L.~R.~\&
  Williams, D.~J.}, \BED{}).
\PrintBackRefs{\CurrentBib}

\bibitem [\protect \citeauthoryear {%
{Malaspina}%
\ \protect \BOthers {.}}{%
{Malaspina}%
\ \protect \BOthers {.}}{%
{\protect \APACyear {2014}}%
}]{%
Malaspina14}
\APACinsertmetastar {%
Malaspina14}%
\begin{APACrefauthors}%
{Malaspina}, D\BPBI M.%
, {Andersson}, L.%
, {Ergun}, R\BPBI E.%
, {Wygant}, J\BPBI R.%
, {Bonnell}, J\BPBI W.%
, {Kletzing}, C.%
\BDBL {}{Larsen}, B\BPBI A.%
\end{APACrefauthors}%
\unskip\
\newblock
\APACrefYearMonthDay{2014}{{\APACmonth{08}}}{}.
\newblock
{\BBOQ}\APACrefatitle {{Nonlinear electric field structures in the inner
  magnetosphere}} {{Nonlinear electric field structures in the inner
  magnetosphere}}.{\BBCQ}
\newblock
\APACjournalVolNumPages{\grl}{41}{}{5693-5701}.
\newblock
\begin{APACrefDOI} \doi{10.1002/2014GL061109} \end{APACrefDOI}
\PrintBackRefs{\CurrentBib}

\bibitem [\protect \citeauthoryear {%
{Malaspina}%
\ \protect \BOthers {.}}{%
{Malaspina}%
\ \protect \BOthers {.}}{%
{\protect \APACyear {2015}}%
}]{%
Malaspina15}
\APACinsertmetastar {%
Malaspina15}%
\begin{APACrefauthors}%
{Malaspina}, D\BPBI M.%
, {Wygant}, J\BPBI R.%
, {Ergun}, R\BPBI E.%
, {Reeves}, G\BPBI D.%
, {Skoug}, R\BPBI M.%
\BCBL {}\ \BBA {} {Larsen}, B\BPBI A.%
\end{APACrefauthors}%
\unskip\
\newblock
\APACrefYearMonthDay{2015}{}{}.
\newblock
{\BBOQ}\APACrefatitle {Electric field structures and waves at plasma boundaries
  in the inner magnetosphere} {Electric field structures and waves at plasma
  boundaries in the inner magnetosphere}.{\BBCQ}
\newblock
\APACjournalVolNumPages{\jgr}{120}{}{n/a--n/a}.
\newblock
\APACrefnote{2015JA021137}
\newblock
\begin{APACrefDOI} \doi{10.1002/2015JA021137} \end{APACrefDOI}
\PrintBackRefs{\CurrentBib}

\bibitem [\protect \citeauthoryear {%
{Meredith}%
\ \protect \BOthers {.}}{%
{Meredith}%
\ \protect \BOthers {.}}{%
{\protect \APACyear {2012}}%
}]{%
Meredith12}
\APACinsertmetastar {%
Meredith12}%
\begin{APACrefauthors}%
{Meredith}, N\BPBI P.%
, {Horne}, R\BPBI B.%
, {Sicard-Piet}, A.%
, {Boscher}, D.%
, {Yearby}, K\BPBI H.%
, {Li}, W.%
\BCBL {}\ \BBA {} {Thorne}, R\BPBI M.%
\end{APACrefauthors}%
\unskip\
\newblock
\APACrefYearMonthDay{2012}{{\APACmonth{10}}}{}.
\newblock
{\BBOQ}\APACrefatitle {{Global model of lower band and upper band chorus from
  multiple satellite observations}} {{Global model of lower band and upper band
  chorus from multiple satellite observations}}.{\BBCQ}
\newblock
\APACjournalVolNumPages{\jgr}{117}{}{10225}.
\newblock
\begin{APACrefDOI} \doi{10.1029/2012JA017978} \end{APACrefDOI}
\PrintBackRefs{\CurrentBib}

\bibitem [\protect \citeauthoryear {%
{Millan}%
\ \BBA {} {Thorne}%
}{%
{Millan}%
\ \BBA {} {Thorne}%
}{%
{\protect \APACyear {2007}}%
}]{%
Millan&Thorne07}
\APACinsertmetastar {%
Millan&Thorne07}%
\begin{APACrefauthors}%
{Millan}, R\BPBI M.%
\BCBT {}\ \BBA {} {Thorne}, R\BPBI M.%
\end{APACrefauthors}%
\unskip\
\newblock
\APACrefYearMonthDay{2007}{{\APACmonth{03}}}{}.
\newblock
{\BBOQ}\APACrefatitle {{Review of radiation belt relativistic electron losses}}
  {{Review of radiation belt relativistic electron losses}}.{\BBCQ}
\newblock
\APACjournalVolNumPages{Journal of Atmospheric and Solar-Terrestrial
  Physics}{69}{}{362-377}.
\newblock
\begin{APACrefDOI} \doi{10.1016/j.jastp.2006.06.019} \end{APACrefDOI}
\PrintBackRefs{\CurrentBib}

\bibitem [\protect \citeauthoryear {%
{Motoba}%
, {Takahashi}%
, {Gjerloev}%
, {Ohtani}%
\BCBL {}\ \BBA {} {Milling}%
}{%
{Motoba}%
\ \protect \BOthers {.}}{%
{\protect \APACyear {2013}}%
}]{%
Motoba13}
\APACinsertmetastar {%
Motoba13}%
\begin{APACrefauthors}%
{Motoba}, T.%
, {Takahashi}, K.%
, {Gjerloev}, J.%
, {Ohtani}, S.%
\BCBL {}\ \BBA {} {Milling}, D\BPBI K.%
\end{APACrefauthors}%
\unskip\
\newblock
\APACrefYearMonthDay{2013}{{\APACmonth{12}}}{}.
\newblock
{\BBOQ}\APACrefatitle {{The role of compressional Pc5 pulsations in modulating
  precipitation of energetic electrons}} {{The role of compressional Pc5
  pulsations in modulating precipitation of energetic electrons}}.{\BBCQ}
\newblock
\APACjournalVolNumPages{Journal of Geophysical Research (Space
  Physics)}{118}{12}{7728-7739}.
\newblock
\begin{APACrefDOI} \doi{10.1002/2013JA018912} \end{APACrefDOI}
\PrintBackRefs{\CurrentBib}

\bibitem [\protect \citeauthoryear {%
{Mourenas}%
\ \protect \BOthers {.}}{%
{Mourenas}%
\ \protect \BOthers {.}}{%
{\protect \APACyear {2021}}%
}]{%
Mourenas21:jgr:ELFIN}
\APACinsertmetastar {%
Mourenas21:jgr:ELFIN}%
\begin{APACrefauthors}%
{Mourenas}, D.%
, {Artemyev}, A\BPBI V.%
, {Zhang}, X\BPBI J.%
, {Angelopoulos}, V.%
, {Tsai}, E.%
\BCBL {}\ \BBA {} {Wilkins}, C.%
\end{APACrefauthors}%
\unskip\
\newblock
\APACrefYearMonthDay{2021}{{\APACmonth{11}}}{}.
\newblock
{\BBOQ}\APACrefatitle {{Electron Lifetimes and Diffusion Rates Inferred From
  ELFIN Measurements at Low Altitude: First Results}} {{Electron Lifetimes and
  Diffusion Rates Inferred From ELFIN Measurements at Low Altitude: First
  Results}}.{\BBCQ}
\newblock
\APACjournalVolNumPages{Journal of Geophysical Research (Space
  Physics)}{126}{11}{e29757}.
\newblock
\begin{APACrefDOI} \doi{10.1029/2021JA029757} \end{APACrefDOI}
\PrintBackRefs{\CurrentBib}

\bibitem [\protect \citeauthoryear {%
{Mourenas}%
\ \protect \BOthers {.}}{%
{Mourenas}%
\ \protect \BOthers {.}}{%
{\protect \APACyear {2018}}%
}]{%
Mourenas18:jgr}
\APACinsertmetastar {%
Mourenas18:jgr}%
\begin{APACrefauthors}%
{Mourenas}, D.%
, {Zhang}, X\BHBI J.%
, {Artemyev}, A\BPBI V.%
, {Angelopoulos}, V.%
, {Thorne}, R\BPBI M.%
, {Bortnik}, J.%
\BDBL {}{Vasiliev}, A\BPBI A.%
\end{APACrefauthors}%
\unskip\
\newblock
\APACrefYearMonthDay{2018}{{\APACmonth{06}}}{}.
\newblock
{\BBOQ}\APACrefatitle {{Electron Nonlinear Resonant Interaction With Short and
  Intense Parallel Chorus Wave Packets}} {{Electron Nonlinear Resonant
  Interaction With Short and Intense Parallel Chorus Wave Packets}}.{\BBCQ}
\newblock
\APACjournalVolNumPages{\jgr}{123}{}{4979-4999}.
\newblock
\begin{APACrefDOI} \doi{10.1029/2018JA025417} \end{APACrefDOI}
\PrintBackRefs{\CurrentBib}

\bibitem [\protect \citeauthoryear {%
{Mourenas}%
\ \protect \BOthers {.}}{%
{Mourenas}%
\ \protect \BOthers {.}}{%
{\protect \APACyear {2022}}%
}]{%
Mourenas22:jgr:ELFIN}
\APACinsertmetastar {%
Mourenas22:jgr:ELFIN}%
\begin{APACrefauthors}%
{Mourenas}, D.%
, {Zhang}, X\BPBI J.%
, {Nunn}, D.%
, {Artemyev}, A\BPBI V.%
, {Angelopoulos}, V.%
, {Tsai}, E.%
\BCBL {}\ \BBA {} {Wilkins}, C.%
\end{APACrefauthors}%
\unskip\
\newblock
\APACrefYearMonthDay{2022}{{\APACmonth{05}}}{}.
\newblock
{\BBOQ}\APACrefatitle {{Short Chorus Wave Packets: Generation Within Chorus
  Elements, Statistics, and Consequences on Energetic Electron Precipitation}}
  {{Short Chorus Wave Packets: Generation Within Chorus Elements, Statistics,
  and Consequences on Energetic Electron Precipitation}}.{\BBCQ}
\newblock
\APACjournalVolNumPages{Journal of Geophysical Research (Space
  Physics)}{127}{5}{e30310}.
\newblock
\begin{APACrefDOI} \doi{10.1029/2022JA030310} \end{APACrefDOI}
\PrintBackRefs{\CurrentBib}

\bibitem [\protect \citeauthoryear {%
{Nakamura}%
\ \protect \BOthers {.}}{%
{Nakamura}%
\ \protect \BOthers {.}}{%
{\protect \APACyear {2002}}%
}]{%
Nakamura02}
\APACinsertmetastar {%
Nakamura02}%
\begin{APACrefauthors}%
{Nakamura}, R.%
, {Baumjohann}, W.%
, {Klecker}, B.%
, {Bogdanova}, Y.%
, {Balogh}, A.%
, {R{\`e}me}, H.%
\BDBL {}{Runov}, A.%
\end{APACrefauthors}%
\unskip\
\newblock
\APACrefYearMonthDay{2002}{{\APACmonth{10}}}{}.
\newblock
{\BBOQ}\APACrefatitle {{Motion of the dipolarization front during a flow burst
  event observed by Cluster}} {{Motion of the dipolarization front during a
  flow burst event observed by Cluster}}.{\BBCQ}
\newblock
\APACjournalVolNumPages{\grl}{29}{20}{200000-1}.
\newblock
\begin{APACrefDOI} \doi{10.1029/2002GL015763} \end{APACrefDOI}
\PrintBackRefs{\CurrentBib}

\bibitem [\protect \citeauthoryear {%
{Nakamura}%
\ \protect \BOthers {.}}{%
{Nakamura}%
\ \protect \BOthers {.}}{%
{\protect \APACyear {2004}}%
}]{%
Nakamura04}
\APACinsertmetastar {%
Nakamura04}%
\begin{APACrefauthors}%
{Nakamura}, R.%
, {Baumjohann}, W.%
, {Mouikis}, C.%
, {Kistler}, L\BPBI M.%
, {Runov}, A.%
, {Volwerk}, M.%
\BDBL {}{Balogh}, A.%
\end{APACrefauthors}%
\unskip\
\newblock
\APACrefYearMonthDay{2004}{{\APACmonth{05}}}{}.
\newblock
{\BBOQ}\APACrefatitle {{Spatial scale of high-speed flows in the plasma sheet
  observed by Cluster}} {{Spatial scale of high-speed flows in the plasma sheet
  observed by Cluster}}.{\BBCQ}
\newblock
\APACjournalVolNumPages{\grl}{31}{}{9804}.
\newblock
\begin{APACrefDOI} \doi{10.1029/2004GL019558} \end{APACrefDOI}
\PrintBackRefs{\CurrentBib}

\bibitem [\protect \citeauthoryear {%
{Ni}%
\ \protect \BOthers {.}}{%
{Ni}%
\ \protect \BOthers {.}}{%
{\protect \APACyear {2014}}%
}]{%
Ni14}
\APACinsertmetastar {%
Ni14}%
\begin{APACrefauthors}%
{Ni}, B.%
, {Bortnik}, J.%
, {Nishimura}, Y.%
, {Thorne}, R\BPBI M.%
, {Li}, W.%
, {Angelopoulos}, V.%
\BDBL {}{Weatherwax}, A\BPBI T.%
\end{APACrefauthors}%
\unskip\
\newblock
\APACrefYearMonthDay{2014}{{\APACmonth{02}}}{}.
\newblock
{\BBOQ}\APACrefatitle {{Chorus wave scattering responsible for the Earth's
  dayside diffuse auroral precipitation: A detailed case study}} {{Chorus wave
  scattering responsible for the Earth's dayside diffuse auroral precipitation:
  A detailed case study}}.{\BBCQ}
\newblock
\APACjournalVolNumPages{\jgr}{119}{}{897-908}.
\newblock
\begin{APACrefDOI} \doi{10.1002/2013JA019507} \end{APACrefDOI}
\PrintBackRefs{\CurrentBib}

\bibitem [\protect \citeauthoryear {%
{Nishimura}%
\ \protect \BOthers {.}}{%
{Nishimura}%
\ \protect \BOthers {.}}{%
{\protect \APACyear {2020}}%
}]{%
Nishimura20:ssr}
\APACinsertmetastar {%
Nishimura20:ssr}%
\begin{APACrefauthors}%
{Nishimura}, Y.%
, {Lessard}, M\BPBI R.%
, {Katoh}, Y.%
, {Miyoshi}, Y.%
, {Grono}, E.%
, {Partamies}, N.%
\BDBL {}{Kurita}, S.%
\end{APACrefauthors}%
\unskip\
\newblock
\APACrefYearMonthDay{2020}{{\APACmonth{01}}}{}.
\newblock
{\BBOQ}\APACrefatitle {{Diffuse and Pulsating Aurora}} {{Diffuse and Pulsating
  Aurora}}.{\BBCQ}
\newblock
\APACjournalVolNumPages{\ssr}{216}{1}{4}.
\newblock
\begin{APACrefDOI} \doi{10.1007/s11214-019-0629-3} \end{APACrefDOI}
\PrintBackRefs{\CurrentBib}

\bibitem [\protect \citeauthoryear {%
{Nunn}%
, {Rycroft}%
\BCBL {}\ \BBA {} {Trakhtengerts}%
}{%
{Nunn}%
\ \protect \BOthers {.}}{%
{\protect \APACyear {2005}}%
}]{%
Nunn05}
\APACinsertmetastar {%
Nunn05}%
\begin{APACrefauthors}%
{Nunn}, D.%
, {Rycroft}, M.%
\BCBL {}\ \BBA {} {Trakhtengerts}, V.%
\end{APACrefauthors}%
\unskip\
\newblock
\APACrefYearMonthDay{2005}{{\APACmonth{12}}}{}.
\newblock
{\BBOQ}\APACrefatitle {{A parametric study of the numerical simulations of
  triggered VLF emissions}} {{A parametric study of the numerical simulations
  of triggered VLF emissions}}.{\BBCQ}
\newblock
\APACjournalVolNumPages{Annales Geophysicae}{23}{12}{3655-3666}.
\newblock
\begin{APACrefDOI} \doi{10.5194/angeo-23-3655-2005} \end{APACrefDOI}
\PrintBackRefs{\CurrentBib}

\bibitem [\protect \citeauthoryear {%
{Nunn}%
, {Santolik}%
, {Rycroft}%
\BCBL {}\ \BBA {} {Trakhtengerts}%
}{%
{Nunn}%
\ \protect \BOthers {.}}{%
{\protect \APACyear {2009}}%
}]{%
Nunn09}
\APACinsertmetastar {%
Nunn09}%
\begin{APACrefauthors}%
{Nunn}, D.%
, {Santolik}, O.%
, {Rycroft}, M.%
\BCBL {}\ \BBA {} {Trakhtengerts}, V.%
\end{APACrefauthors}%
\unskip\
\newblock
\APACrefYearMonthDay{2009}{{\APACmonth{06}}}{}.
\newblock
{\BBOQ}\APACrefatitle {{On the numerical modelling of VLF chorus dynamical
  spectra}} {{On the numerical modelling of VLF chorus dynamical
  spectra}}.{\BBCQ}
\newblock
\APACjournalVolNumPages{Annales Geophysicae}{27}{}{2341-2359}.
\newblock
\begin{APACrefDOI} \doi{10.5194/angeo-27-2341-2009} \end{APACrefDOI}
\PrintBackRefs{\CurrentBib}

\bibitem [\protect \citeauthoryear {%
{Nunn}%
, {Zhang}%
, {Mourenas}%
\BCBL {}\ \BBA {} {Artemyev}%
}{%
{Nunn}%
\ \protect \BOthers {.}}{%
{\protect \APACyear {2021}}%
}]{%
Nunn21}
\APACinsertmetastar {%
Nunn21}%
\begin{APACrefauthors}%
{Nunn}, D.%
, {Zhang}, X\BPBI J.%
, {Mourenas}, D.%
\BCBL {}\ \BBA {} {Artemyev}, A\BPBI V.%
\end{APACrefauthors}%
\unskip\
\newblock
\APACrefYearMonthDay{2021}{{\APACmonth{04}}}{}.
\newblock
{\BBOQ}\APACrefatitle {{Generation of Realistic Short Chorus Wave Packets}}
  {{Generation of Realistic Short Chorus Wave Packets}}.{\BBCQ}
\newblock
\APACjournalVolNumPages{\grl}{48}{7}{e92178}.
\newblock
\begin{APACrefDOI} \doi{10.1029/2020GL092178} \end{APACrefDOI}
\PrintBackRefs{\CurrentBib}

\bibitem [\protect \citeauthoryear {%
{O'Brien}%
, {Looper}%
\BCBL {}\ \BBA {} {Blake}%
}{%
{O'Brien}%
\ \protect \BOthers {.}}{%
{\protect \APACyear {2004}}%
}]{%
OBrien04}
\APACinsertmetastar {%
OBrien04}%
\begin{APACrefauthors}%
{O'Brien}, T\BPBI P.%
, {Looper}, M\BPBI D.%
\BCBL {}\ \BBA {} {Blake}, J\BPBI B.%
\end{APACrefauthors}%
\unskip\
\newblock
\APACrefYearMonthDay{2004}{{\APACmonth{02}}}{}.
\newblock
{\BBOQ}\APACrefatitle {{Quantification of relativistic electron microburst
  losses during the GEM storms}} {{Quantification of relativistic electron
  microburst losses during the GEM storms}}.{\BBCQ}
\newblock
\APACjournalVolNumPages{\grl}{31}{4}{L04802}.
\newblock
\begin{APACrefDOI} \doi{10.1029/2003GL018621} \end{APACrefDOI}
\PrintBackRefs{\CurrentBib}

\bibitem [\protect \citeauthoryear {%
{Omura}%
}{%
{Omura}%
}{%
{\protect \APACyear {2021}}%
}]{%
Omura21:review}
\APACinsertmetastar {%
Omura21:review}%
\begin{APACrefauthors}%
{Omura}, Y.%
\end{APACrefauthors}%
\unskip\
\newblock
\APACrefYearMonthDay{2021}{{\APACmonth{04}}}{}.
\newblock
{\BBOQ}\APACrefatitle {{Nonlinear wave growth theory of whistler-mode chorus
  and hiss emissions in the magnetosphere}} {{Nonlinear wave growth theory of
  whistler-mode chorus and hiss emissions in the magnetosphere}}.{\BBCQ}
\newblock
\APACjournalVolNumPages{Earth, Planets and Space}{73}{1}{95}.
\newblock
\begin{APACrefDOI} \doi{10.1186/s40623-021-01380-w} \end{APACrefDOI}
\PrintBackRefs{\CurrentBib}

\bibitem [\protect \citeauthoryear {%
{Ross}%
, {Glauert}%
, {Horne}%
, {Watt}%
\BCBL {}\ \BBA {} {Meredith}%
}{%
{Ross}%
\ \protect \BOthers {.}}{%
{\protect \APACyear {2021}}%
}]{%
Ross21}
\APACinsertmetastar {%
Ross21}%
\begin{APACrefauthors}%
{Ross}, J\BPBI P\BPBI J.%
, {Glauert}, S\BPBI A.%
, {Horne}, R\BPBI B.%
, {Watt}, C\BPBI E\BPBI J.%
\BCBL {}\ \BBA {} {Meredith}, N\BPBI P.%
\end{APACrefauthors}%
\unskip\
\newblock
\APACrefYearMonthDay{2021}{}{}.
\newblock
{\BBOQ}\APACrefatitle {On the Variability of EMIC Waves and the Consequences
  for the Relativistic Electron Radiation Belt Population} {On the variability
  of emic waves and the consequences for the relativistic electron radiation
  belt population}.{\BBCQ}
\newblock
\APACjournalVolNumPages{Journal of Geophysical Research: Space
  Physics}{126}{}{e2975426}.
\newblock
\begin{APACrefDOI} \doi{10.1029/2021JA029754} \end{APACrefDOI}
\PrintBackRefs{\CurrentBib}

\bibitem [\protect \citeauthoryear {%
{Runov}%
\ \protect \BOthers {.}}{%
{Runov}%
\ \protect \BOthers {.}}{%
{\protect \APACyear {2011}}%
}]{%
Runov11jgr}
\APACinsertmetastar {%
Runov11jgr}%
\begin{APACrefauthors}%
{Runov}, A.%
, {Angelopoulos}, V.%
, {Zhou}, X\BHBI Z.%
, {Zhang}, X\BHBI J.%
, {Li}, S.%
, {Plaschke}, F.%
\BCBL {}\ \BBA {} {Bonnell}, J.%
\end{APACrefauthors}%
\unskip\
\newblock
\APACrefYearMonthDay{2011}{{\APACmonth{05}}}{}.
\newblock
{\BBOQ}\APACrefatitle {{A THEMIS multicase study of dipolarization fronts in
  the magnetotail plasma sheet}} {{A THEMIS multicase study of dipolarization
  fronts in the magnetotail plasma sheet}}.{\BBCQ}
\newblock
\APACjournalVolNumPages{\jgr}{116}{}{5216}.
\newblock
\begin{APACrefDOI} \doi{10.1029/2010JA016316} \end{APACrefDOI}
\PrintBackRefs{\CurrentBib}

\bibitem [\protect \citeauthoryear {%
{Santol{\'\i}k}%
, {Gurnett}%
\BCBL {}\ \BBA {} {Pickett}%
}{%
{Santol{\'\i}k}%
\ \protect \BOthers {.}}{%
{\protect \APACyear {2004}}%
}]{%
Santolik04}
\APACinsertmetastar {%
Santolik04}%
\begin{APACrefauthors}%
{Santol{\'\i}k}, O.%
, {Gurnett}, D.%
\BCBL {}\ \BBA {} {Pickett}, J.%
\end{APACrefauthors}%
\unskip\
\newblock
\APACrefYearMonthDay{2004}{{\APACmonth{07}}}{}.
\newblock
{\BBOQ}\APACrefatitle {{Multipoint investigation of the source region of
  storm-time chorus}} {{Multipoint investigation of the source region of
  storm-time chorus}}.{\BBCQ}
\newblock
\APACjournalVolNumPages{Annales Geophysicae}{22}{7}{2555-2563}.
\newblock
\begin{APACrefDOI} \doi{10.5194/angeo-22-2555-2004} \end{APACrefDOI}
\PrintBackRefs{\CurrentBib}

\bibitem [\protect \citeauthoryear {%
{Santol{\'\i}k}%
\ \BBA {} {Gurnett}%
}{%
{Santol{\'\i}k}%
\ \BBA {} {Gurnett}%
}{%
{\protect \APACyear {2003}}%
}]{%
Santolik&Gurnett03}
\APACinsertmetastar {%
Santolik&Gurnett03}%
\begin{APACrefauthors}%
{Santol{\'\i}k}, O.%
\BCBT {}\ \BBA {} {Gurnett}, D\BPBI A.%
\end{APACrefauthors}%
\unskip\
\newblock
\APACrefYearMonthDay{2003}{{\APACmonth{01}}}{}.
\newblock
{\BBOQ}\APACrefatitle {{Transverse dimensions of chorus in the source region}}
  {{Transverse dimensions of chorus in the source region}}.{\BBCQ}
\newblock
\APACjournalVolNumPages{\grl}{30}{2}{1031}.
\newblock
\begin{APACrefDOI} \doi{10.1029/2002GL016178} \end{APACrefDOI}
\PrintBackRefs{\CurrentBib}

\bibitem [\protect \citeauthoryear {%
{Santol{\'{\i}}K}%
, {Gurnett}%
, {Pickett}%
, {Parrot}%
\BCBL {}\ \BBA {} {Cornilleau-Wehrlin}%
}{%
{Santol{\'{\i}}K}%
\ \protect \BOthers {.}}{%
{\protect \APACyear {2003}}%
}]{%
Santolik03:storm}
\APACinsertmetastar {%
Santolik03:storm}%
\begin{APACrefauthors}%
{Santol{\'{\i}}K}, O.%
, {Gurnett}, D\BPBI A.%
, {Pickett}, J\BPBI S.%
, {Parrot}, M.%
\BCBL {}\ \BBA {} {Cornilleau-Wehrlin}, N.%
\end{APACrefauthors}%
\unskip\
\newblock
\APACrefYearMonthDay{2003}{{\APACmonth{07}}}{}.
\newblock
{\BBOQ}\APACrefatitle {{Spatio-temporal structure of storm-time chorus}}
  {{Spatio-temporal structure of storm-time chorus}}.{\BBCQ}
\newblock
\APACjournalVolNumPages{\jgr}{108}{}{1278}.
\newblock
\begin{APACrefDOI} \doi{10.1029/2002JA009791} \end{APACrefDOI}
\PrintBackRefs{\CurrentBib}

\bibitem [\protect \citeauthoryear {%
{Schulz}%
\ \BBA {} {Lanzerotti}%
}{%
{Schulz}%
\ \BBA {} {Lanzerotti}%
}{%
{\protect \APACyear {1974}}%
}]{%
bookSchulz&anzerotti74}
\APACinsertmetastar {%
bookSchulz&anzerotti74}%
\begin{APACrefauthors}%
{Schulz}, M.%
\BCBT {}\ \BBA {} {Lanzerotti}, L\BPBI J.%
\end{APACrefauthors}%
\unskip\
\newblock
\APACrefYear{1974}.
\newblock
\APACrefbtitle {{Particle diffusion in the radiation belts}} {{Particle
  diffusion in the radiation belts}}.
\newblock
\APACaddressPublisher{}{Springer, New York}.
\PrintBackRefs{\CurrentBib}

\bibitem [\protect \citeauthoryear {%
{Shklyar}%
}{%
{Shklyar}%
}{%
{\protect \APACyear {2021}}%
}]{%
Shklyar21}
\APACinsertmetastar {%
Shklyar21}%
\begin{APACrefauthors}%
{Shklyar}, D\BPBI R.%
\end{APACrefauthors}%
\unskip\
\newblock
\APACrefYearMonthDay{2021}{{\APACmonth{02}}}{}.
\newblock
{\BBOQ}\APACrefatitle {{A Theory of Interaction Between Relativistic Electrons
  and Magnetospherically Reflected Whistlers}} {{A Theory of Interaction
  Between Relativistic Electrons and Magnetospherically Reflected
  Whistlers}}.{\BBCQ}
\newblock
\APACjournalVolNumPages{Journal of Geophysical Research (Space
  Physics)}{126}{2}{e28799}.
\newblock
\begin{APACrefDOI} \doi{10.1029/2020JA028799} \end{APACrefDOI}
\PrintBackRefs{\CurrentBib}

\bibitem [\protect \citeauthoryear {%
{Shklyar}%
\ \BBA {} {Matsumoto}%
}{%
{Shklyar}%
\ \BBA {} {Matsumoto}%
}{%
{\protect \APACyear {2009}}%
}]{%
Shklyar09:review}
\APACinsertmetastar {%
Shklyar09:review}%
\begin{APACrefauthors}%
{Shklyar}, D\BPBI R.%
\BCBT {}\ \BBA {} {Matsumoto}, H.%
\end{APACrefauthors}%
\unskip\
\newblock
\APACrefYearMonthDay{2009}{{\APACmonth{04}}}{}.
\newblock
{\BBOQ}\APACrefatitle {{Oblique Whistler-Mode Waves in the Inhomogeneous
  Magnetospheric Plasma: Resonant Interactions with Energetic Charged
  Particles}} {{Oblique Whistler-Mode Waves in the Inhomogeneous Magnetospheric
  Plasma: Resonant Interactions with Energetic Charged Particles}}.{\BBCQ}
\newblock
\APACjournalVolNumPages{Surveys in Geophysics}{30}{}{55-104}.
\newblock
\begin{APACrefDOI} \doi{10.1007/s10712-009-9061-7} \end{APACrefDOI}
\PrintBackRefs{\CurrentBib}

\bibitem [\protect \citeauthoryear {%
{Shumko}%
, {Blum}%
\BCBL {}\ \BBA {} {Crew}%
}{%
{Shumko}%
\ \protect \BOthers {.}}{%
{\protect \APACyear {2021}}%
}]{%
Shumko21}
\APACinsertmetastar {%
Shumko21}%
\begin{APACrefauthors}%
{Shumko}, M.%
, {Blum}, L\BPBI W.%
\BCBL {}\ \BBA {} {Crew}, A\BPBI B.%
\end{APACrefauthors}%
\unskip\
\newblock
\APACrefYearMonthDay{2021}{{\APACmonth{09}}}{}.
\newblock
{\BBOQ}\APACrefatitle {{Duration of Individual Relativistic Electron
  Microbursts: A Probe Into Their Scattering Mechanism}} {{Duration of
  Individual Relativistic Electron Microbursts: A Probe Into Their Scattering
  Mechanism}}.{\BBCQ}
\newblock
\APACjournalVolNumPages{\grl}{48}{17}{e93879}.
\newblock
\begin{APACrefDOI} \doi{10.1029/2021GL093879} \end{APACrefDOI}
\PrintBackRefs{\CurrentBib}

\bibitem [\protect \citeauthoryear {%
{Shumko}%
\ \protect \BOthers {.}}{%
{Shumko}%
\ \protect \BOthers {.}}{%
{\protect \APACyear {2020}}%
}]{%
Shumko20}
\APACinsertmetastar {%
Shumko20}%
\begin{APACrefauthors}%
{Shumko}, M.%
, {Johnson}, A\BPBI T.%
, {Sample}, J\BPBI G.%
, {Griffith}, B\BPBI A.%
, {Turner}, D\BPBI L.%
, {O'Brien}, T\BPBI P.%
\BDBL {}{Claudepierre}, S\BPBI G.%
\end{APACrefauthors}%
\unskip\
\newblock
\APACrefYearMonthDay{2020}{{\APACmonth{03}}}{}.
\newblock
{\BBOQ}\APACrefatitle {{Electron Microburst Size Distribution Derived With
  AeroCube-6}} {{Electron Microburst Size Distribution Derived With
  AeroCube-6}}.{\BBCQ}
\newblock
\APACjournalVolNumPages{Journal of Geophysical Research (Space
  Physics)}{125}{3}{e27651}.
\newblock
\begin{APACrefDOI} \doi{10.1029/2019JA027651} \end{APACrefDOI}
\PrintBackRefs{\CurrentBib}

\bibitem [\protect \citeauthoryear {%
{Tao}%
, {Bortnik}%
, {Albert}%
, {Thorne}%
\BCBL {}\ \BBA {} {Li}%
}{%
{Tao}%
\ \protect \BOthers {.}}{%
{\protect \APACyear {2013}}%
}]{%
Tao13}
\APACinsertmetastar {%
Tao13}%
\begin{APACrefauthors}%
{Tao}, X.%
, {Bortnik}, J.%
, {Albert}, J\BPBI M.%
, {Thorne}, R\BPBI M.%
\BCBL {}\ \BBA {} {Li}, W.%
\end{APACrefauthors}%
\unskip\
\newblock
\APACrefYearMonthDay{2013}{{\APACmonth{07}}}{}.
\newblock
{\BBOQ}\APACrefatitle {{The importance of amplitude modulation in nonlinear
  interactions between electrons and large amplitude whistler waves}} {{The
  importance of amplitude modulation in nonlinear interactions between
  electrons and large amplitude whistler waves}}.{\BBCQ}
\newblock
\APACjournalVolNumPages{Journal of Atmospheric and Solar-Terrestrial
  Physics}{99}{}{67-72}.
\newblock
\begin{APACrefDOI} \doi{10.1016/j.jastp.2012.05.012} \end{APACrefDOI}
\PrintBackRefs{\CurrentBib}

\bibitem [\protect \citeauthoryear {%
{Tao}%
\ \protect \BOthers {.}}{%
{Tao}%
\ \protect \BOthers {.}}{%
{\protect \APACyear {2011}}%
}]{%
Tao11}
\APACinsertmetastar {%
Tao11}%
\begin{APACrefauthors}%
{Tao}, X.%
, {Thorne}, R\BPBI M.%
, {Li}, W.%
, {Ni}, B.%
, {Meredith}, N\BPBI P.%
\BCBL {}\ \BBA {} {Horne}, R\BPBI B.%
\end{APACrefauthors}%
\unskip\
\newblock
\APACrefYearMonthDay{2011}{{\APACmonth{04}}}{}.
\newblock
{\BBOQ}\APACrefatitle {{Evolution of electron pitch angle distributions
  following injection from the plasma sheet}} {{Evolution of electron pitch
  angle distributions following injection from the plasma sheet}}.{\BBCQ}
\newblock
\APACjournalVolNumPages{\jgr}{116}{}{A04229}.
\newblock
\begin{APACrefDOI} \doi{10.1029/2010JA016245} \end{APACrefDOI}
\PrintBackRefs{\CurrentBib}

\bibitem [\protect \citeauthoryear {%
{Tao}%
, {Zonca}%
, {Chen}%
\BCBL {}\ \BBA {} {Wu}%
}{%
{Tao}%
\ \protect \BOthers {.}}{%
{\protect \APACyear {2020}}%
}]{%
Tao20}
\APACinsertmetastar {%
Tao20}%
\begin{APACrefauthors}%
{Tao}, X.%
, {Zonca}, F.%
, {Chen}, L.%
\BCBL {}\ \BBA {} {Wu}, Y.%
\end{APACrefauthors}%
\unskip\
\newblock
\APACrefYearMonthDay{2020}{{\APACmonth{01}}}{}.
\newblock
{\BBOQ}\APACrefatitle {{Theoretical and numerical studies of chorus waves: A
  review}} {{Theoretical and numerical studies of chorus waves: A
  review}}.{\BBCQ}
\newblock
\APACjournalVolNumPages{Science China Earth Sciences}{63}{1}{78-92}.
\newblock
\begin{APACrefDOI} \doi{10.1007/s11430-019-9384-6} \end{APACrefDOI}
\PrintBackRefs{\CurrentBib}

\bibitem [\protect \citeauthoryear {%
{Teng}%
, {Tao}%
\BCBL {}\ \BBA {} {Li}%
}{%
{Teng}%
\ \protect \BOthers {.}}{%
{\protect \APACyear {2019}}%
}]{%
Teng19}
\APACinsertmetastar {%
Teng19}%
\begin{APACrefauthors}%
{Teng}, S.%
, {Tao}, X.%
\BCBL {}\ \BBA {} {Li}, W.%
\end{APACrefauthors}%
\unskip\
\newblock
\APACrefYearMonthDay{2019}{{\APACmonth{04}}}{}.
\newblock
{\BBOQ}\APACrefatitle {{Typical Characteristics of Whistler Mode Waves
  Categorized by Their Spectral Properties Using Van Allen Probes
  Observations}} {{Typical Characteristics of Whistler Mode Waves Categorized
  by Their Spectral Properties Using Van Allen Probes Observations}}.{\BBCQ}
\newblock
\APACjournalVolNumPages{\grl}{46}{7}{3607-3614}.
\newblock
\begin{APACrefDOI} \doi{10.1029/2019GL082161} \end{APACrefDOI}
\PrintBackRefs{\CurrentBib}

\bibitem [\protect \citeauthoryear {%
{Teng}%
\ \protect \BOthers {.}}{%
{Teng}%
\ \protect \BOthers {.}}{%
{\protect \APACyear {2017}}%
}]{%
Teng17}
\APACinsertmetastar {%
Teng17}%
\begin{APACrefauthors}%
{Teng}, S.%
, {Tao}, X.%
, {Xie}, Y.%
, {Zonca}, F.%
, {Chen}, L.%
, {Fang}, W\BPBI B.%
\BCBL {}\ \BBA {} {Wang}, S.%
\end{APACrefauthors}%
\unskip\
\newblock
\APACrefYearMonthDay{2017}{Dec}{}.
\newblock
{\BBOQ}\APACrefatitle {{Analysis of the Duration of Rising Tone Chorus
  Elements}} {{Analysis of the Duration of Rising Tone Chorus
  Elements}}.{\BBCQ}
\newblock
\APACjournalVolNumPages{\grl}{44}{24}{12,074-12,082}.
\newblock
\begin{APACrefDOI} \doi{10.1002/2017GL075824} \end{APACrefDOI}
\PrintBackRefs{\CurrentBib}

\bibitem [\protect \citeauthoryear {%
{Thorne}%
, {Bortnik}%
, {Li}%
\BCBL {}\ \BBA {} {Ma}%
}{%
{Thorne}%
\ \protect \BOthers {.}}{%
{\protect \APACyear {2021}}%
}]{%
Thorne21:AGU}
\APACinsertmetastar {%
Thorne21:AGU}%
\begin{APACrefauthors}%
{Thorne}, R\BPBI M.%
, {Bortnik}, J.%
, {Li}, W.%
\BCBL {}\ \BBA {} {Ma}, Q.%
\end{APACrefauthors}%
\unskip\
\newblock
\APACrefYearMonthDay{2021}{}{}.
\newblock
{\BBOQ}\APACrefatitle {Wave–Particle Interactions in the Earth's
  Magnetosphere} {Wave–particle interactions in the earth's
  magnetosphere}.{\BBCQ}
\newblock
\BIn{} \APACrefbtitle {Magnetospheres in the Solar System} {Magnetospheres in
  the solar system}\ (\BPG~93-108).
\newblock
\APACaddressPublisher{}{American Geophysical Union (AGU)}.
\newblock
\begin{APACrefDOI} \doi{https://doi.org/10.1002/9781119815624.ch6}
  \end{APACrefDOI}
\PrintBackRefs{\CurrentBib}

\bibitem [\protect \citeauthoryear {%
{Thorne}%
, {O'Brien}%
, {Shprits}%
, {Summers}%
\BCBL {}\ \BBA {} {Horne}%
}{%
{Thorne}%
\ \protect \BOthers {.}}{%
{\protect \APACyear {2005}}%
}]{%
Thorne05}
\APACinsertmetastar {%
Thorne05}%
\begin{APACrefauthors}%
{Thorne}, R\BPBI M.%
, {O'Brien}, T\BPBI P.%
, {Shprits}, Y\BPBI Y.%
, {Summers}, D.%
\BCBL {}\ \BBA {} {Horne}, R\BPBI B.%
\end{APACrefauthors}%
\unskip\
\newblock
\APACrefYearMonthDay{2005}{{\APACmonth{09}}}{}.
\newblock
{\BBOQ}\APACrefatitle {{Timescale for MeV electron microburst loss during
  geomagnetic storms}} {{Timescale for MeV electron microburst loss during
  geomagnetic storms}}.{\BBCQ}
\newblock
\APACjournalVolNumPages{\jgr}{110}{}{9202}.
\newblock
\begin{APACrefDOI} \doi{10.1029/2004JA010882} \end{APACrefDOI}
\PrintBackRefs{\CurrentBib}

\bibitem [\protect \citeauthoryear {%
{Tsai}%
, {Artemyev}%
, {Zhang}%
\BCBL {}\ \BBA {} {Angelopoulos}%
}{%
{Tsai}%
\ \protect \BOthers {.}}{%
{\protect \APACyear {2022}}%
}]{%
Tsai22}
\APACinsertmetastar {%
Tsai22}%
\begin{APACrefauthors}%
{Tsai}, E.%
, {Artemyev}, A.%
, {Zhang}, X\BHBI J.%
\BCBL {}\ \BBA {} {Angelopoulos}, V.%
\end{APACrefauthors}%
\unskip\
\newblock
\APACrefYearMonthDay{2022}{{\APACmonth{05}}}{}.
\newblock
{\BBOQ}\APACrefatitle {{Relativistic Electron Precipitation Driven by Nonlinear
  Resonance With Whistler-Mode Waves}} {{Relativistic Electron Precipitation
  Driven by Nonlinear Resonance With Whistler-Mode Waves}}.{\BBCQ}
\newblock
\APACjournalVolNumPages{Journal of Geophysical Research (Space
  Physics)}{127}{5}{e30338}.
\newblock
\begin{APACrefDOI} \doi{10.1029/2022JA030338} \end{APACrefDOI}
\PrintBackRefs{\CurrentBib}

\bibitem [\protect \citeauthoryear {%
{Turner}%
, {Fennell}%
\BCBL {}\ \protect \BOthers {.}}{%
{Turner}%
, {Fennell}%
\BCBL {}\ \protect \BOthers {.}}{%
{\protect \APACyear {2017}}%
}]{%
Turner17:injection}
\APACinsertmetastar {%
Turner17:injection}%
\begin{APACrefauthors}%
{Turner}, D\BPBI L.%
, {Fennell}, J\BPBI F.%
, {Blake}, J\BPBI B.%
, {Claudepierre}, S\BPBI G.%
, {Clemmons}, J\BPBI H.%
, {Jaynes}, A\BPBI N.%
\BDBL {}{Reeves}, G\BPBI D.%
\end{APACrefauthors}%
\unskip\
\newblock
\APACrefYearMonthDay{2017}{{\APACmonth{11}}}{}.
\newblock
{\BBOQ}\APACrefatitle {{Multipoint Observations of Energetic Particle
  Injections and Substorm Activity During a Conjunction Between Magnetospheric
  Multiscale (MMS) and Van Allen Probes}} {{Multipoint Observations of
  Energetic Particle Injections and Substorm Activity During a Conjunction
  Between Magnetospheric Multiscale (MMS) and Van Allen Probes}}.{\BBCQ}
\newblock
\APACjournalVolNumPages{Journal of Geophysical Research (Space
  Physics)}{122}{11}{11,481-11,504}.
\newblock
\begin{APACrefDOI} \doi{10.1002/2017JA024554} \end{APACrefDOI}
\PrintBackRefs{\CurrentBib}

\bibitem [\protect \citeauthoryear {%
{Turner}%
, {Lee}%
\BCBL {}\ \protect \BOthers {.}}{%
{Turner}%
, {Lee}%
\BCBL {}\ \protect \BOthers {.}}{%
{\protect \APACyear {2017}}%
}]{%
Turner17:scales}
\APACinsertmetastar {%
Turner17:scales}%
\begin{APACrefauthors}%
{Turner}, D\BPBI L.%
, {Lee}, J\BPBI H.%
, {Claudepierre}, S\BPBI G.%
, {Fennell}, J\BPBI F.%
, {Blake}, J\BPBI B.%
, {Jaynes}, A\BPBI N.%
\BDBL {}{Santolik}, O.%
\end{APACrefauthors}%
\unskip\
\newblock
\APACrefYearMonthDay{2017}{{\APACmonth{11}}}{}.
\newblock
{\BBOQ}\APACrefatitle {{Examining Coherency Scales, Substructure, and
  Propagation of Whistler Mode Chorus Elements With Magnetospheric Multiscale
  (MMS)}} {{Examining Coherency Scales, Substructure, and Propagation of
  Whistler Mode Chorus Elements With Magnetospheric Multiscale (MMS)}}.{\BBCQ}
\newblock
\APACjournalVolNumPages{Journal of Geophysical Research (Space
  Physics)}{122}{11}{11,201-11,226}.
\newblock
\begin{APACrefDOI} \doi{10.1002/2017JA024474} \end{APACrefDOI}
\PrintBackRefs{\CurrentBib}

\bibitem [\protect \citeauthoryear {%
{Watt}%
\ \protect \BOthers {.}}{%
{Watt}%
\ \protect \BOthers {.}}{%
{\protect \APACyear {2011}}%
}]{%
Watt11}
\APACinsertmetastar {%
Watt11}%
\begin{APACrefauthors}%
{Watt}, C\BPBI E\BPBI J.%
, {Degeling}, A\BPBI W.%
, {Rankin}, R.%
, {Murphy}, K\BPBI R.%
, {Rae}, I\BPBI J.%
\BCBL {}\ \BBA {} {Singer}, H\BPBI J.%
\end{APACrefauthors}%
\unskip\
\newblock
\APACrefYearMonthDay{2011}{{\APACmonth{10}}}{}.
\newblock
{\BBOQ}\APACrefatitle {{Ultralow-frequency modulation of whistler-mode wave
  growth}} {{Ultralow-frequency modulation of whistler-mode wave
  growth}}.{\BBCQ}
\newblock
\APACjournalVolNumPages{Journal of Geophysical Research (Space
  Physics)}{116}{A10}{A10209}.
\newblock
\begin{APACrefDOI} \doi{10.1029/2011JA016730} \end{APACrefDOI}
\PrintBackRefs{\CurrentBib}

\bibitem [\protect \citeauthoryear {%
{Wright}%
\ \BBA {} {Elsden}%
}{%
{Wright}%
\ \BBA {} {Elsden}%
}{%
{\protect \APACyear {2020}}%
}]{%
Wright&Elsden20}
\APACinsertmetastar {%
Wright&Elsden20}%
\begin{APACrefauthors}%
{Wright}, A\BPBI N.%
\BCBT {}\ \BBA {} {Elsden}, T.%
\end{APACrefauthors}%
\unskip\
\newblock
\APACrefYearMonthDay{2020}{}{}.
\newblock
{\BBOQ}\APACrefatitle {Simulations of MHD Wave Propagation and Coupling in a
  3-D Magnetosphere} {Simulations of mhd wave propagation and coupling in a 3-d
  magnetosphere}.{\BBCQ}
\newblock
\APACjournalVolNumPages{Journal of Geophysical Research: Space
  Physics}{125}{2}{e2019JA027589}.
\newblock
\begin{APACrefURL}
  \url{https://agupubs.onlinelibrary.wiley.com/doi/abs/10.1029/2019JA027589}
  \end{APACrefURL}
\newblock
\APACrefnote{e2019JA027589 10.1029/2019JA027589}
\newblock
\begin{APACrefDOI} \doi{10.1029/2019JA027589} \end{APACrefDOI}
\PrintBackRefs{\CurrentBib}

\bibitem [\protect \citeauthoryear {%
{Xia}%
\ \protect \BOthers {.}}{%
{Xia}%
\ \protect \BOthers {.}}{%
{\protect \APACyear {2016}}%
}]{%
Xia16:ulf&chorus}
\APACinsertmetastar {%
Xia16:ulf&chorus}%
\begin{APACrefauthors}%
{Xia}, Z.%
, {Chen}, L.%
, {Dai}, L.%
, {Claudepierre}, S\BPBI G.%
, {Chan}, A\BPBI A.%
, {Soto-Chavez}, A\BPBI R.%
\BCBL {}\ \BBA {} {Reeves}, G\BPBI D.%
\end{APACrefauthors}%
\unskip\
\newblock
\APACrefYearMonthDay{2016}{{\APACmonth{09}}}{}.
\newblock
{\BBOQ}\APACrefatitle {{Modulation of chorus intensity by ULF waves deep in the
  inner magnetosphere}} {{Modulation of chorus intensity by ULF waves deep in
  the inner magnetosphere}}.{\BBCQ}
\newblock
\APACjournalVolNumPages{\grl}{43}{}{9444-9452}.
\newblock
\begin{APACrefDOI} \doi{10.1002/2016GL070280} \end{APACrefDOI}
\PrintBackRefs{\CurrentBib}

\bibitem [\protect \citeauthoryear {%
X.~{Zhang}%
, {Angelopoulos}%
, {Artemyev}%
\BCBL {}\ \BBA {} {Liu}%
}{%
X.~{Zhang}%
\ \protect \BOthers {.}}{%
{\protect \APACyear {2018}}%
}]{%
Zhang18:whistlers&injections}
\APACinsertmetastar {%
Zhang18:whistlers&injections}%
\begin{APACrefauthors}%
{Zhang}, X.%
, {Angelopoulos}, V.%
, {Artemyev}, A\BPBI V.%
\BCBL {}\ \BBA {} {Liu}, J.%
\end{APACrefauthors}%
\unskip\
\newblock
\APACrefYearMonthDay{2018}{{\APACmonth{09}}}{}.
\newblock
{\BBOQ}\APACrefatitle {{Whistler and Electron Firehose Instability Control of
  Electron Distributions in and Around Dipolarizing Flux Bundles}} {{Whistler
  and Electron Firehose Instability Control of Electron Distributions in and
  Around Dipolarizing Flux Bundles}}.{\BBCQ}
\newblock
\APACjournalVolNumPages{\grl}{45}{}{9380-9389}.
\newblock
\begin{APACrefDOI} \doi{10.1029/2018GL079613} \end{APACrefDOI}
\PrintBackRefs{\CurrentBib}

\bibitem [\protect \citeauthoryear {%
X\BPBI J.~{Zhang}%
, {Agapitov}%
\BCBL {}\ \protect \BOthers {.}}{%
X\BPBI J.~{Zhang}%
, {Agapitov}%
\BCBL {}\ \protect \BOthers {.}}{%
{\protect \APACyear {2020}}%
}]{%
Zhang20:grl:phase}
\APACinsertmetastar {%
Zhang20:grl:phase}%
\begin{APACrefauthors}%
{Zhang}, X\BPBI J.%
, {Agapitov}, O.%
, {Artemyev}, A\BPBI V.%
, {Mourenas}, D.%
, {Angelopoulos}, V.%
, {Kurth}, W\BPBI S.%
\BDBL {}{Hospodarsky}, G\BPBI B.%
\end{APACrefauthors}%
\unskip\
\newblock
\APACrefYearMonthDay{2020}{{\APACmonth{10}}}{}.
\newblock
{\BBOQ}\APACrefatitle {{Phase Decoherence Within Intense Chorus Wave Packets
  Constrains the Efficiency of Nonlinear Resonant Electron Acceleration}}
  {{Phase Decoherence Within Intense Chorus Wave Packets Constrains the
  Efficiency of Nonlinear Resonant Electron Acceleration}}.{\BBCQ}
\newblock
\APACjournalVolNumPages{\grl}{47}{20}{e89807}.
\newblock
\begin{APACrefDOI} \doi{10.1029/2020GL089807} \end{APACrefDOI}
\PrintBackRefs{\CurrentBib}

\bibitem [\protect \citeauthoryear {%
X\BPBI J.~{Zhang}%
, {Angelopoulos}%
, {Artemyev}%
, {Hartinger}%
\BCBL {}\ \BBA {} {Bortnik}%
}{%
X\BPBI J.~{Zhang}%
, {Angelopoulos}%
\BCBL {}\ \protect \BOthers {.}}{%
{\protect \APACyear {2020}}%
}]{%
Zhang20:jgr:ulf}
\APACinsertmetastar {%
Zhang20:jgr:ulf}%
\begin{APACrefauthors}%
{Zhang}, X\BPBI J.%
, {Angelopoulos}, V.%
, {Artemyev}, A\BPBI V.%
, {Hartinger}, M\BPBI D.%
\BCBL {}\ \BBA {} {Bortnik}, J.%
\end{APACrefauthors}%
\unskip\
\newblock
\APACrefYearMonthDay{2020}{{\APACmonth{10}}}{}.
\newblock
{\BBOQ}\APACrefatitle {{Modulation of Whistler Waves by Ultra-Low-Frequency
  Perturbations: The Importance of Magnetopause Location}} {{Modulation of
  Whistler Waves by Ultra-Low-Frequency Perturbations: The Importance of
  Magnetopause Location}}.{\BBCQ}
\newblock
\APACjournalVolNumPages{Journal of Geophysical Research (Space
  Physics)}{125}{10}{e28334}.
\newblock
\begin{APACrefDOI} \doi{10.1029/2020JA028334} \end{APACrefDOI}
\PrintBackRefs{\CurrentBib}

\bibitem [\protect \citeauthoryear {%
X\BHBI J.~{Zhang}%
, {Angelopoulos}%
\BCBL {}\ \protect \BOthers {.}}{%
X\BHBI J.~{Zhang}%
, {Angelopoulos}%
\BCBL {}\ \protect \BOthers {.}}{%
{\protect \APACyear {2022}}%
}]{%
Zhang22:microbursts}
\APACinsertmetastar {%
Zhang22:microbursts}%
\begin{APACrefauthors}%
{Zhang}, X\BHBI J.%
, {Angelopoulos}, V.%
, {Mourenas}, D.%
, {Artemyev}, A.%
, {Tsai}, E.%
\BCBL {}\ \BBA {} {Wilkins}, C.%
\end{APACrefauthors}%
\unskip\
\newblock
\APACrefYearMonthDay{2022}{{\APACmonth{05}}}{}.
\newblock
{\BBOQ}\APACrefatitle {{Characteristics of Electron Microburst Precipitation
  Based on High-Resolution ELFIN Measurements}} {{Characteristics of Electron
  Microburst Precipitation Based on High-Resolution ELFIN
  Measurements}}.{\BBCQ}
\newblock
\APACjournalVolNumPages{Journal of Geophysical Research (Space
  Physics)}{127}{5}{e30509}.
\newblock
\begin{APACrefDOI} \doi{10.1029/2022JA030509} \end{APACrefDOI}
\PrintBackRefs{\CurrentBib}

\bibitem [\protect \citeauthoryear {%
X\BHBI J.~{Zhang}%
, {Artemyev}%
\BCBL {}\ \protect \BOthers {.}}{%
X\BHBI J.~{Zhang}%
, {Artemyev}%
\BCBL {}\ \protect \BOthers {.}}{%
{\protect \APACyear {2022}}%
}]{%
Zhang22:natcom}
\APACinsertmetastar {%
Zhang22:natcom}%
\begin{APACrefauthors}%
{Zhang}, X\BHBI J.%
, {Artemyev}, A.%
, {Angelopoulos}, V.%
, {Tsai}, E.%
, {Wilkins}, C.%
, {Kasahara}, S.%
\BDBL {}{Matsuoka}, A.%
\end{APACrefauthors}%
\unskip\
\newblock
\APACrefYearMonthDay{2022}{{\APACmonth{03}}}{}.
\newblock
{\BBOQ}\APACrefatitle {{Superfast precipitation of energetic electrons in the
  radiation belts of the Earth}} {{Superfast precipitation of energetic
  electrons in the radiation belts of the Earth}}.{\BBCQ}
\newblock
\APACjournalVolNumPages{Nature Communications}{13}{}{1611}.
\newblock
\begin{APACrefDOI} \doi{10.1038/s41467-022-29291-8} \end{APACrefDOI}
\PrintBackRefs{\CurrentBib}

\bibitem [\protect \citeauthoryear {%
X\BHBI J.~{Zhang}%
, {Chen}%
, {Artemyev}%
, {Angelopoulos}%
\BCBL {}\ \BBA {} {Liu}%
}{%
X\BHBI J.~{Zhang}%
\ \protect \BOthers {.}}{%
{\protect \APACyear {2019}}%
}]{%
Zhang19:jgr:modulation}
\APACinsertmetastar {%
Zhang19:jgr:modulation}%
\begin{APACrefauthors}%
{Zhang}, X\BHBI J.%
, {Chen}, L.%
, {Artemyev}, A\BPBI V.%
, {Angelopoulos}, V.%
\BCBL {}\ \BBA {} {Liu}, X.%
\end{APACrefauthors}%
\unskip\
\newblock
\APACrefYearMonthDay{2019}{{\APACmonth{11}}}{}.
\newblock
{\BBOQ}\APACrefatitle {{Periodic Excitation of Chorus and ECH Waves Modulated
  by Ultralow Frequency Compressions}} {{Periodic Excitation of Chorus and ECH
  Waves Modulated by Ultralow Frequency Compressions}}.{\BBCQ}
\newblock
\APACjournalVolNumPages{Journal of Geophysical Research (Space
  Physics)}{124}{11}{8535-8550}.
\newblock
\begin{APACrefDOI} \doi{10.1029/2019JA027201} \end{APACrefDOI}
\PrintBackRefs{\CurrentBib}

\bibitem [\protect \citeauthoryear {%
X\BPBI J.~{Zhang}%
, {Mourenas}%
\BCBL {}\ \protect \BOthers {.}}{%
X\BPBI J.~{Zhang}%
, {Mourenas}%
\BCBL {}\ \protect \BOthers {.}}{%
{\protect \APACyear {2020}}%
}]{%
Zhang20:grl:frequency}
\APACinsertmetastar {%
Zhang20:grl:frequency}%
\begin{APACrefauthors}%
{Zhang}, X\BPBI J.%
, {Mourenas}, D.%
, {Artemyev}, A\BPBI V.%
, {Angelopoulos}, V.%
, {Kurth}, W\BPBI S.%
, {Kletzing}, C\BPBI A.%
\BCBL {}\ \BBA {} {Hospodarsky}, G\BPBI B.%
\end{APACrefauthors}%
\unskip\
\newblock
\APACrefYearMonthDay{2020}{{\APACmonth{08}}}{}.
\newblock
{\BBOQ}\APACrefatitle {{Rapid Frequency Variations Within Intense Chorus Wave
  Packets}} {{Rapid Frequency Variations Within Intense Chorus Wave
  Packets}}.{\BBCQ}
\newblock
\APACjournalVolNumPages{\grl}{47}{15}{e88853}.
\newblock
\begin{APACrefDOI} \doi{10.1029/2020GL088853} \end{APACrefDOI}
\PrintBackRefs{\CurrentBib}

\bibitem [\protect \citeauthoryear {%
{Zhelavskaya}%
, {Aseev}%
\BCBL {}\ \BBA {} {Shprits}%
}{%
{Zhelavskaya}%
\ \protect \BOthers {.}}{%
{\protect \APACyear {2021}}%
}]{%
Zhelavskaya21}
\APACinsertmetastar {%
Zhelavskaya21}%
\begin{APACrefauthors}%
{Zhelavskaya}, I\BPBI S.%
, {Aseev}, N\BPBI A.%
\BCBL {}\ \BBA {} {Shprits}, Y\BPBI Y.%
\end{APACrefauthors}%
\unskip\
\newblock
\APACrefYearMonthDay{2021}{{\APACmonth{03}}}{}.
\newblock
{\BBOQ}\APACrefatitle {{A Combined Neural Network and Physics Based Approach
  for Modeling Plasmasphere Dynamics}} {{A Combined Neural Network and Physics
  Based Approach for Modeling Plasmasphere Dynamics}}.{\BBCQ}
\newblock
\APACjournalVolNumPages{Journal of Geophysical Research (Space
  Physics)}{126}{3}{e28077}.
\newblock
\begin{APACrefDOI} \doi{10.1029/2020JA028077} \end{APACrefDOI}
\PrintBackRefs{\CurrentBib}

\end{thebibliography}

\newpage

\end{document}